\documentclass[fleqn,usenatbib]{mnras}
\usepackage{capt-of}
\usepackage{caption}
\usepackage[bottom]{footmisc}
\usepackage{mathtools}
 \usepackage{multirow}
\usepackage{newtxtext,newtxmath}

\usepackage[T1]{fontenc}

\DeclareRobustCommand{\VAN}[3]{#2}
\let\VANthebibliography\thebibliography
\def\thebibliography{\DeclareRobustCommand{\VAN}[3]{##3}\VANthebibliography}

\usepackage{graphicx}	
\usepackage{amsmath}	
\usepackage{comment}

\title[Elemental abundances in open clusters using priors]{The GALAH survey: Elemental abundances in open clusters using joint effective temperature and surface gravity photometric priors}

\author[K. L. Beeson et al.]{Kevin L. Beeson$^{1}$\thanks{E-mail: klbeeson@gmail.com},
Janez Kos$^{1}$,
Richard de Grijs $^{2,3,4,5}$,
Sarah L. Martell$^{5,6}$,
Sven Bunder $^{5,7}$,
Gregor Traven$^{1}$,
\newauthor
Geraint F. Lewis$^{8}$,
Tayyaba Zafar$^{2,3}$, 
Joss Bland-Hawthorn$^{5,8}$,
Ken C. Freeman$^{5,7}$,
Michael Hayden$^{5,8}$,
\newauthor
Sanjib Sharma$^{9}$,
Gayandhi M. De Silva$^{2,3,5}$
\\
$^{1}$Faculty of Mathematics and Physics,Jadranska ulica 19, Ljubljana 1000, Slovenia\\
$^{2}$School of Mathematical and Physical Sciences, Macquarie University, Balaclava Road, Sydney, NSW 2109, Australia\\
$^{3}$Astrophysics and Space Technologies Research Centre, Macquarie University, Balaclava Road, Sydney, NSW 2109, Australia\\
$^{4}$International Space Science Institute--Beijing, 1 Nanertiao, Zhongguancun, Hai Dian District, Beijing 100190, China\\
$^{5}$ARC Centre of Excellence for All Sky Astrophysics in 3 Dimensions (ASTRO 3D), Australia\\
$^{6}$School of Physics, University of New South Wales, Sydney, NSW 2052, Australia\\
$^{7}$Research School of Astronomy and Astrophysics, The Australian National University, Canberra, ACT 2611, Australia\\
$^{8}$Sydney Institute for Astronomy, School of Physics, A28, The University of Sydney, NSW 2006, Australia\\
$^{9}$Space Telescope Science Institute, 3700 San Martin Drive, Baltimore, MD 21218\\
}

\date{Accepted XXX. Received YYY; in original form ZZZ}

\pubyear{2023}

\begin{document}
\label{firstpage}
\pagerange{\pageref{firstpage}--\pageref{lastpage}}
\maketitle

\begin{abstract}
The ability to measure precise and accurate stellar effective temperatures ($T_{\rm{eff}}$) and surface gravities ($\log(g)$) is essential in determining accurate and precise abundances of chemical elements in stars. Measuring $\log(g)$ from isochrones fitted to colour-magnitude diagrams of open clusters is significantly more accurate and precise compared to spectroscopic $\log(g)$. By determining the ranges of ages, metallicity, and extinction of isochrones that fit the colour-magnitude diagram, we constructed a joint probability distribution of $T_{\rm{eff}}$ and $\log(g)$. The joint photometric probability shows the complex correlations between $T_{\rm{eff}}$ and $\log(g)$, which depend on the evolutionary stage of the star. We show that by using this photometric prior while fitting spectra, we can acquire more precise spectroscopic stellar parameters and abundances of chemical elements. This reveals higher-order abundance trends in open clusters like traces of atomic diffusion. We used photometry and astrometry provided by the \textit{Gaia} DR3 catalogue, Padova isochrones, and Galactic Archaeology with HERMES (GALAH) DR4 spectra. We analysed the spectra of 1979 stars in nine open clusters, using MCMC to fit the spectroscopic abundances of 26 elements, $T_{\rm{eff}}$, $\log(g)$, $v_{\rm{mic}}$, and $v_{\rm{broad}}$. We found that using photometric priors improves the accuracy of abundances and $\log(g)$, which enables us to view higher-order trends of abundances caused by atomic diffusion in M67 and Ruprecht 147.

\end{abstract}




\section{Introduction}

Open clusters provide a unique environment to study the chemical abundances of stars. With open clusters, we are given a set of tens to a few thousand stars that were born in the same star forming region, thus, they are chemically homogeneous \citep{desilvaChemicalHomogeneityHyades2006,desilvaChemicalHomogeneityCollinder2007,fengEarlyTurbulentMixing2014}. Open clusters are quickly disrupted but they remain kinematically homogeneous for a few hundred million years, with the oldest open clusters observed being 2-3 Gyrs \citep{bubarSpectroscopicAbundancesMembership2010,rangwalAstrometricPhotometricStudy2019,sariyaAstrometricPhotometricInvestigation2021}. This allows for stars in the open cluster core to be easily distinguished from field stars using their kinematics. The kinematic and chemical homogeneity means open clusters have been used as a place to benchmark distances \citep{bailer-jones_estimating_2021}, measure their ages with high precision \citep{jefferyBAYESIANANALYSISAGES2016}, study the role of physical processes such as atomic diffusion that will affect the chemical abundances of stars \citep{liu_chemical_2019}, examine the possibility of strong chemical tagging \citep{casamiquela_abundance-age_2021}, study stellar rotation rates of entire stellar populations \citep{sunTidallockinginducedStellarRotation2019,heRoleTidalInteractions2023}, and study the chemical trends of the galactic disk \citep{neyerTHESANProjectConnecting2023}.

 Chemical tagging is the idea that by measuring chemical abundances, one can group stars with similar chemical signatures into physically meaningful populations \citep{freeman_new_2002}. There are two types of chemical tagging: weak and strong. Weak chemical tagging groups together stars that originated from the same galactic environment. An example is tagging to find the remnants of disrupted satellite galaxies \citep{freeman_new_2002,buderGALAHSurveyChemical2022}. The second type is strong chemical tagging, the grouping of stars that originated from the same birth cluster. The possibilities of strong chemical tagging have been explored in \citet{casamiquela_differential_2020,dotter_influence_2017,ting_payne_2019,casamiquelaImPossibilityStrong2021}. If strong chemical tagging can be done, it will give major insights into the dynamical history of stars in our Galactic disk. 

The extent of chemical homogeneity and the uniqueness of each open cluster's abundance pattern are essential in determining how viable chemical tagging is. Open clusters are believed to be chemically homogeneous when initially created, as any chemical signatures in the star forming region have time to homogenise and star formation has finished before any supernovas can occur, which can change the chemical composition of the star forming region \citep{bland-hawthornLongtermEvolutionGalactic2010,fengEarlyTurbulentMixing2014}. However, during the star's life, some effects can cause chemical inhomogeneities to occur. One of the causes of chemical inhomogeneities is the existence of planets around the star. As rocky planets are formed, they will be made out of elements with a high condensation temperature, causing the abundance of these elements in the star to be lower. As some of the planets have a highly eccentric orbit, they will be reabsorbed during the star's life cycle, which causes the surface abundances of the elements with a high condensation temperature to increase \citep{melendezPeculiarSolarComposition2009,liuDetailedChemicalComposition2016,spinaChemicalInhomogeneitiesPleiades2018}.

Atomic diffusion is another factor contributing to the chemical inhomogeneities of stars within open clusters. Atomic diffusion is a physical process where gravitational settling driven by temperature and pressure gradients concentrates heavier elements towards the centre of the star while radiative acceleration pushes elements away from the centre \citep{michaudLithiumAbundanceConstraints1984,michaudAtomicDiffusionStars2015}. The effects of atomic diffusion are most significant as a star reaches the turn-off-point on the colour-magnitude diagram. As a star reaches its turn-off point, the effects of gravitational settling will be greater compared to the radiative acceleration, and consequently the concentration of heavy elements will decrease as they sink from the surface layer to the convection region. After the star has passed the turn-off point, the convection zone will deepen. As a result, the elements that have sunk to the convection region will now be found in the surface layer. This change will increase the concentration of these elements once again on the surface layer. Atomic diffusion is one of the drivers that reduces the chemical homogeneity of stars in open clusters. In M67, atomic diffusion will create inhomogeneity to a 0.2 dex level for certain elements \citep{liu_chemical_2019}.

Surface gravity is used as a marker of how far along the star is in its evolutionary stage. The uncertainty of spectroscopic measurements of $\log(g)$ is typically on the order of 0.1-0.25 dex \citep{jofre_accuracy_2019}. This is because $\log(g)$ only has a weak effect on the observed spectra. An uncertainty of 0.25 dex in $\log(g)$ will significantly affect where we view the star in its evolutionary stage. If the measured $\log(g)$ is imprecise, the effect of atomic diffusion will be observed as scatter in abundances rather than a trend in abundances with respect to $\log(g)$.

Photometry has been used to measure much more accurate and precise $\log(g)$ than spectroscopy for stars in open clusters. In \citet{kosDiscovery21Myr2019,kos_galah_2021}, they fitted Padova \citep{bressanPARSECStellarTracks2012a,chenImprovingPARSECModels2014,chenParsecEvolutionaryTracks2015,marigoNewGenerationPARSECCOLIBRI2017,pastorelliConstrainingThermallyPulsing2019,tangNewPARSECEvolutionary2014} isochrones to HR diagrams of open clusters. The isochrones were fitted very precisely, as the open clusters had stars at various evolutionary stages. The nearest point in \textit{Gaia}'s three photometric bands ($G$, $G_{BP}$, and $G_{RP}$) space from the star to the isochrone is used to measure $T_{\rm{eff}}$ and $\log(g)$. Using the photometric $T_{\rm{eff}}$ and $\log(g)$ as priors when fitting abundances and stellar parameters to spectra increased the accuracy of $\log(g)$ and the precision of abundances.

\textit{Gaia} \citep{gaia_collaboration_gaia_2016,2023A&A...674A...1G} has transformed the study of open clusters. The precise astrometry has significantly improved cluster membership identification, leading to the rejection of 50\% of the pre-\textit{Gaia} era open cluster listings \citep{cantat-gaudinGaiaDR2View2018a}. Furthermore, \textit{Gaia}'s precise kinematic data has provided unique insights into the three-dimensional structure of open clusters \citep{armstrongStructure3DKinematics2022}. By combining precise photometry and astrometry, the mission has enabled accurate age determinations for open clusters \citep{bossini_age_2019}. In the third data release, \textit{Gaia} has released the full astrometric solution (positions on the sky, parallax, and proper motion) for 1.46 billion sources and photometry in its three bands for 1.54 billion sources. 

Many large spectroscopic surveys with a high resolution (R$> 20 \, 000$) have targeted open clusters as part of their observing time. A notable example is the \textit{Gaia}-ESO survey \citep{randich_gaia-eso_2013,gilmore_gaia-eso_2012} which has dedicated a significant fraction of time to target open clusters. \textit{Gaia}-ESO has provided a homogeneous set of 65 open clusters \citep{bragagliaGaiaESOSurveyTarget2022}. GALAH \citep{de_silva_galah_2015} and APOGEE \citep{majewski_apache_2017} have targeted open clusters as part of their calibration process \citep{kosGALAHSurveyChemical2018,donorOpenClusterChemical2018a} and have often observed open cluster stars serendipitously. These large spectroscopic surveys produce cutting-edge abundance measurements and spectral parameters for these stars. Consequently, they enable the comprehensive exploration of abundance trends and the assessment of homogeneity within a diverse range of open clusters.

This paper aims to explore the effect of using a joint $T_{\rm{eff}}$ and $\log(g)$ photometric prior on spectroscopic $T_{\rm{eff}}$, $\log(g)$, micro-turbulence ($v_{\rm{mic}}$), broadening velocity ($v_{\rm{broad}}$), and 26 elemental abundances. We first explore how to calculate better photometric $T_{\rm{eff}}$ and $\log(g)$ by improving the calculated absolute magnitude. The absolute magnitude is improved by optimising the precision and accuracy of how we measure individual distances to stars in open clusters. Then, we examine how the joint photometric prior changes between stars at different evolutionary stages. To determine the effect of using a joint $T_{\rm{eff}}$ and $\log(g)$ for spectroscopic fitting, we fit the spectra of 1979 stars in nine open clusters with and without a photometric prior. We compare the scatter in abundances in these two methods and determine whether higher-order trends such as atomic diffusion are visible in both methods. We examine the validity of the precision of abundances and stellar parameters calculated using Markov chain Monte Carlo (MCMC) and we examine how well GALAH DR4 measures its precision.

The structure of the paper is as follows: Sec. \ref{Theory} describes how we selected the members of the open clusters using astrometry and photometry from the Gaia DR3 catalogue, our method to improve \textit{Gaia} DR3 distances to stars in open clusters using a cluster centre distance prior, our technique to calculate the $T_{\rm{eff}}$ and $\log(g)$ joint probability from isochrones fitted to HR diagrams, and our approach to generate synthetic spectra and fit stellar parameters and elemental abundances to GALAH spectra using the joint $T_{\rm{eff}}$ and $\log(g)$ photometric prior. Sec. \ref{Results} describes the distance improvements after using the cluster centre prior, the best-fitting isochrones, and the resulting photometric $T_{\rm{eff}}$ and $\log(g)$ joint probability distributions, and the effect of the photometric $T_{\rm{eff}}$ and $\log(g)$ joint probability prior on spectroscopic abundances. The section also includes a comparison between our abundances and GALAH DR4 (Buder et al. 2023, in preparation). Sec. \ref{spectroscopic uncertainties} describes our spectroscopic precisions and the ways we verified them, we also examine how well GALAH DR4 estimates their abundance precision. Sec. \ref{discussion} is our discussion of the implications of our results, and Sec. \ref{congclusion} collects our conclusions.

\section{Methods and data}\label{Theory}
To create a joint $T_{\rm{eff}}$ and $\log(g)$ photometric prior to get better abundances for stars in open clusters, we first have to tackle the problem of how to acquire the best photometry, specifically the best absolute magnitudes given the measured observed magnitudes and parallaxes. This process will be elaborated upon in the next section, \ref{Calculating better distances to open clusters}. We describe selecting the open cluster stars through \textit{Gaia} and which priors we used to make the distance measurements more accurate and precise. After acquiring these improved absolute magnitudes, we need to be able to get representative stellar parameters and their uncertainties. In \ref{calculating photometric stellar parameters}, we describe how we calculated photometric stellar parameters and their precision from individual isochrones and how to construct a joint $T_{\rm{eff}}$ and $\log(g)$ probability distribution from the ranges of ages, metallicities, and extinction obtained from the fitting isochrones to HR diagrams of open clusters. Lastly in Sec. \ref{spectroscopic fitting}, we describe how we synthesised spectra and how we fitted spectra using photometric stellar parameters as priors.

\subsection{Calculating better distances to open clusters}\label{Calculating better distances to open clusters}

Using the \textit{Gaia} DR3 catalogue, we get access to photometry in the $G$, $G_{BP}$, and $G_{RP}$ bands, and the astrometry (ranging from proper motions and parallax to positions) of over 1.46 billion objects using 34 months of observation. The catalogue gives us an excellent way to probe open clusters with high completeness of objects brighter than $G=19$ \citep{fabriciusGaiaEarlyData2021}. The \textit{Gaia} DR3 catalogue provides excellent photometry with an average uncertainty of 1, 12, and 6 mmag for the $G$, $G_{BP}$, and $G_{RP}$ bands respectively at $G=17$ mag \citep{rielloGaiaEarlyData2021}. The parallax provided has a median uncertainty of 0.07 mas at $G=17$ \citep{lindegrenGaiaEarlyData2021}.

Using absolute magnitudes of stars in open clusters, we can make HR diagrams of said clusters and fit isochrones to them. Using the isochrones, we can calculate the $T_{\rm eff}$ and $\log(g)$ of the stars. The two sources of uncertainties in the absolute magnitude come from the measurement uncertainty of apparent magnitudes and the uncertainty in the distance of the star (the parallax). The latter is the main contributor to the absolute magnitude uncertainty. Even in a nearby cluster such as Melotte 22 ($\sim$ 130 pc), the contribution from the distance uncertainty is around 13 times larger than the contribution from the apparent magnitude measurement.

In the next subsection, we explain our selection function to acquire \textit{Gaia} DR3 open cluster stars, and in \ref{distances} we explain our updated cluster distance prior to increase the accuracy and precision of the parallax.

\subsubsection{Selecting open cluster members}\label{selecting gaia}
 
We decided to only use the stars that are easily differentiable from field stars in proper motion space to minimise contamination. The population obtained will mainly comprise stars from the open cluster core. We selected the stars that belong to each open cluster from the \textit{Gaia} DR3 archive. To do this, we use several criteria, we represent the searched parameters in brackets:
\begin{itemize}
  \item Using a conical search around the centre of the cluster to limit ($\alpha$ (\texttt{ra}), $\delta$ (\texttt{dec})) of each star. The size of the conical search was determined by eye using the \textit{Gaia} data.
  \item Constraining the parallax by introducing a minimum and a maximum parallax. The parallaxes and their maximum and minimum values were selected by eye using the \textit{Gaia} data.
  \item Limiting $\mu_\alpha\cos(\delta)$ (\texttt{pmra}) and $\mu_\delta$ (\texttt{pmdec}) using maximum $\mu_\delta$ and $\mu_\alpha\cos(\delta)$. This is done as open cluster members still share the same kinematic values. The $\mu_\alpha\cos(\delta)$ and $\mu_\delta$ and their maximum and minimum values were selected by eye using the \textit{Gaia} data.
\end{itemize}

\begin{table*}
    \caption{Our parameters used for finding cluster members. The centre of our positional conical limits is indicated by the $\alpha$ and $\delta$ values, and the radius of the conical search by $d$. Our centre values and limits for pmra, pmdec, and parallax are indicated in the $\mu_\alpha\cos(\delta)$, $\mu_\delta$, and $\varpi$ columns. The values of the centres of the clusters are taken from \citet{wengerSIMBADAstronomicalDatabase2000} except for ASCC 16 which was taken from \citet{paunzenWEBDAToolCP2008}. $N_{\it{Gaia}}$ and $N_{\rm{GALAH}}$ are the number of observations from \textit{Gaia} and GALAH respectively. The last column indicates the number of unique stars observed by GALAH. Note that the parallax centres are not the actual centres of the cluster; they are just used as a boundary to exclude stars that are not in the cluster.}
    \centering
    \begin{tabular}{lccccccccc}
        \hline
        Name & $\alpha$ & $\delta$ & $d$ & $\mu_\alpha\cos(\delta)$ & $\mu_\delta$ & $\varpi$ & $N_{\it{Gaia}}$ & $N_\mathrm{GALAH}$& unique GALAH stars \\ \hline
         & deg. & deg. & deg. & $\mathrm{mas}\, \mathrm{yr}^{-1}$ & $\mathrm{mas}\, \mathrm{yr}^{-1}$ & mas & & &\\
        \hline\hline
ASCC 16 & 81.2 & 1.8 & 0.5 & 1.4 $\pm$ 0.9 & $-0.2$$\pm$1.5 & 2.9$\pm$0.5 & 247 & 173 & 63 \\ 
        Blanco 1 & 0.85 & $-30.0$ & 1.0 & 18.6$\pm$1.1 & 2.7$\pm$1.0 & 4.19$\pm$0.3 & 328 & 109 & 63 \\ 
        Melotte 22 & 56.6 & 24.1 & 1.5 & 19.8$\pm$ 3.5 & $-45.3$$\pm$3.2 & 7.29$\pm$0.3 & 552 & 91 & 91 \\ 
        NGC 2516 & 119.5 & $-60.8$ & 0.8 & $-4.6$$\pm$1.9 & 11.3$\pm$1.7 & 2.41$\pm$0.4 & 1686 & 204 & 146 \\ 
        NGC 2632 & 130.1 & 19.6 & 1.1 & $-35.8$$\pm$5.5 & $-12.8$$\pm$4.1 & 5.35$\pm$0.4 & 913 & 147 & 147 \\ 
        M67 & 132.8 & 11.8 & 0.33 & $-11.0$$\pm$0.7 & $-3.0$$\pm$0.55 & 1.1$\pm$0.4 & 1389 & 922 & 291 \\ 
        Ruprecht 147 & 289.1 & $-16.3$ & 0.3 & $-0.9$$\pm$1.4 & $-26.6$$\pm$1.6 & 3.0$\pm$0.3 & 118 & 130 & 59 \\
        Trumpler 10 & 131.9 & $-42.6$ & 2.25 & $-12.4$$\pm$2.0 & 6.5$\pm$1.2 & 2.5$\pm$1.0 & 1190 & 119 & 119 \\ 
        Upper Scorpius & 243.0 & $-23.4$ & 2.0 & $-12$$\pm$12.0 & $-25.0$$\pm$5.0 & 1.5$\pm$2.0 & 401 & 84 & 61 \\ \hline
    \end{tabular}
    \label{tab:cluster_centres}

\end{table*}

We also implement quality cuts to make sure the chosen observations of cluster members have the required quality:
\begin{itemize}
    \item Visibility periods ($\texttt{visibility\_periods\_used}$) $\geq$ 7 as recommend in \citet{lindegren_gaia_2018}.
    \item Mean flux over error (\texttt{phot\_g\_mean\_flux\_over\_error}, \texttt{phot\_bp\_mean\_flux\_over\_error}, and \texttt{phot\_rp\_mean\_flux\_over\_error}) $\geq$ 25,10,10 for the $G$, $G_{RP}$, and $G_{BP}$ bands respectively.
    \item Limits on precision on the blue and red magnitudes (G$_{\rm BP}$ and G$_{RP}$) excess factors (\texttt{phot\_bp\_rp\_excess\_factor})  as recommend in \citet{gaia_collaboration_gaia_2018}.
    \item Limits on the goodness of fit of the parallaxes using the chi squared fit on the astrometry (\texttt{astrometric\_chi2\_al}) as recommend in \citet{gaia_collaboration_gaia_2018}.
\end{itemize}

The ADQL code for selection is provided in Appendix \ref{section:ADQL code}. The values used in the selection criteria and quality cuts are shown in table \ref{tab:cluster_centres}. To create the list of open clusters, we selected the nine clusters with the highest number of observed stars in GALAH DR4.

\subsubsection{Calculating distances to open clusters}\label{distances}

The recommended method to calculate distances for individual stars is to use the likelihood and priors in \cite{bailer-jones_estimating_2021}. The likelihood is shown in equation \ref{eqn:Original likelihood}. The paper introduces a geometric prior and a photometric prior. The geometric prior uses our knowledge of the empirical distribution of stars based on their celestial position, and the photometric prior uses the fact that the colour space of the stars is not uniformly distributed with respect to distance. We have decided to only use the geometric prior. The photometric prior, as described in \citet{bailer-jones_estimating_2021} applies to individual stars rather than a cluster.

The following equations are used for estimating distances to individual stars using the method outlined in \citet{bailer-jones_estimating_2021}.

\begin{equation}
\label{eqn:Original likelihood}
	\begin{aligned}
	P(r|\varpi,\sigma_\varpi)=\begin{cases} K\mathcal{N}(\frac{1}{r}-\varpi+\varpi_0,\sigma_\varpi^2)&  \text{if }r>0\\
	0& \text{if } r<0 
	\end{cases}
	\end{aligned}
\end{equation}
\begin{equation}
\label{eqn:individual distance equation}
P(r|\varpi,\sigma_\varpi\alpha,\beta,L)=	P(r|\varpi,\sigma_\varpi)P(r|\alpha,\beta,L)
\end{equation}
\begin{equation}
\label{eqn:geometric prior}
	\begin{aligned}
    P(r|\alpha,\beta,L)=\begin{cases}Kr^{\beta}\text{exp}(-r/L)^\alpha&  \text{if }r>0\\
	0& \text{if } r<0 
	\end{cases}
    \end{aligned}
\end{equation}
Equation \ref{eqn:individual distance equation} describes the probability of a star being at a distance $r$, given its parallax($\varpi$), uncertainty in parallax ($\sigma_{\varpi}$), $\alpha$, $\beta$, and $L$. $\alpha$, $\beta$, and $L$ are the parameters of the empirical probability distribution of stars in the galaxy 
 modelled using a three-parameter Generalised Gamma Distribution, their values depend on the galactic longitude and latitude. Thus, for every star with a different line of sight their $\alpha$, $\beta$, and $L$ values will be different. A fuller description of $\alpha$, $\beta$, and $L$ can be found in \citet{bailer-jones_estimating_2021}. The first component of equation \ref{eqn:individual distance equation} is likelihood shown in equation \ref{eqn:Original likelihood} for an individual star at $r$, with $\varpi$ and $\varpi_0$ (zero point correction), $\mathcal{N}$ is a normal distribution, and $K$ is the normalising constant. The zero point correction is used as the zero point of the parallax in \textit{Gaia} is not at 0. The zero point correction is dependent on the celestial position and the number of astrometry points. It varies between $-0.150$ and $+0.130$ mas \citep{lindegrenGaiaEarlyData2021}, and it varies as a function of stellar type \citep{renGaiaEDR3Parallax2021}. Geometric prior is shown in equation \ref{eqn:geometric prior}.

To decrease the distance uncertainty we will use the fact that the stars are in the open cluster's core. If we assume the distribution of stars in the cluster is a normal distribution around the cluster centre and if we know the distance to the centre of the cluster ($c$) and the radial size of the cluster ($s$), we will be able to create a cluster distance prior for the cluster. We decided to use a normal distribution to model the open cluster core rather than a more complicated distribution such as the Elson, Fall, and Freeman profile \citep{elsonStructureYoungStar1987a}, as the difference between these profiles in the resulting cluster centre distance and size will not be large enough to significantly affect the resulting absolute magnitude. 

Therefore we shall first calculate the probability distribution of $s$ and $c$. We introduce another prior for the radial size of the cluster. This is done so the radial cluster size can not be negative or too large; the prior used is a gamma distribution as proposed in \citet{luri_gaia_2018}. The cluster distance prior and the cluster size prior are shown below.

\begin{equation}\label{eqn:open cluster star distribution}
P(r|s,c)=\mathcal{N}(r-c,s^2)
\end{equation}

\begin{equation}\label{eqn:cluster size prior}
\text{Cluster size prior}= \chi(s|k,\theta)=\frac{x^{k-1} \exp^{-\frac{x}{\theta}}}{\Gamma(k)\theta^{k}} 
\end{equation}
Where shape ($k$) and scale ($\theta$) are the parameters of the cumulative distribution function of the Gamma distribution. We have chosen a $k$ and $\theta$ that make the mean of the Gamma distribution 10 pc and a variance of 0.05 pc, which are the recommended values in \citet{luri_gaia_2018}.

By combining equation \ref{eqn:individual distance equation}, \ref{eqn:open cluster star distribution}, and \ref{eqn:cluster size prior}, marginalising over $r$ we can get the probability distribution of $s$ and $c$ for a single star:

\begin{equation}
\begin{split}
\mathbf{P}&(s,c|\varpi,\sigma_{\varpi},L,k,\theta)=\\ 
 & \int_{0}^{\infty}\mathbf{P}(r_i|\varpi,\sigma_{\varpi_i},\alpha,\beta,L) \cdot \mathcal{N}(r_{i}-c,s^2) dr_i \cdot \chi(s|k,\theta)
\end{split}
\end{equation}

To calculate the probability distribution of $s$ and $c$ for a cluster, we convolve the probability distribution of each star to make:

\begin{equation}
\label{eqn:likelihood cluster}
\begin{split}
\mathbf{L}&(s,c|\boldsymbol{\varpi},\boldsymbol{\sigma_{\varpi}},\boldsymbol{\alpha},\boldsymbol{\beta},\mathbf{L},k,\theta)=\\ 
 &\prod_{i} \int_{0}^{\infty}\mathbf{P}(r|\varpi_i,\sigma_{\varpi_i},\alpha_i,\beta_i,L_i) \cdot \mathcal{N}(r_{i}-c,s^2) dr \cdot \chi(s|k,\theta)
\end{split}
\end{equation}
With said probability distribution, we can now find more accurate individual distances by combining equations \ref{eqn:individual distance equation} and \ref{eqn:open cluster star distribution} with $L(s,c)$ and marginalising a star's radial position through $s$ and $c$.

\begin{equation}
\label{eqn:final distance equation}
\begin{aligned}
    \mathbf{L}(r|\varpi,\sigma_\varpi)&=\begin{cases}\qquad\qquad\text{if }r>0:\\
    \iint \mathbf{L}(s,c)P(r|s,c)\mathbf{L}(r|\varpi,\sigma_\varpi)\, dc \,ds& \\
    \qquad\qquad\text{if } r<0:\\
	0 &
	\end{cases} 
	\end{aligned}
\end{equation}
To implement these equations, we used the MCMC package emcee \citep{emcee}. We first sampled the probability space of $s$ and $c$ using equation \ref{eqn:likelihood cluster}. We used four walkers in total and multi-processed the integration process for each star. To know if the space was sampled correctly, we required the sampling iteration over auto-correlation time to be at least 17 iterations. With a 64-core computing cluster, even with the most numerous open clusters, it takes around 2 hours to sample. For equation \ref{eqn:final distance equation} we used the same convergence limits as for the previous equations. Like the distance probability distributions found in \citet{bailer-jones_estimating_2021}, we found that they do not follow a normal distribution; however, as the spread of the distances is so small, we still treated them as normal distributions and found our uncertainty in distance and our peak using a normal distribution fit.

\subsection{Calculating photometric stellar parameters}
\label{calculating photometric stellar parameters}

In this section, we describe the theory and implementation behind calculating photometric $T_{\rm{eff}}$ and $\log(g)$ and their errors for individual stars from an isochrone. In section \ref{subsec: photometric errors}, we explain how by determining the ranges of metallicity, ages, and extinctions of isochrones that will fit the HR diagrams of an open cluster; we can generate the joint $T_{\rm{eff}}$ and $\log(g)$ probability distribution of an individual star.

\subsubsection{Calculating $T_{\rm eff}$ and $\log(g)$ from individual isochrones}\label{subsec: calculating photometric stellar parameters}

Using the more precise and accurate absolute magnitudes, we now better fit isochrones to the open cluster HR diagrams. We have decided to fit the Padova CMD 3.7 isochrones by eye; this method has been shown to be highly effective for fitting in \citet{kos_galah_2021,kosDiscovery21Myr2019}. We chose to do this rather than using Bayesian fitting programs like BASE-9 (\citet{hippelInvertingColorMagnitudeDiagrams2006} and see also \citet{jefferyBAYESIANANALYSISAGES2016}) as we could not provide accurate complementary data, such as a binary star indication or individual reddening values. Another problem is that the Padova isochrones still do not fit the HR-diagram in certain sections, such as the lower part of the main sequence, even when using Bayesian fitting programs \citep{brandnerAstrophysicalProperties6002023}. Irrespective we are confident in the fitting and the probability distribution that comes with it. When fitting by eye, we use isochrones with a spacing of 0.01 dex for both log ages and metallicity.

Using an isochrone we can find the $T_{\rm eff}$ and $\log(g)$ for an individual star by using:

\begin{equation}
\label{eq:HR weighted teff}
\mathbfit{T}_{\rm eff}= \frac{1}{\textbf{K}}\int T_{\rm eff}(m)\prod_{\mathclap{i=G,BP-RP}}\mathcal{N}(M_i-M_i(m),\sigma_{M_i})n(m)\thinspace dm
\end{equation}
Where $m$ is the initial mass of a star in the open cluster it is used as a variable to parameterise the shape of the isochrone. $n(m)$ is the density of stars per unit $m$ that we should observe at that point in the isochrone and $K$ is the normalisation constant. $M$ is the absolute $G$ magnitude or the $BP-RP$ colour, depending on the subscript. $\sigma_M$ is the uncertainty in absolute magnitude. $n(m)$ is included as the theoretical density of stars along the isochrone is not constant. We integrate along the isochrone. An analogous equation is used for $\log(g)$. Note that in the equation, we are combining $G_{BP}$ and $G_{RP}$ bands into $G_{BP-RP}$ unlike in \citet{kosDiscovery21Myr2019,kos_galah_2021}, where they used the nearest point in all three photometric bands. We did this as our testing shows at the turn-off point, some stars' absolute magnitude's can be close to one of the $G_{BP}$ and $G_{RP}$ bands but not close to the $G_{BP}-G_{RP}$ axis, therefore creating anomalous results when viewed on a colour-magnitude diagram. This line integral takes into account the photometric uncertainty and the number density of stars one should observe at that point in the isochrone.

After acquiring the $T_{\rm{eff}}$ and $\log(g)$, we can calculate the precision of the derived parameters from the magnitude uncertainties by modifying the equation above:

\begin{equation}
\label{eqn:errors isochrone individual teff}
	\begin{aligned}
        \sigma_T^2= \frac{1}{\textbf{K}}\int (\mathbfit{T}_{\rm eff}-T_{\rm eff}(m))^2\prod_{\mathclap{i=G,BP-RP}}\mathcal{N}(M_i-M_i(m),\sigma_{M_i})n(m)\thinspace dm
    \end{aligned}  
\end{equation}
The Padova isochrone is sampled at discrete intervals. To make sure that the sampling along the isochrone does not affect our measurement, we interpolated the provided isochrone. We measured the change in $T_{\rm{eff}}$ or $\log(g)$ between the interpolated isochrone and non-interpolated one. If the change was below 1\% of the precision, we consider the isochrone to be well sampled. If not, the interpolation process is repeated using the integrated isochrone as a starting isochrone. When interpolating we do it with respect to the initial starting mass of the star. For a cluster of 1000 stars, it takes, on average, 10 to 20 minutes to calculate the parameters and their associated precision on a 144-core computer cluster.

\subsubsection{Using the uncertainty in the isochrone fitting to create the joint $T_{\rm{eff}}$ and $\log(g)$ probability distribution}\label{subsec: photometric errors}

\begin{figure}
    \centering
    \includegraphics{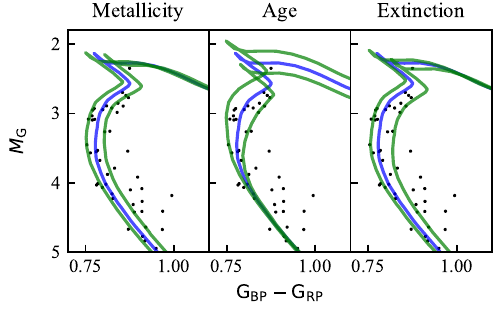}
    \caption{Isochrones when changing the metallicity, age, and extinction for Ruprecht 147. Blue line shows the best fit and the green lines show the maximum and minimum values parameters we have chosen. The chosen ranges are Age$=9.40^{+0.04}_{-0.05}$ log(age), metallicity $=0.08^{+0.12}_{-0.08}$ dex, and extinction $\it{A_{\rm{V}}}$ $=0.40 ^{+0.10}_{-0.05}$ mag.}
    \label{fig:Ranges_ruprehct}
\end{figure}

When fitting the isochrone, we vary three main parameters: age, metallicity, and extinction. We start by finding the optimum values of parameters, and then we vary parameters one by one until around 85\% of the stars that are in the area of high sensitivity to the change are encompassed (an example is shown in Fig. \ref{fig:Ranges_ruprehct}). These areas are normally around the turn-off point or the red giant clump. The best-fitted isochrone and the two extreme isochrones define an asymmetric normal distribution of parameters. We use an asymmetric normal distribution as the range of metallicities, ages, and extinctions can differ depending on whether you are increasing or decreasing the parameter. These ranges are wide enough to encompass the extended main sequence turn-off stars.

By using the range of isochrones, we can sample the probability density ($\rho (T_{\rm{eff}},\rm{log}(g))$), as each isochrone will generate a particular $T_{\rm{eff}}$ and $\rm{log}(g)$. Each of these samples will have its own precision associated with it, which will differ depending on the shape of the isochrone which depends on the metallicity, age, and extinction. Thus when calculating the probability of the star having a $T_{\rm{eff}}=T_1$ and $\rm{log}(g)=g_1$ from the probability density ($\rho (T_{\rm{eff}},\rm{log}(g))$) each sample will contribute a probability of 
\begin{equation}
    P(T_1,g_1)=\mathcal{N}(T_1-T_{\rm{eff}},\sigma_{T_{\rm{eff}}})\mathcal{N}(g_1-\rm{log}(g),\sigma_{\rm{log}(g)}).
\end{equation} 

To find the overall probability we marginalise over $T_{\rm{eff}}$ and $\rm{log}(g)$:

\begin{equation}\label{eqn:theoretical stellar spread}
\begin{split}
      P(T_1,g_1)=& \iint \rho  (T_{\rm{eff}},\rm{log}(g))\mathcal{N}(T_1-T_{\rm{eff}},\sigma_{T_{\rm{eff}}})\\
      &\mathcal{N}(g_1-\rm{log}(g),\sigma_{\rm{log}(g)})dT_{\rm{eff}}d\rm{log}(g)  
\end{split}
\end{equation}

 We found that the uncertainty we got from each isochrone is much smaller than the spread of $T_{\rm{eff}}$ and $\rm{log}(g)$ from varying ages, metallicities, and extinctions. As a result the normal distributions act like delta functions around $T_1$ and $g_1$ so we can approximate the equation as:
\begin{equation}\label{eqn:final photometric equation}
P(T_1,g_1)= \rho  (T_1,g_1)
\end{equation}
where $\rho$ is obtained using \texttt{scipy's} Gaussian KDE fitting function \citep{2020SciPy-NMeth}. To take into account the individual precisions obtained during fitting, we used them as weights during the Gaussian KDE fitting. Using equation \ref{eqn:final photometric equation}, we have a fast way to sample the probability space of photometric stellar parameters. We decided to use 1000 isochrones to sample the space generated from the asymmetric normal distribution of ages, metallicity, and extinction. It took 2 hours for a 1000-star open cluster to be fully sampled on a 72 core computing cluster.

\subsection{Spectroscopic fitting}\label{spectroscopic fitting}

We have chosen to use the GALAH survey as our source of spectroscopic data. GALAH is a spectroscopic survey employing the HERMES fibre-fed multi-object spectrograph, which is installed on the 3.9-meter Anglo-Australian Telescope (AAT). With the release of GALAH DR4, the survey has observed more than $900\,000$ objects and provided stellar parameters $T_{\rm{eff}}$, $\log(g)$, [Fe/H], $v_{\rm{mic}}$, $v_{\rm{broad}}$ and radial velocity ($v_{\rm{rad}}$), and abundances of 30 elements. The survey observes in 4 bands: 4718-4903 \AA \,(blue band), 5649-5873 \AA \,(green band), 6481-6739 \AA \,(red band), and 7590-7890 \AA \,(IR band). Each band has a nominal resolving power of $R=28 \,000$.

\subsubsection{Creating synthetic spectra}\label{subsection:creating_spectra}

This subsection gives an overview of creating the spectra, from preparing the training data to training the neural network and evaluating the neural network's performance and quality. 

To maximise the speed of synthesising spectra, we used the \textit{Payne} algorithm \citep{ting_payne_2019}. This algorithm relies on an artificial neural network to generate synthetic spectra. After it is trained, it can generate a high-resolution spectrum with an average error of $\lesssim 10^{-3}$ in normalised flux \citep{ting_payne_2019}. With PySME, we need around 4 minutes to synthesise all four HERMES bands, but using \textit{Payne}, it only takes $0.01\ \mathrm{s}$ while preserving the accuracy needed for spectroscopic fitting.

We chose to use \textit{Payne} and did not use the pre-made grids method used in \citet{kos_galah_2021} as it would be computationally too intensive to compute a grid that spans the 36 labels (stellar parameters and elemental abundances) we would like to optimise. We chose not to use a generative model system as used in \citet{rixConstructingPolynomialSpectral2016,nessCannonDatadrivenApproach2015} as it has been shown in \citet{ting_payne_2019} that it cannot generate precise spectra (relative error of $<10^{-3}$) of normalised flux when generating spectra over 25 labels (stellar parameters and elemental abundances).

By using a neural network, we can swiftly and precisely generate spectra to effectively address the issue of blended lines. While other methods involve restricting the analysis to specific segments of spectra with no line blending issues \citep[e.g., ][]{buder_galah_2021,kos_galah_2021}. The approach runs into some problems. First, we are not using all the available information in the observed spectra. Second, if that a specific point on the observed spectra is contaminated (for example, by cosmic rays). In that case the ability to fit that element will be lost, and lastly, a problem will be encountered if there is a strong line just out of our specific region. This line will affect the line inside the region, but it is not taken into account. Hence, for these reasons, we will be fitting the majority of the wavelength range and all the parameters at once.

We produced the training spectra using PySME \citep{wehrhahn_pysme_2022} which is a Python package based on the SME (Spectroscopy Made Easy; \citet{valenti_spectroscopy_1996, piskunov_spectroscopy_2017}). PySME is open source, free to use, and as powerful as the old SME package.

To make sure we account for as many non-local thermodynamic equilibrium (non-LTE) effects as practically possible, we used all the available non-LTE elemental model grids in the PySME large data server{\footnote{\href{https://pysme-astro.readthedocs.io/en/latest/usage/lfs.html}{https://pysme-astro.readthedocs.io/en/latest/usage/lfs.html}}}. The list of the models is shown in table \ref{tab:model table}. An explanation of the grids can be found in \citet{amarsi_galah_2020}. In short, these non-NLTE models contain grids of pre-calculated line-by-line intensity corrections that are used to correct the intensity of synthesised LTE lines. 

Our line list is acquired from VALD \citep{ryabchikovaMajorUpgradeVALD2015}\footnote{\href{http://vald.astro.uu.se/}{http://vald.astro.uu.se/}} and a full list of references of the line list used is available in Appendix \ref{appendix: line list}. To accurately generate synthetic spectra while cutting down on computing time, we wanted a complete but not superfluous line list. To reduce the number of superfluous lines, we requested a line list with an intensity of at least 0.005 in normalised flux (this value was chosen so the errors of the GALAH spectra are always significantly larger than that), and to make sure our line list is complete, we submitted this request using multiple stellar parameters that will cover the parameter space of the stars that we wish to analyse. We requested a line list using stellar parameters ($T_{\rm{eff}}$, $\log(g)$): (4500 $K$, 4.5), (4750 $K$, 2.4), (6000 $K$, 4.1), and (6000 $K$, 4.1); for each stellar $T_{\rm{eff}}$, $\log(g)$ pair, we then requested them with three different metallicities ($-1$, 0, 1). Thus, we had 12 line lists in total that we joined to make our final line list. 

\begin{figure*}
    \centering
    \includegraphics[width=\linewidth]{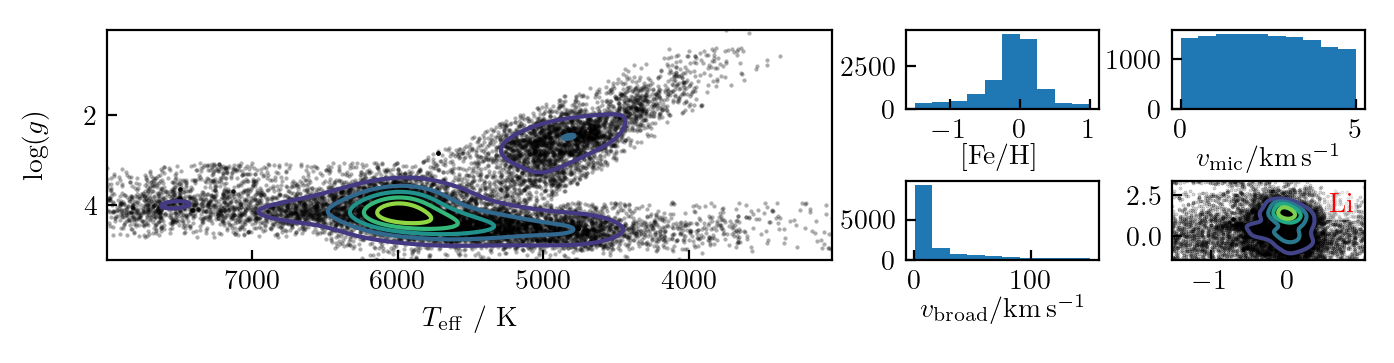}
    \includegraphics[width=\linewidth]{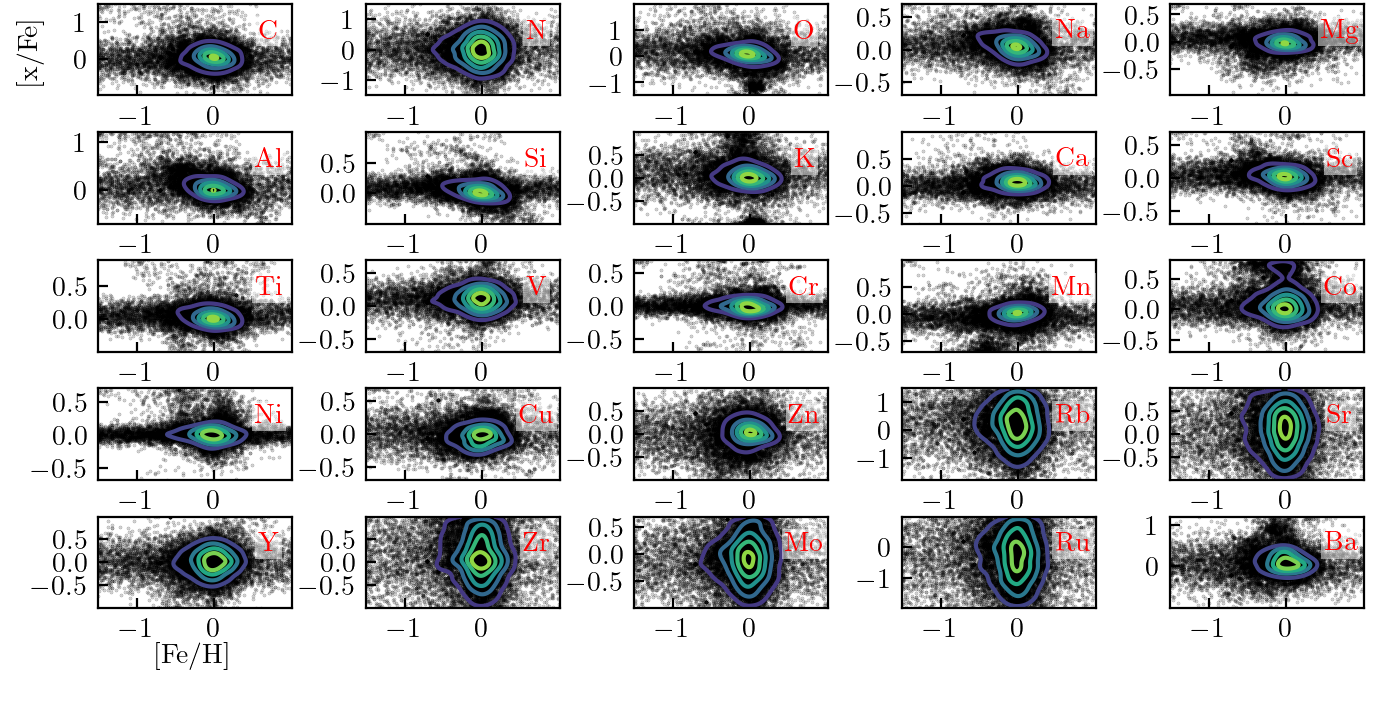}
    \caption{The parameter space of the $10\,000$ spectra used for the training set. Each black dot represents a value used to generate a training spectrum. The bottom 26 panels are a scatter plot of [Fe/H] against [x/Fe], with the element indicated in the top right of the panels.}
    \label{fig:parameter_training_space}
\end{figure*}

We decided to train the \textit{Payne} on four stellar parameters: $T_{\rm{eff}}$, $\log(g)$, $v_{\rm{mic}}$, $v_{\rm{broad}}$ and 32 abundances ([Fe/H], [Li/Fe], [C/Fe], [N/Fe], [O/Fe], [Na/Fe], [Mg/Fe], [Al/Fe], [Si/Fe], [K/Fe], [Ca/Fe], [Sc/Fe], [Ti/Fe], [V/Fe], [Cr/Fe], [Mn/Fe], [Co/Fe], [Ni/Fe], [Cu/Fe], [Zn/Fe], [Rb/Fe], [Sr/Fe], [Y/Fe], [Zr/Fe], [Mo/Fe], [Ru/Fe], [Ba/Fe], [La/Fe], [Ce/Fe], [Nd/Fe], [Sm/Fe], [Eu/Fe]). As in \citet{buder_galah_2021} and DR4, we have set macro-turbulence ($v_{\rm{mac}}$) to 0 as the resolution of GALAH $v_{\rm{broad}}$ and $v_{\rm{mac}}$ are degenerate broadening influences. We created a training set that is representative of the parameter space of the stars that we will analyse. As we are fitting GALAH spectra, we chose to use GALAH DR4 to create the parameter space of our training set. By using the values and uncertainties measured in GALAH DR4 of the stars in our open clusters, we were able to generate the probability space of the training set. Our training set is more densely sampled in areas in which there are more stars, as we would like our model to be better trained at these points. When creating the training set, we selected maximum and minimum values for abundances and stellar parameters, as we are not expecting to fit any stars that have extreme abundances or stellar parameters. To make sure we have created a training set that will span the parameter space of stars that we want to observe, we overlay the Kiel diagrams of the fitted isochrone for the open cluster we want to analyse. The parameter space of our training set is shown in Fig. \ref{fig:parameter_training_space}, with contours representing the density of the points made with a Gaussian kernel density estimator.

The number of training spectra used was $10\,000$ as recommended in \citet{ting_payne_2019}. We synthesised spectra with $0.004\ $\AA \, steps, so we have a sample rate of at least 10 times finer than the GALAH spectra. We chose wavelength limits for each band to allow for at least $\pm 200\ \mathrm{km\, s^{-1}}$ of stellar radial velocity.

The complexity of the neural networks depends on parameters called hyperparameters. One of the hyperparameters that we can change is the number of neurons per layer. If we have a number that is too large, we will overfit the model, or if we do not give enough neurons, the model cannot simulate the complexity of the training set. So, we have done hyperparameter tuning (i.e., trying to find the best value for a hyperparameter that will create a model that fits the validation data the best) on the number of neurons per layer and found that the default 300 neurons per layer is the optimum number. Hyperparameter tuning of the number of layers was already done in \citet{ting_payne_2019}, and it was found that extra layers only led to overfitting of the spectra.

From our overall training data, 20\% was used for our validation set and the rest for our training set. The actual training took around 20 thousand iterations to converge, which took around 2 hours per band to finish. The training was done on NVIDIA Tesla P100 GPU provided by the free online computing cluster Kaggle \footnote{\href{https://www.kaggle.com/}{https://www.kaggle.com/}}. We found that the average difference in flux between our neural network model and our validation set is $0.0056$ in the blue band, $0.0060$ in the green band, $0.0069$ in the red band, and $0.0046$ in the IR band. This is in line with the expected values in \citet{ting_payne_2019} of $0.007$.

These high-resolution synthetic spectra then need to be degraded to the resolution of the observed spectra. The observed spectra have a nominal resolution of 28000. However, this varies with wavelength, fibres, and time of observation. To precisely degrade the spectra, we used the resolution profile for each observed spectrum in each band. This contains information on the resolution at each wavelength. The resolution profile is of a generalised Gaussian profile. By using photonic combs and the Thorium-Xenon arc lamp, \citet{kosGALAHSurveyData2017} was able to measure the resolution profile. A photonic comb can generate spectra that are predictably spaced out with a known Lorentzian profile. The time needed to degrade the resolution of the synthetic spectra to that of the observed spectra is similar to the time needed to synthesise the high resolution spectra with \textit{Payne}. Some observed spectra do not have the resolution profile available or have an invalid resolution profile (e.g., they have values that are too large or have a negative value); thus, we used the resolution profile of the closest observation in time done on the same fibre as the spectrum in question. This incorrect resolution profile happens in around 8\% of spectra. For more information on GALAH resolution profiles, see Buder et al. (2023, in preparation).

\begin{table}
	\centering
	\caption{Non-LTE and atmospheric models used in the creation of the training set.}
	\label{tab:model table}
	\begin{tabular}{lccr} 
		\hline
		Type & Model\\
		\hline\hline
		Atmospheric & \citet{gustafsson_grid_2008} \\
		Ba & \citet{mashonkina_barium_1999} \\
		Ca & \citet{mashonkina_non-lte_2007} \\
		Fe & \citet{amarsi3DNonLTEIron2022} \\
		O & \citet{amarsi_non-lte_2016} \\
		Li & \citet{lind_departures_2009} \\
		Na & \citet{lind_departures_2009} \\
		Mg & \citet{osorio_mg_2016} \\
		Si & \citet{amarsi_solar_2017} \\
		Ti & \citet{sitnova_non-local_2016} \\
		\hline
	\end{tabular}
\end{table}

\subsubsection{Fitting procedure}\label{subsection:fitting_spectra}

To combine both spectroscopic and photometric stellar parameters, we decided to use a Bayesian fitting scheme. \citet{pinsonneault_distances_2004} have shown how photometric $T_{\rm{eff}}$ and spectroscopic $T_{\rm{eff}}$ derived without using a photometric prior might differ. However, in our case, we found that there were no overall strong trends in the difference between spectroscopic temperatures derived without using a photometric prior and photometric temperatures, so we decided not to shift the temperatures. The spread of the temperature differences is lower than in \citet{kos_galah_2021} where the shifting of the temperature scale was necessary.

The likelihood used to fit stellar parameters and abundances to spectra is expressed as:
\begin{equation}
\rm{log}(\mathcal{L}(\mathbfit{x}))\propto -\sum_{i}\frac{(s_{\rm{i }}-Payne(\mathbfit{x}))^2}{\sigma_i^2},
\end{equation}
where $\mathbfit{x}$ is the vector of stellar parameters and abundances, $s_{\rm{i}}$ is the flux of an observed spectrum at wavelength $i$, and $\sigma_i$ is the observational error of the normalised flux at wavelength $i$.

We have included two priors in the fitting. The first one is the photometric stellar parameters prior, derived from equation \ref{eqn:theoretical stellar spread}, and the other is a mask. The mask will prevent the program from fitting wavelength ranges that we do not want to fit as we know our models do not accurately model them, while ensuring we always fit wavelength ranges of strong non blended absorption lines. A full description of the mask is found in Appendix \ref{appendix: masks}.

The fitting is done in four steps: The first step is an initial check of whether the GALAH reduction process produced an estimate for $T_{\rm{eff}}$ and $\log(g)$. Estimates for $T_{\rm{eff}}$ and $\log(g)$ are made for each band of each spectrum during the GALAH data reduction process unless the data quality is too poor. We do not use any band that does not have these initial estimates. Then, if there is no valid resolution profile, we acquire one. For the infrared band, we discard the first 25\% of the wavelength range as it is prone to having extreme values in it, as the infrared is susceptible to be completely dominated by the telluric A band. 

In the second step, we find the radial velocities of the individual bands. We synthesise stellar spectra using GALAH DR4 values as starting parameters, and if there are no abundances available in the GALAH DR4 catalogue, we use solar abundances. We vary the radial velocities of each band individually to find which radial velocity for each band has the highest likelihood value. These radial velocity values are kept for the rest of the analysis process. In this work, we found that the average difference in radial velocities between the bands is 0.89 $\mathrm{km\, s^{-1}}$, this is the expected value as it is around the expected radial velocity uncertainty of GALAH DR4. 

The third step is to do an initial fit of the spectra using GALAH DR4 as the starting value. After an initial test, we chose not to fit [Rb/Fe], [Sr/Fe], [Mo/Fe], [Ru/Fe], [La/Fe], and [Ce/Fe]. The fitting is done by using Bayesian sampling from the MCMC package emcee. As we have 30 parameters, we run the sampler using the minimum recommended 60 walkers.

 While the GALAH reduction process provides us with pre-normalised spectra, we opt to normalise spectra using our synthetic spectra. We did this as our method to generate spectra differs from the reduction process, and the values used to generate the synthetic spectra to normalise the observed spectra are different. We normalise the spectra at the beginning of this step. The normalisation is done using the same methods described in the reduction process of GALAH \citep{kosGALAHSurveyData2017}. In short, we use a univariate spline function over the entire spectrum.
 
 The fitting process ends when the burn-in has finished for $T_{\rm{eff}}$, $\log(g)$, $v_{\rm{mic}}$, $v_{\rm{broad}}$, and [Fe/H] plus an additional 50 iterations. To check whether the burn-in has finished, we fitted a 1D polynomial to each walker's samples (chain) for these five parameters. As the values of the parameters are vastly different, we normalised them using their mean value before fitting the polynomial. We checked every 40 iterations to see if the gradient was below 0.01, which corresponds to a 1\% change every 100 iterations. If 90\% of the chains have achieved this value, we consider the sampler to be burned in. This first process takes around 410 iterations to finish, which is around 6-7 minutes when using 18 cores. We then create an additional mask, as described in Appendix \ref{appendix: masks}. 

Using the three combined masks, we do the final fitting run. In this run, we re-normalise the spectra once at the beginning using the values acquired during the mask run. Our starting parameters for the main fit come from the last iteration of the mask fitting run. As we have just introduced a new mask, we let the sampler run until burn-in is achieved again. We detect when burn-in is achieved using the same method as in the previous step. 

To check whether the sampler has converged, we use auto-correlation time. When using an MCMC sampler, each sample is not independent from one another; auto-correlation time is a measure of how independent the samples are from one another. To calculate the number of independent samples, we divide the total number of samples by the auto-correlation time. When the sampler has converged, it means it has successfully explored the underlying distribution and thus the samples are representative of the underlying distribution. By accurately calculating the number of independent samples, we have a quantitative measure of how well the sampler has sampled the underlying distribution. We use an auto-regression model to calculate our auto-correlation time. A full discussion of why is available in Appendix \ref{appendix:auto-correlation}. We found that 400 independent samples were sufficient for calculating the mean value of the parameter and the uncertainties associated with it. Thus, the program will stop sampling once a mean average of 400 independent samples across the 30 parameters has been achieved. However, even with 100 independent samples, it was already sufficient to sample the mean value of the parameter \citep[as the number of independent samples only contributes an extra 10\% to the uncertainty;][]{goodmanEnsembleSamplersAffine2010a}.

After the initial burn-in, we estimated the auto-correlation for the first time after 100 iterations and recalculated the auto-correlation time for every 80 independent samples. We did not recalculate the auto-correlation time after each iteration, as it was computationally expensive. This process continued until an average of 400 independent samples was obtained. After completing the sampling, we calculated the number of independent samples for each parameter, which served as a quality metric. The sampling process averaged $15\, 000$ iterations and took approximately 4 hours to compute using 18 cores. We had set a maximum iteration limit of $40\,000$. If the maximum iteration is reached, we assume that the fitting method will not converge, so we will stop sampling. After the fitting process is finished a mean value and precision are calculated for each fitted parameter.

\section{Results}\label{Results}

The results section is divided into three parts. The first section, \ref{subsection:distance improvements}, discusses how well the distance measurements are using a cluster centre distance prior. The next section \ref{subsection:isochorne fitting} discusses the photometric stellar parameters and the fits of the isochrone. The last section discusses the effect of using photometric stellar parameters as priors on the spectroscopic stellar parameters and abundances of stars in open clusters.

\subsection{Distance improvements}\label{subsection:distance improvements}

\begin{table*}
\caption{Comparison between our cluster centres and cluster sizes to cluster distances and sizes calculated directly from geometric and photo-geometric distances.}
\label{tab:Distance vs geometric vs photogeometric}
\begin{tabular}{lp{2.2cm}p{2.2cm}p{2.2cm}p{2.2cm}p{2.2cm}p{2.2cm}}
\hline
Cluster        & This work's distance to the cluster  centre & This work's radial cluster size & Geometric distance to cluster centre   & Geometric cluster size & Photo-geometric distance to cluster core & Photo-geometric cluster size \\ \hline
               & pc                                          & pc                              & pc    & pc       & pc       & pc            \\ \hline\hline
ASCC 16        & 362.4 $\substack{+3.5 \\-3.5}$              & 4.2 $\substack{+3.3\\ -2.6}$    & 343.9 & 18.0     & 308.7    & 10.7        \\
Blanco 1       & 234.7$\substack{+0.2\\ -0.2}$               & 2.7$\substack{+0.2 \\ -0.2}$    & 235.4 & 6.4      & 234.9    & 6.3           \\
Melotte 22     & 135.2 $\substack{+0.1\\ -0.1}$              & 1.9$\substack{+0.1 \\ -0.1}$    & 135.3 & 2.3      & 135.2    & 2.3           \\
NGC 2516       & 407.3 $\substack{+0.2\\ -0.3}$              & 7.6$\substack{+0.3 \\ -0.2 }$   & 411.9 & 22.3     & 410.8    & 25.1          \\
NGC 2632       & 183.9 $\substack{+0.1\\-0.1}$               & 3.7$\substack{+0.1 \\ -0.1}$    & 184.4 & 5.1      & 184.1    & 5.2           \\
M67            & 838.7$\substack{+1.3\\-1.2}$                & 24.5$\substack{+1.3 \\ -1.0}$   & 865.0 & 89.3     & 833.6    & 53.8          \\
Ruprecht 147   & 307.1 $\substack{+0.7\\ -0.6}$              & 6.4$\substack{+0.5 \\ -0.5}$    & 309.4 & 11.3     & 308.7    & 10.7          \\
Trumpler 10    & 430.9 $\substack{+0.5\\ -0.5}$              & 7.9$\substack{+0.6 \\ -0.7}$    & 457.7 & 71.6     & 455.3    & 71.9         \\
Upper Scorpius & 146 $\substack{+0.4\\ -0.5}$                & 8.7$\substack{+0.3 \\ -0.3}$    & 146.0 & 8.9      & 146.0    & 8.9           \\ \hline
\end{tabular}
\end{table*}
\begin{figure}
    \centering
    \includegraphics{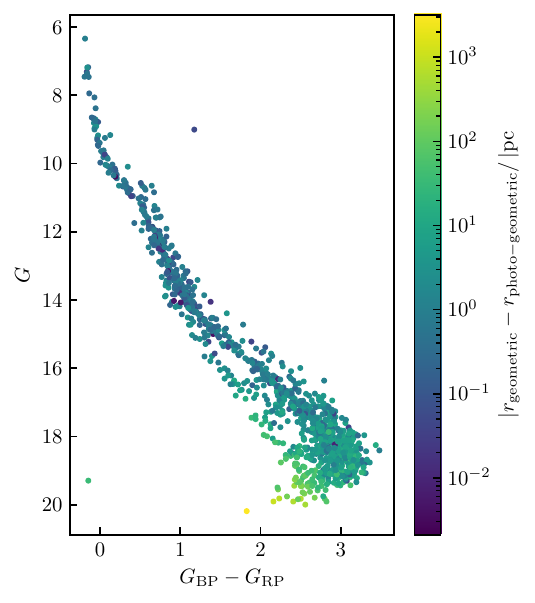}
    \caption{Colour-magnitude diagram of Trumpler 10, with the colours indicating geometric distance $-$ photo-geometric distance.}
    \label{fig:Trumpler 10 delta phot vs  geo}
\end{figure}
\addtolength{\tabcolsep}{-0.5em}
\begin{table}
	\centering
	\caption{Comparison between our cluster centre, the size, and the values given in BO19, HR23.}
	\label{tab:Distance comparison table}
	\begin{tabular}{lp{1.2cm}p{1cm}p{1.2cm}p{1.2cm}p{0.8cm}p{0.8cm}} 
		\hline
Cluster  & This work's distance to the cluster  centre & This work's radial cluster size & BO19 distance to the cluster centre&HR23 distance to the cluster centre&HR23  core cluster size& HR23 tidal cluster size  \\
\hline
&pc&pc&pc&pc&pc&pc\\
		\hline
  \hline
ASCC 16                              & 362.4 $\substack{+3.5 \\-3.5}$ & 4.2 $\substack{+3.3\\ -2.6}$ & 317.7 $\substack{+3.5\\-8.7}$&341.8$\substack{+0.3 \\-0.3}$&1.9 &5.9\\
Blanco 1                               & 234.7$\substack{+0.2\\ -0.2}$ & 2.7$\substack{+0.2 \\ -0.2}$ & 237.2 $\substack{+0.0\\ -0.0}$&234.3$\substack{+0.1 \\-0.1}$&2.4&7.4 \\
Melotte 22                             & 135.2 $\substack{+0.1\\ -0.1}$ & 1.9$\substack{+0.1 \\ -0.1}$& 136.0 $\substack{+0.0\\ -0.0}$&134.8$\substack{+0.0 \\-0.0}$&2.7&13.7\\
NGC 2516                               & 407.3 $\substack{+0.2\\ -0.3}$ & 7.6$\substack{+0.3 \\ -0.2 }$ & 415.1 $\substack{+0.0\\ -0.0}$              &407.1$\substack{+0.1 \\-0.1}$&3.2&13.5\\
NGC 2632                               & 183.9 $\substack{+0.1\\-0.1}$ & 3.7$\substack{+0.1 \\ -0.1}$ & 186.2 $\substack{+0.0\\ -0.0}$               &183.5$\substack{+0.0 \\-0.0}$& 3.2&10.5\\
M67                               & 838.7$\substack{+1.3\\-1.2}$ & 24.5$\substack{+1.3 \\ -1.0}$ & 881.4 $\substack{+0.0\\ -0.0}$&836.6$\substack{+0.6 \\-0.6}$&2.5&8.7 \\
		Ruprecht 147                               & 307.1 $\substack{+0.7\\ -0.6}$ & 6.4$\substack{+0.5 \\ -0.5}$  & -&302.5 $\substack{+0.1 \\-0.1}$&3.4&15.2\\
		Trumpler 10                               & 430.9 $\substack{+0.5\\ -0.5}$ & 7.9$\substack{+0.6 \\ -0.7}$  & 365.8 $\substack{+4.7 \\-3.6}$&428.0$\substack{+0.2 \\-0.2}$&4.5&10.3\\
				Upper Scorpius                              & 146 $\substack{+0.4\\ -0.5}$ & 8.7$\substack{+0.3 \\ -0.3}$  & -&-&-&-\\
		\hline
	\end{tabular}
\end{table}

\begin{table}
\caption{Membership numbers from our work compared to the membership numbers in HR23, showing the possible field stars. The second and fifth columns shows the number of member stars in our study obtained from \textit{Gaia} and GALAH catalogues, respectively. The third column shows the number of open cluster members from HR23. The fourth and last columns show the possible field stars in our study from \textit{Gaia} and GALAH catalogues, respectively.}
\label{tab:membership}
\begin{tabular}{lccp{1.2cm}cp{1.2cm}}
\hline\hline
 Cluster  & $N_{\it{Gaia}}$    & $N_{\mathrm{HR23}}$   & Possible $N_{\it{Gaia}}$ field stars & $N_{\mathrm{GALAH}}$ & Possible $N_{\mathrm{GALAH}}$ field stars \\ \hline
ASCC 16      & 247  & 171  & 126         & 173                & 75                                 \\
Blanco 1     & 328  & 841  & 0           & 109                & 0                                  \\
M67           & 913  & 1314 & 4           & 147                & 0                                  \\
Melotte 22   & 552  & 1721 & 0           & 91                 & 0                                  \\
NGC 2516     & 1686 & 3784 & 29          & 204                & 2                                  \\
NGC 2682     & 1389 & 1844 & 7           & 922                & 0                                  \\
Ruprecht 147 & 118  & 279  & 4           & 130                & 0                                  \\
Trumpler 10  & 1190 & 1425 & 466         & 119                & 18                                \\ \hline
\end{tabular}
\end{table}

We compared our cluster distance centres and sizes obtained by using equation \ref{eqn:likelihood cluster} to geometric and photo-geometric distances obtained in \citet{bailer-jones_estimating_2021}. We calculated the cluster distance centres and sizes from the distances in \citet{bailer-jones_estimating_2021} by fitting a normal distribution to the geometric and photo-geometric distances of the open cluster members. The results are shown in Table \ref{tab:Distance vs geometric vs photogeometric}. We found that faint stars' photo-geometric and geometric distances differ by a significant amount, with some stars having up to $\sim 7\,000 \, \rm{pc}$ in difference. An example of this difference is shown in Fig. \ref{fig:Trumpler 10 delta phot vs  geo}. Thus, we have excluded stars that have more than a 100 pc difference between their photo-geometric and geometric distances when calculating the distance and size of the open cluster.

We observed that the geometric and the photo-geometric distances to the cluster centre were within one cluster size for all open clusters except for ASCC 16. We find that our distances to the cluster centre are within one cluster size compared to the cluster distances and sizes calculated directly using the geometric and photo-geometric distances for all the open clusters except for ASCC 16.

We found that the cluster sizes calculated directly from geometric and photo-geometric distances are larger compared to the cluster sizes calculated using equation \ref{eqn:likelihood cluster}. For Trumpler 10, M67, and NGC 2516 we find that our cluster sizes are significantly smaller, we think this is caused by our prior on cluster size (i.e., equation \ref{eqn:cluster size prior}).

We Further compared our cluster centre distances and sizes obtained by using equation \ref{eqn:likelihood cluster} to two different catalogues \citet{bossini_age_2019} and \citet{hunt_improving_2023}; called BO19 and HR23 hereafter. We have chosen these two catalogues because BO19 uses an automated isochrone fitter to get distances; hence, it is relevant to see how different our distance results and isochrone parameters are. We chose HR23 as it uses a clustering algorithm on \textit{Gaia} DR3 data to find distances; hence, we should find similar results as we have the same data. The tabulated distance results with these two catalogues are shown in table \ref{tab:Distance comparison table}. We found that the majority of our calculated cluster centres are within 3 pc of BO19 and HR23 values. In BO19, M67 and Trumpler 10 were outliers; this could be caused by the difference in priors, methods, and the raw data used (as we used \textit{Gaia} DR3 rather than \textit{Gaia} DR2, which BO19 used, which has 20\% smaller uncertainties than \textit{Gaia} DR2 \citep{lindegrenGaiaEarlyData2021}). M67 is also the furthest cluster we studied in this paper at 839 pc. Our values correspond more with HR23, as we both use the same data from \textit{Gaia} DR3. The cluster with the largest difference was ASCC 16.

We decided not to study a specific group in the Upper Scorpius region but to study it as a whole. Our Upper Scorpius distances are within the distances of the groups inside the Upper Scorpius region. The distances of the groups inside the Upper Scorpius regions range from 142 pc to 159 pc in \citet{ratzenbockSignificanceModeAnalysis2022} or ranges from 126 pc to 153 pc in \citet{miret-roigStarFormationHistory2022}. Our distance of 146 pc lies within both of these ranges.

\begin{figure}
\centering
\includegraphics{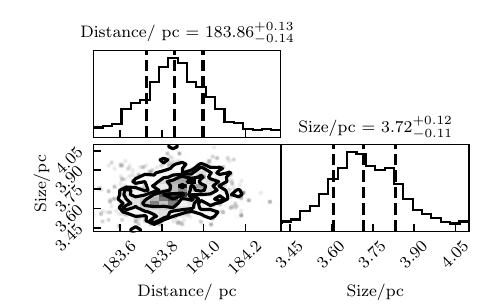}
\caption{Corner plot of the probability distribution of the cluster distance prior calculated for NGC 2632. The vertical dashed lines in the histogram show the mean and $\pm 1 \sigma$ for each parameter}
\label{fig:corner ngc 2632}
\end{figure}
\begin{figure}
    \centering
	\includegraphics{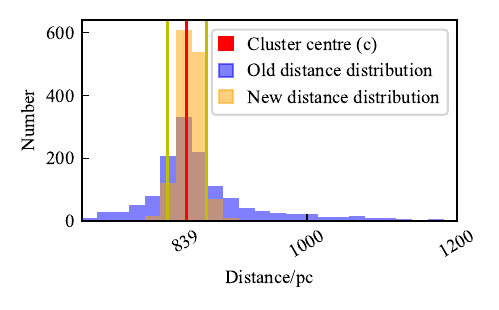}
	\caption{Histogram showing the distance distribution of stars in M67 without the cluster distance prior (i.e., geometric distances, shown in equation \ref{eqn:individual distance equation}) shown in blue and after using the cluster prior (i.e., equation \ref{eqn:final distance equation}) shown in yellow. The red line indicates  the centre of the cluster according to the cluster distance prior and the yellow lines are the radial size of the cluster according to the cluster distance prior}
    \label{fig:Cluster distance improvements}
    
\end{figure}
Our radial cluster sizes are as expected, with all but one of the clusters being around 10 pc. M67 does have a much larger cluster size; this could be caused by the pencil effect that happens when the distance to the cluster centre increases. The increase in distance increases the uncertainty in the parallax. Comparing our radial cluster sizes with the HR23 size values, we found that the sizes of our clusters are within the bounds of the core radius and the tidal radius for all but M67.

One of the reasons that our distances and cluster sizes could differ is the different membership of the open clusters. We have compared our membership with the membership in HR23 in table \ref{tab:membership}. We expect HR23 to have a higher membership count than us, as their membership includes stars in the tidal tail, whereas ours focuses on stars in the open cluster core. The possible field stars are the stars in our study but not found in HR23. The amount of possible field stars does not exceed more than 4 \% of the total member stars in our study in all open clusters except for ASCC 16 and Trumpler 10. In ASCC 16, the possible field stars lie on the best-fitting isochrone; thus, we think these possible field stars are open cluster members. However, in Trumpler 10, these possible field stars do not lie on the best-fitting isochrone; thus, we think our membership criteria for Trumpler 10 are not strict enough. Even though 39 \% of stars in Trumpler 10 are possible field stars, we found that our distance to the cluster core is within 3 pc compared to the distance in HR23. Thus, we do not think that the field stars have impacted the distance in a significant way. We also confirmed that none of these possible field stars in ASCC 16 and Trumpler 10 are members of any other cluster studied in HR23.

Overall, we found that we estimated our cluster core distances and sizes well for our purpose of calculating absolute magnitudes. Even though some of our cluster core distances and sizes are not within the ranges in BO19 and HR23, the resulting distance distribution of the stars in the open clusters will be similar to our values. Furthermore the difference between our cluster core distances and the catalogues are within one cluster size. 

The corner plots of the probability distribution of the cluster distance priors (e.g., Fig. \ref{fig:corner ngc 2632}) are as expected, with the probability centred around one point. If there was a lot of contamination with other field stars or other clusters, we could have found multiple peaks in the PDF.

After the introduction of the cluster distance prior, the stars are normally distributed within the calculated size of the cluster; this is shown in Fig. \ref{fig:Cluster distance improvements}. The new distribution of stars is in line with our expected results when using the cluster distance and shows how our cluster distance prior is working as intended.

By using the distance cluster prior we saw an average increase in the precision of the individual star distance of 31\% compared to the distances calculated without using the distance cluster prior. i.e., the geometric distances calculated in \citet{bailer-jones_estimating_2021}. We calculated the geometric distance precision of individual stars by using:
\begin{equation}
    \frac{\texttt{r\_hi\_geo}-\texttt{r\_lo\_geo}}{2}
\end{equation}
Where \texttt{r\_hi\_geo} and \texttt{r\_lo\_geo} are the 84th and 16th percentiles of the distance posterior calculated using the geometric prior.

The stars in Upper Scorpius were the only ones that did not see an average increase in precision. We think that this is caused by the size of the region estimated by the cluster distance prior (8.7 pc), which simulates well the original size of the region, thus leading to the prior having a negligible effect on this cluster. The radial cluster size was expected, as Upper Scorpius is a large region with multiple groups in it, and we did not choose a specific group to study. In \citet{miret-roigStarFormationHistory2022} they estimates the groups having sizes of $\sim 1 $ pc to $\sim 10$ pc. Our estimation of 8.7 pc is within the ranges of these groups.

Our method of using a cluster distance prior to fit individual distances has increased the accuracy of the individual distances as the distances now lie inside the expected size of an open cluster core. The method has increased the precision of the individual distance measurement. With both improvements to the distance measurements, it will improve our absolute magnitudes, thus improving the photometric $T_{\rm{eff}}$ and $\log(g)$ measurements.

\subsection{Isochrone fitting and photometric stellar parameters}\label{subsection:isochorne fitting}

Using the method explained in Sec. \ref{subsec: calculating photometric stellar parameters}, we fitted isochrones to the HR diagrams and selected appropriate ranges for the ages, metallicities, and extinctions for each cluster. The ranges are in the expected bounds for an open cluster. The ages are younger than $10^{10}$ years, and the metallicities are around solar metallicity. The ranges selected are shown in table \ref{tab:Isochrone parameters table}. The HR diagrams fitted best fitting isochrone are shown in Appendix \ref{appendix: all HR diagrams}.

We did extensive checks of the validity of the ranges of ages, metallicities, and extinctions for each cluster, as demonstrated in Appendix \ref{appendix: validity of ranges}. We also verified the method of calculating photometric $T_{\rm{eff}}$ and $\log(g)$ (i.e., equation \ref{eq:HR weighted teff}) in Appendix \ref{appendix: photometric diagnostic plots}.

\begin{figure*}
    \centering
    \includegraphics[width=\linewidth]{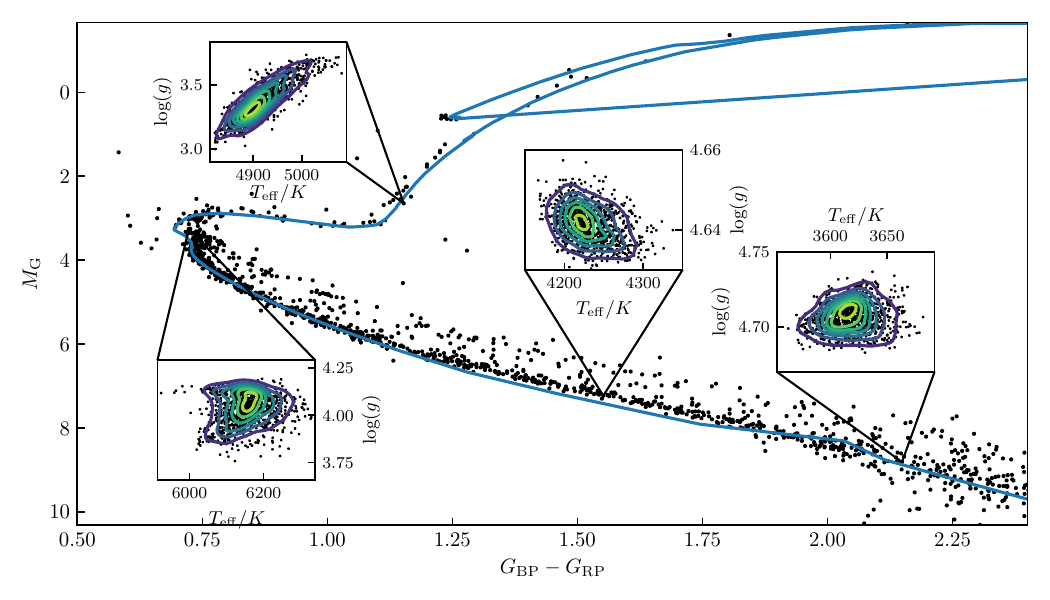}
    \caption{$T_\mathrm{eff}$ and $\log(g)$ joint probability distribution for four different stars in M67. The four stars represent the typical PDF for different types of stars in the cluster. The blue line is the best-fitting isochrone.}
    \label{fig:Parameter spread}
\end{figure*}

We constructed a joint PDF for $T_{\rm{eff}}$ and $\log(g)$ for each star in the cluster. An example for M67 is shown in Fig. \ref{fig:Parameter spread}. We found that for stars that are in the lower half of the main sequence, the PDFs were uncorrelated in $T_{\rm{eff}}$ vs. $\log(g)$. This is as expected because when we plot a Kiel diagram, the gradient of $T_{\rm{eff}}$ and $\log(g)$ with respect to the starting masses of the stars at that point along the isochrone is small, so when we change the isochrone parameters, the change in $T_{\rm{eff}}$ and $\log(g)$ will be quite small. Another factor is that the isochrone changes shape in a symmetrical way when increasing or decreasing a parameter in the isochrone. Further along the isochrone, the correlation of $T_{\rm{eff}}$ and $\log(g)$ becomes stronger. Stars at the turn-off point have a very unsymmetrical PDF shape, which is because of the complex shape of the isochrone at the turn-off point. Therefore, when we change the parameters of the isochrones, the newest closest point to the star can have a very different $T_{\rm{eff}}$ and $\log(g)$. In the lower red giant branch, the trends become very strong as shifts in the isochrone parameters cause large changes in the shape of the isochrone, thus causing a large change in $T_{\rm{eff}}$ and $\log(g)$.

The average photometric precision of $T_{\rm{eff}}$ and $\log(g)$ is 38.2 K and 0.031 dex. This precision is calculated from PDF in equation \ref{eqn:final photometric equation}. The average precision measured when using an individual isochrone (i.e., equation \ref{eqn:errors isochrone individual teff}) is 3.15 K and 0.0016 dex. As the final photometric precision is much worse than the precision measured using an individual isochrone, it shows that the ranges of metallicities, ages, and extinctions are a much larger source of imprecision than the absolute magnitudes.

Spectroscopic uncertainty in GALAH is 70 K in $T_{\rm{eff}}$ and 0.064 dex in $\log(g)$. The size of GALAH uncertainty is comparable to the size of the joint $T_{\rm{eff}}$ and $\log(g)$ probability distribution. This comparable size is an indication that the shape of the photometric prior will have an impact on our spectroscopic fit.

\begin{table}
\setlength{\tabcolsep}{8pt}
	\centering
	\caption{The parameters and their ranges for isochrones for each open cluster. Age is given in $\log(\rm{age})$.}
	\label{tab:Isochrone parameters table}
\begin{tabular}{lccr} 
		\hline
Cluster  & Age & Metallicity & Extinction ($\it{A_{\rm{V}}}$)\\
		\hline
  &$\log\,\rm{age \,(yr)}$&dex& mag \\
  \hline
            \hline
ASCC 16                              & 7.10 $\substack{+0.09 \\-0.08}$ & 0.05 $\substack{+0.20\\ -0.15}$ & 0.25 $\substack{+0.15\\-0.15}$ \\
Blanco 1                           & 7.98 $\substack{+0.12 \\-0.08}$ & 0.10 $\substack{+0.08\\ -0.10}$ & 0.05 $\substack{+0.05\\-0.05}$ \\
Melotte 22                           & 8.00 $\substack{+0.15 \\-0.20}$ & 0.15 $\substack{+0.15\\ -0.15}$ & 0.25 $\substack{+0.10\\-0.15}$ \\
NGC 2516                           & 8.30 $\substack{+0.15 \\-0.15}$ & 0.20 $\substack{+0.15\\ -0.15}$ & 0.3 $\substack{+0.1\\-0.1}$ \\
NGC 2632                           & 8.90 $\substack{+0.05 \\-0.09}$ & 0.19 $\substack{+0.11\\ -0.08}$ & 0.08 $\substack{+0.07\\-0.07}$ \\
M67                           & 9.59 $\substack{+0.06 \\-0.05}$ & 0.11 $\substack{+0.07\\ -0.05}$ & 0.09 $\substack{+0.06\\-0.04}$ \\
Ruprecht 147                              & 9.40 $\substack{+0.04 \\-0.05}$ & 0.08 $\substack{+0.12\\ -0.08}$ & 0.40 $\substack{+0.1\\-0.05}$ \\
Trumpler 10                           & 7.65 $\substack{+0.12 \\-0.08}$ & 0.10 $\substack{+0.20\\ -0.20}$ & 0.15 $\substack{+0.15\\-0.15}$ \\
Upper Scorpous                           & 6.90 $\substack{+0.15 \\-0.15}$ & 0.15 $\substack{+0.30\\ -0.25}$ & 0.60 $\substack{+0.20\\-0.20}$ \\

		\hline
	\end{tabular}
\end{table}

\subsection{Abundances}\label{subsection:abundances results}
\begin{table}
    \centering
    \caption{An overview of the number of stars successfully fitted by the cluster fitting process. $N_\mathrm{GALAH}$ is the total number of spectra we fitted from the DR4 GALAH catalogue. $N_{100\rm{np}>}$ and $N_{400\rm{np}>}$ are the number of stars with more than 100 and 400 independent samples on average for each fitted parameter without using the prior. $N_{100\rm{p}>}$ and $N_{400\rm{p}>}$ are the number of stars with more than 100 and 400 independent samples on average for each fitted parameter using the prior. }
    \begin{tabular}{lcccccr}
\hline \\
Cluster  & $N_\mathrm{GALAH}$ & $N_{100\rm{np}>}$ & $N_{100\rm{p}>}$ & $N_{400\rm{np}>}$ & $N_{400\rm{p}>}$ &SNR \\
\hline\hline\\
        ASCC 16 & 173 & 18 & 27 & 2 & 4 & 40.0 \\ 
        Blanco 1 & 109 & 53 & 52 & 21 & 22 & 94.0 \\ 
        Melotte 22 & 91 & 26 & 27 & 11 & 9 & 75.4 \\ 
        NGC 2516 & 204 & 32 & 25 & 6 & 10 & 40.3 \\ 
        NGC 2632 & 147 & 58 & 49 & 18 & 19 & 60.3 \\ 
        NGC 2682 & 922 & 608 & 580 & 152 & 163 & 61.9 \\
        Ruprecht 147 & 130 & 85 & 86 & 16 & 17 & 91.6 \\ 
        Trumpler 10 & 119 & 58 & 57 & 8 & 8 & 32.2 \\ 
        Upper Scorpius & 84 & 9 & 4 & 1 & 0 & 31.8  \\ \hline  
    \end{tabular}
\label{tab:summary of fitting}
\end{table}

\begin{table*}
\caption{Precision values of spectroscopic abundances and stellar parameters calculated using different methods and a comparison with GALAH DR4. The MCMC columns show the average uncertainty acquired from MCMC fit when fitting using the photometric prior, and no photometric prior. The overall scatter shows the mean scatter of the parameters when fitted using the prior in each cluster weighted with the number of stars. M67 scatter is the spread of the parameters in M67 with $4.0<\log(g)<4.1$. Precision GALAH is the precision given in GALAH DR3. Accuracy GALAH is the estimated accuracy of the parameters in GALAH DR4. Uncertainty GALAH is the total uncertainty estimation in GALAH DR4 acquired by adding in quadrature the precision and the accuracy.}
\setlength{\tabcolsep}{6pt}
\begin{tabular}{lccccccccl}
\hline
\multirow{2}{*}{Parameter} & \multirow{2}{*}{Units} & \multicolumn{2}{c}{MCMC} & \multicolumn{2}{c}{Scatter} & \multirow{2}{*}{Repeat observations} & \multirow{2}{*}{Precision GALAH} & \multirow{2}{*}{Accuracy GALAH} & \multirow{2}{*}{Uncertainty GALAH} \\
                           &                        & Prior     & No prior     & Overall       & M67         &                                      &                                  &                                 &                                     \\ \hline \hline
$T_{\rm{eff}}$             & K                      & 10.27     & 10.284       & -             & -           & 20.975                               & 26.254                           & 66                              & 71.03                               \\
$\log(g)$                   & dex                    & 0.026     & 0.031        & -             & -           & 0.037                                & 0.057                            & 0.042                           & 0.071                               \\
{[}Fe/H{]}                 & dex                    & 0.01      & 0.01         & 0.078         & 0.059       & 0.019                                & 0.038                           & 0.044                           & 0.058                               \\
$v_{\rm{mic}}$             & $\mathrm{km\, s^{-1}}$                   & 0.04      & 0.04         & -             & -           & 0.071                                & 0.075                            & 0.28                            & 0.29                                \\
$v_{\rm{broad}}$               & $\mathrm{km\, s^{-1}}$                   & 0.135     & 0.133        & -             & -           & 0.212                                & 0.555                            & 1.4                             & 1.506                               \\
{[}Li/Fe{]}                & dex                    & 0.291     & 0.244        & 1.302         & 0.100       & 0.025                                & 0.154                            & -                               & 0.154                               \\
{[}C/Fe{]}                 & dex                    & 0.041     & 0.032        & 0.089         & 0.052       & 0.028                                & 0.086                            & -                               & 0.086                               \\
{[}N/Fe{]}                 & dex                    & 0.063     & 0.063        & 0.251         & 0.120       & 0.093                                & 0.181                            & -                               & 0.181                               \\
{[}O/Fe{]}                 & dex                    & 0.055     & 0.056        & 0.16          & 0.096       & 0.039                                & 0.084                            & -                               & 0.084                               \\
{[}Na/Fe{]}                & dex                    & 0.048     & 0.051        & 0.097         & 0.076       & 0.035                                & 0.037                            & -                               & 0.037                               \\
{[}Mg/Fe{]}                & dex                    & 0.028     & 0.028        & 0.086         & 0.042       & 0.036                                & 0.035                            & -                               & 0.035                               \\
{[}Al/Fe{]}                & dex                    & 0.052     & 0.055        & 0.135         & 0.091       & 0.090                                & 0.07                             & -                               & 0.07                                \\
{[}Si/Fe{]}                & dex                    & 0.02      & 0.02         & 0.054         & 0.028       & 0.037                                & 0.025                            & -                               & 0.025                               \\
{[}K/Fe{]}                 & dex                    & 0.059     & 0.06         & 0.263         & 0.293       & 0.053                                & 0.062                            & -                               & 0.062                               \\
{[}Ca/Fe{]}                & dex                    & 0.032     & 0.034        & 0.069         & 0.041       & 0.019                                & 0.041                            & -                               & 0.041                               \\
{[}Sc/Fe{]}                & dex                    & 0.036     & 0.037        & 0.074         & 0.061       & 0.036                                & 0.04                             & -                               & 0.04                                \\
{[}Ti/Fe{]}                & dex                    & 0.025     & 0.026        & 0.064         & 0.032       & 0.022                                & 0.033                            & -                               & 0.033                               \\
{[}V/Fe{]}                 & dex                    & 0.043     & 0.044        & 0.112         & 0.063       & 0.046                                & 0.057                            & -                               & 0.057                               \\
{[}Cr/Fe{]}                & dex                    & 0.022     & 0.024        & 0.049         & 0.037       & 0.025                                & 0.024                            & -                               & 0.024                               \\
{[}Mn/Fe{]}                & dex                    & 0.052     & 0.057        & 0.11          & 0.046       & 0.038                                & 0.037                            & -                               & 0.037                               \\
{[}Co/Fe{]}                & dex                    & 0.044     & 0.043        & 0.112         & 0.070       & 0.031                                & 0.055                            & -                               & 0.055                               \\
{[}Ni/Fe{]}                & dex                    & 0.018     & 0.018        & 0.042         & 0.021       & 0.014                                & 0.018                            & -                               & 0.018                               \\
{[}Cu/Fe{]}                & dex                    & 0.099     & 0.107        & 0.244         & 0.195       & 0.113                                & 0.056                            & -                               & 0.056                               \\
{[}Zn/Fe{]}                & dex                    & 0.096     & 0.095        & 0.194         & 0.155       & 0.082                                & 0.096                            & -                               & 0.096                               \\
{[}Y/Fe{]}                 & dex                    & 0.074     & 0.075        & 0.144         & 0.121       & 0.090                                & 0.074                            & -                               & 0.074                               \\
{[}Zr/Fe{]}                & dex                    & 0.209     & 0.208        & 0.493         & 0.402       & 0.247                                & 0.564                            & -                               & 0.564                               \\
{[}Ba/Fe{]}                & dex                    & 0.064     & 0.065        & 0.16          & 0.083       & 0.055                                & 0.073                            & -                               & 0.073                               \\
{[}Nd/Fe{]}                & dex                    & 0.08      & 0.078        & 0.191         & 0.199       & 0.098                                & 0.228                            & -                               & 0.228                               \\
{[}Sm/Fe{]}                & dex                    & 0.185     & 0.191        & 0.449         & 0.391       & 0.312                                & 0.294                            & -                               & 0.294                               \\
{[}Eu/Fe{]}                & dex                    & 0.299     & 0.252        & 0.59          & 0.477       & 0.111                                & 0.223                            & -                               & 0.223   \\                           
\hline
\end{tabular}
\label{tab:spreads and uncertainties}
\end{table*}

Table \ref{tab:summary of fitting} shows the overall number of stars that were fitted for each cluster and how many stars were successfully fitted. To see whether the star is successfully fitted, we used some basic cuts. The cuts we used are $v_\mathrm{mic}>0 \, \mathrm{km\, s^{-1}}$, $15\, \mathrm{km\, s^{-1}} \,>v_\mathrm{broad}>0 \,\mathrm{km\, s^{-1}}$, $5$ dex $>\log(g)>$ $2$ dex. We chose these values for the quality cuts as the stars outside of these ranges have unreliable abundances, thus contributing erroneous scatter to our data. As we are looking at the trends of abundances, we decided to only use stars that was fitted with more than 100 independent samples on average, as we found it to be a sufficient number to compare abundance trends. In ASCC 16 and Upper Scorpius, our fitting software struggled to fit spectra successfully. This is because these two clusters are our two youngest clusters at $10^{7.10}$ and $10^{6.90}$ years, respectively; thus, the stars will be more likely to have a higher rotation \citep{nielsenRotationPeriods122013,godoy-riveraStellarRotationGaia2021}, as a result not pass our $v_\mathrm{broad}$ of our quality cuts. The other reason is that the SNRs for those clusters are among the lowest, at 47.0 and 35.5, respectively. SNR is defined as the average ratio of the uncertainty over normalised flux in the spectra.

\begin{figure*}
\includegraphics[scale=1]{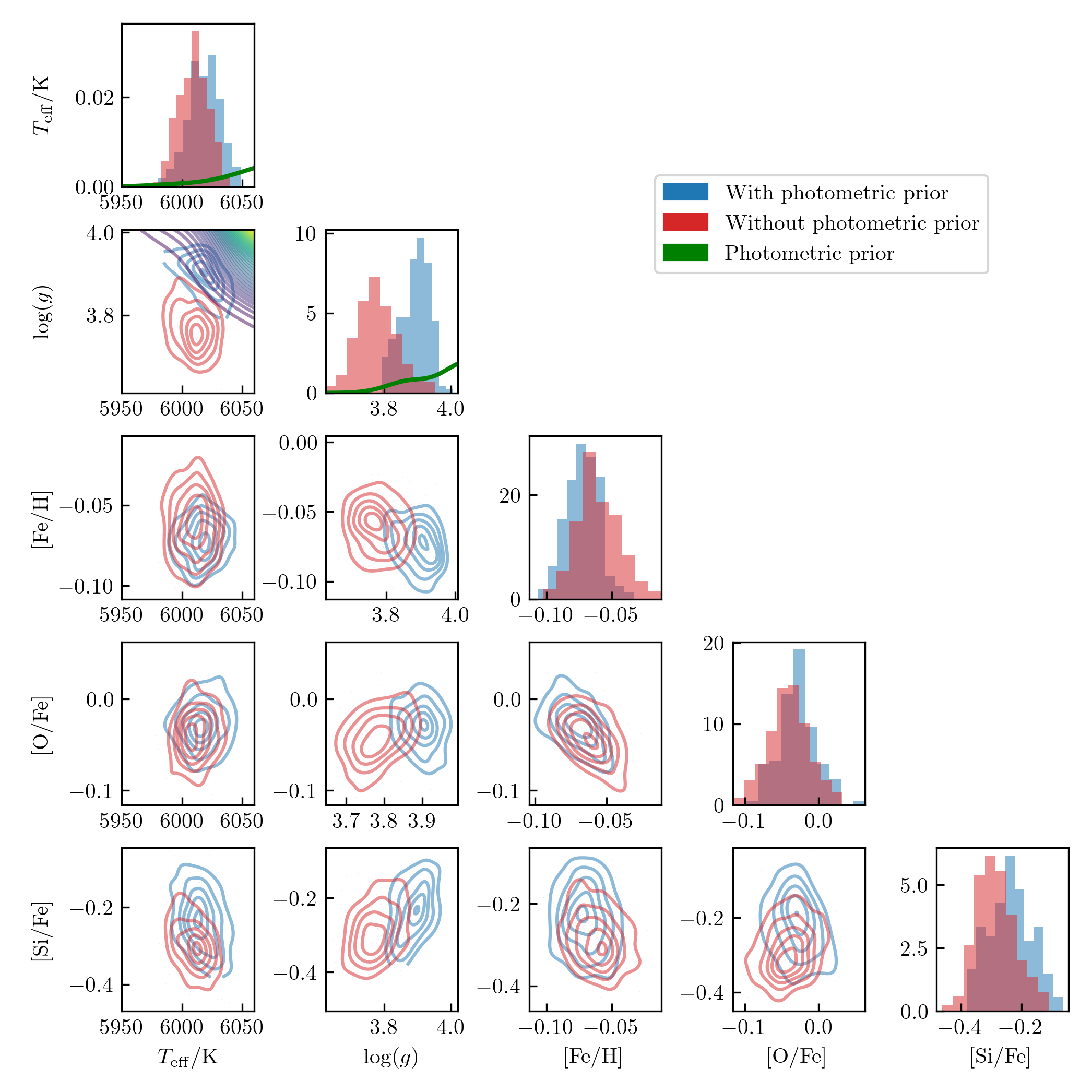}
\caption{Corner plot of a star with a \textit{Gaia} id $=604907110979352576$. The photometric prior is shown in green with it being marginalised values on the $T_{\rm{eff}}$ and $\log(g)$ histogram plots. The probability distribution of the fit with and without the photometric prior is shown in blue and red respectively.}
\label{fig:corner abundances}
\end{figure*}

In Fig. \ref{fig:corner abundances}, we can see an example of the effect of using a joint $T_{\rm{eff}}$ and $\log(g)$ prior when fitting spectra. The probability distributions are displayed using a Gaussian KDE, of a star that fitted with more than an average of 400 individual samples for each parameter. We can clearly see that the photometric prior affects the value of $T_{\rm{eff}}$ and $\log(g)$. As the fit using the prior has a different $T_{\rm{eff}}$ and $\log(g)$ to the fit without using the prior, we can see that the photometric prior has affected the probably distribution of [Fe/H], [O/Fe], and [Si/Fe]. We chose these three abundances as they show dependence on $T_{\rm{eff}}$ and $\log(g)$ most clearly. Fig. \ref{fig:corner abundances} also demonstrates that the $T_{\rm{eff}}$ and $\log(g)$ prior cannot be approximated by a multivariate normal distribution.

\begin{figure}
    \centering
    \includegraphics{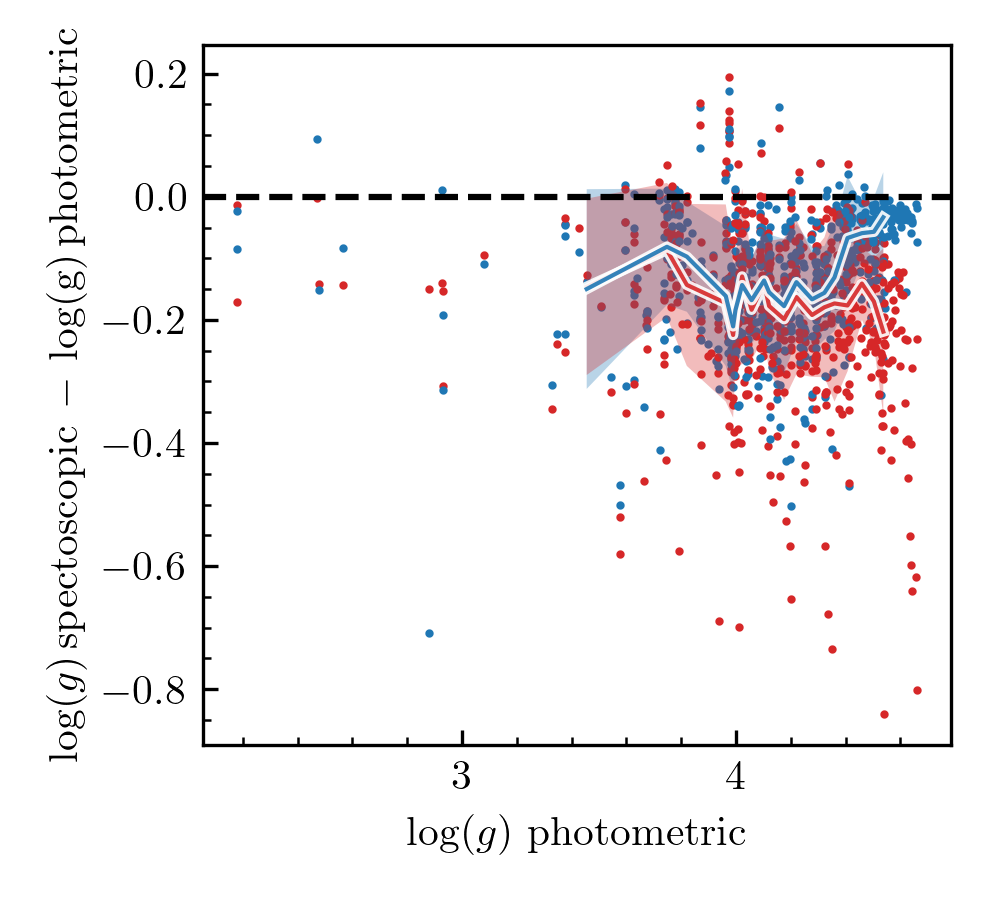}
    \includegraphics{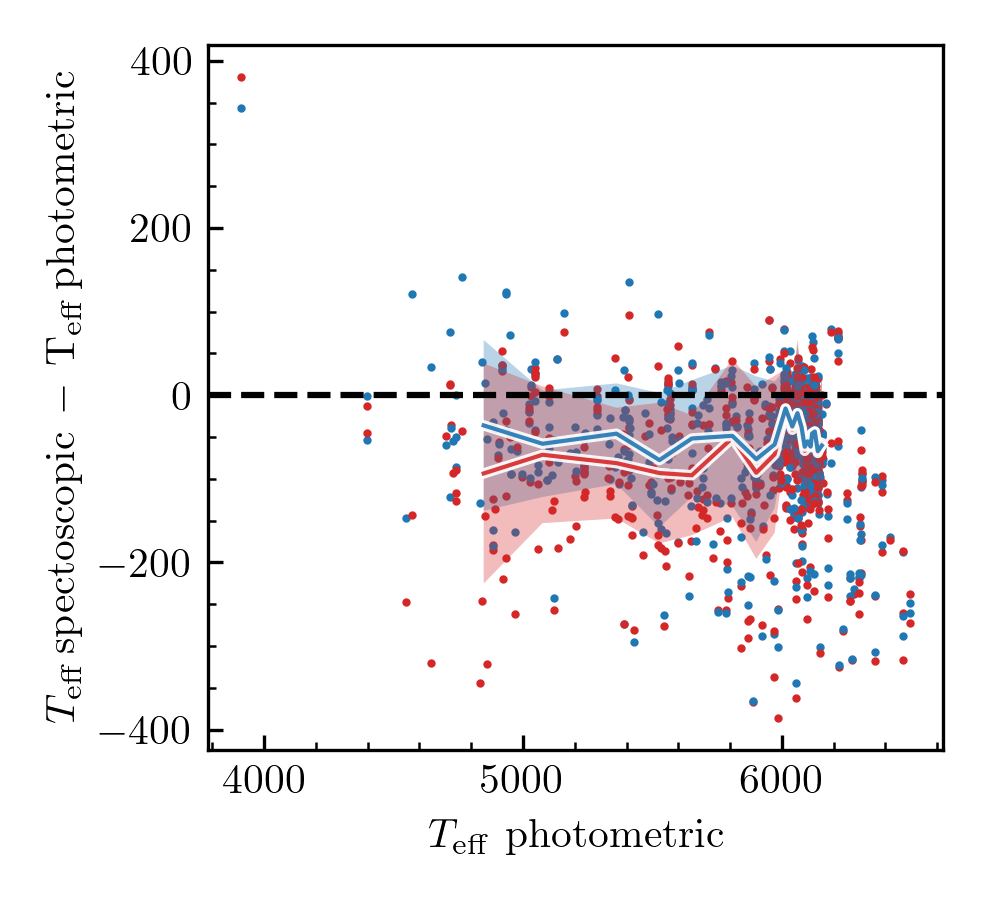}
    \includegraphics{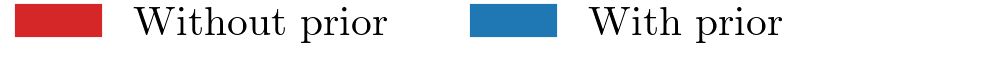}
    \caption{The top panel is a scatter plot of photometric $\log(g)$ vs spectroscopic $\log(g)$ minus photometric $\log(g)$. The bottom is a scatter plot of photometric $T_{\rm{eff}}$ vs spectroscopic $T_{\rm{eff}}$ minus  photometric $T_{\rm{eff}}$. For both panels the spectroscopic $\log(g)$ in the blue points are calculated using the photometric prior and the red points without. For both panels we used all the stars from all open clusters that passed the quality cuts.  The solid lines are the median of a bin containing 5\% of the stars in the cluster. The coloured area shows $\pm$ 1 $\sigma$}
    \label{fig: overall Residual logg teff}
\end{figure}

\begin{figure}
    \centering
    \includegraphics{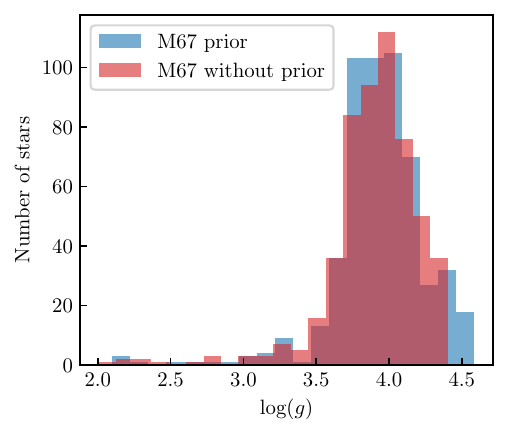}
    \includegraphics{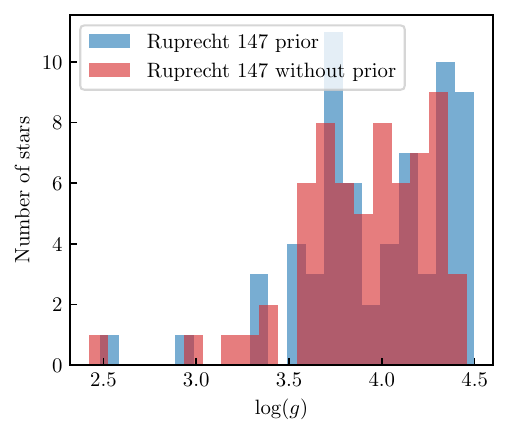}
    \includegraphics{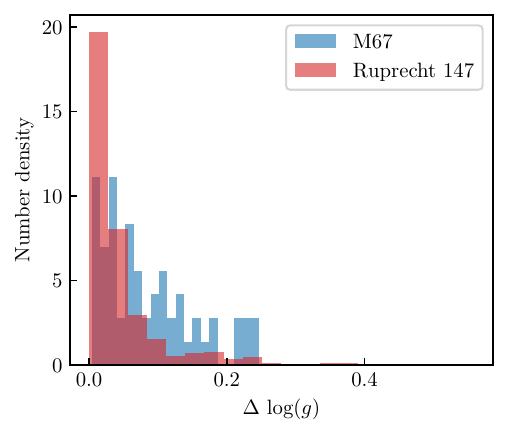}
    \caption{The top and middle panel shows the spread of $\log(g)$ in M67 and Ruprecht 147 when calculated with and without the photometric prior. The bottom panel shows the change in $\log(g)$ between the stars fitted with and without the photometric prior.}
    \label{fig:m67 logg}
\end{figure}

M67 and Ruprecht 147 fitted abundances and stellar parameters can be seen in Fig. \ref{fig:M67 Abundance spread} and Fig. \ref{fig:Ruprecht 147 abundance spread}, the other clusters can be seen in Appendix \ref{appendix: abundances}. In the figures comparing stellar parameters and abundances fitted with and without the photometric prior, we only show stars that have passed the quality cuts of both types of fitting. We see a large effect on the accuracy of $\log(g)$, as seen in Fig. \ref{fig:M67 Abundance spread}. This is especially true in areas that are not most sensitive to isochrone changes (i.e., not at the turn-off point). This increase in accuracy causes a new population of stars located at $4.20-4.55$ dex to appear, as shown in Fig. \ref{fig:m67 logg}. The $\log(g)$ change in M67 is mostly below a $\log(g)$ of 0.1, with a small population of stars that has a change in $\log(g)$ above 0.1 dex. This shift in $\log(g)$ will be the source of the different abundance trends in the abundances fitted using the photometric prior and without. Thus, we do not expect a change in abundance in stars between $3.5<\log(g)<4.20$ as the majority of the stars have the same $\log(g)$ when fitted with or without the photometric prior. The effect of the photometric prior on the fitting regime will be most apparent in the $4.20<\log(g)<4.55$ range as there lies the shifted population. From the Kiel diagrams, we found that our fitting regime struggled to fit stars with $T_{\rm{eff}}$<$4\,500$ K and a $\log(g)$ >4.20 dex. Our model tended to vastly underestimate the $\log(g)$ of those stars. This underestimation is also shown in Fig. \ref{fig: overall Residual logg teff}. The figure shows the difference between photometric $\log(g)$ and $T_{\rm{eff}}$ to spectroscopic $\log(g)$ and $T_{\rm{eff}}$ calculated with and without the prior, in all open clusters. We see that when calculating spectroscopic $\log(g)$ without using the photometric prior, we underestimate $\log(g)$ of stars with a $\log(g)$ >4.20 dex. We found that for stars that have a photometric temperature of lower than $5\,800$ K, the photometric prior had a more substantial effect on spectroscopic $T_{\rm{eff}}$ compared to stars with a photometric temperature of more than $5\,800$ K. 

\begin{figure*}
    \centering
    \includegraphics[height=0.90\textheight,width=\textwidth]{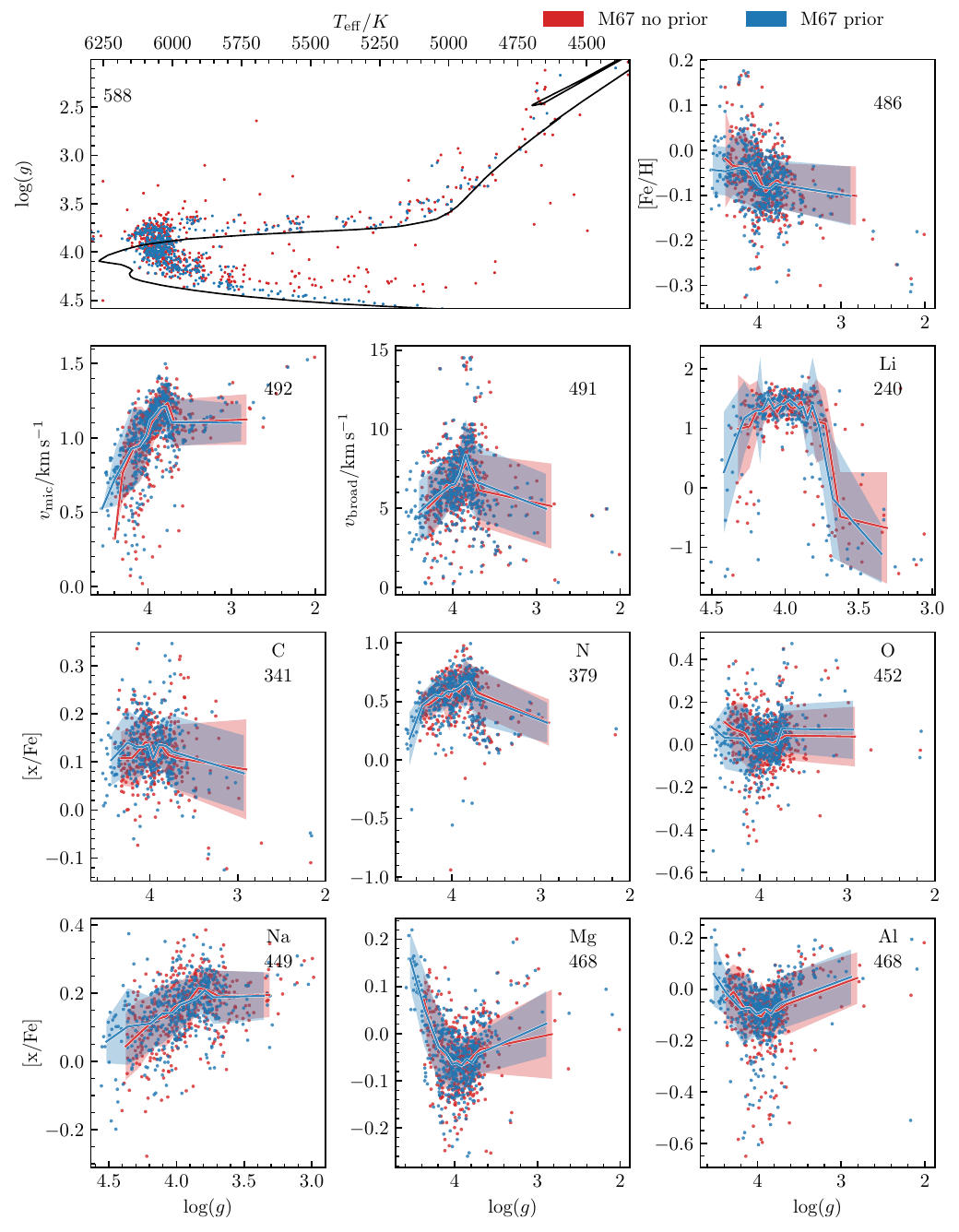}
    \caption{Stellar parameters and abundance spread with respect to $\log(g)$ for M67. Blue dots are the stars that were analysed using the photometric prior and red without the prior. The number on the top right corner of each graph indicates the number of stars present in each graph. The Kiel diagram panel has the best fitting isochrone used in getting the photometric parameters. The solid lines are the median of a bin containing 10\% of the stars in the cluster, except for the Li panel where it is 5\% of the stars in the cluster. The coloured area shows $\pm$ 1 $\sigma$.}
    \label{fig:M67 Abundance spread}
\end{figure*}
\begin{figure*}
\addtocounter{figure}{-1}
    \centering
    \includegraphics[height=0.96\textheight,width=\textwidth]{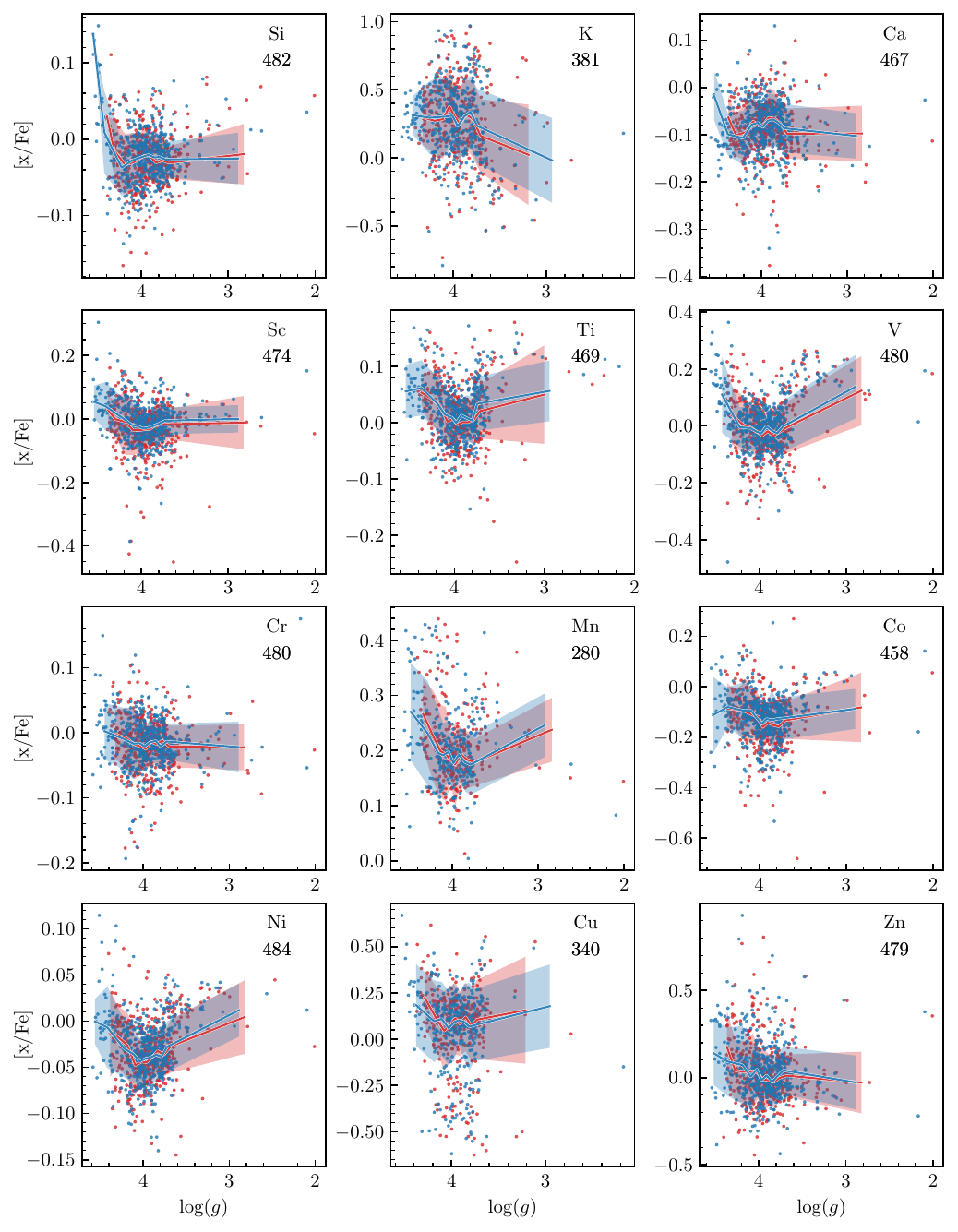}
    \caption{Continuation of Fig. \ref{fig:M67 Abundance spread}}
    \label{fig:M67 Abundance spread page 2}
\end{figure*}
\begin{figure*}
\addtocounter{figure}{-1}
    \centering   \includegraphics[width=\textwidth]{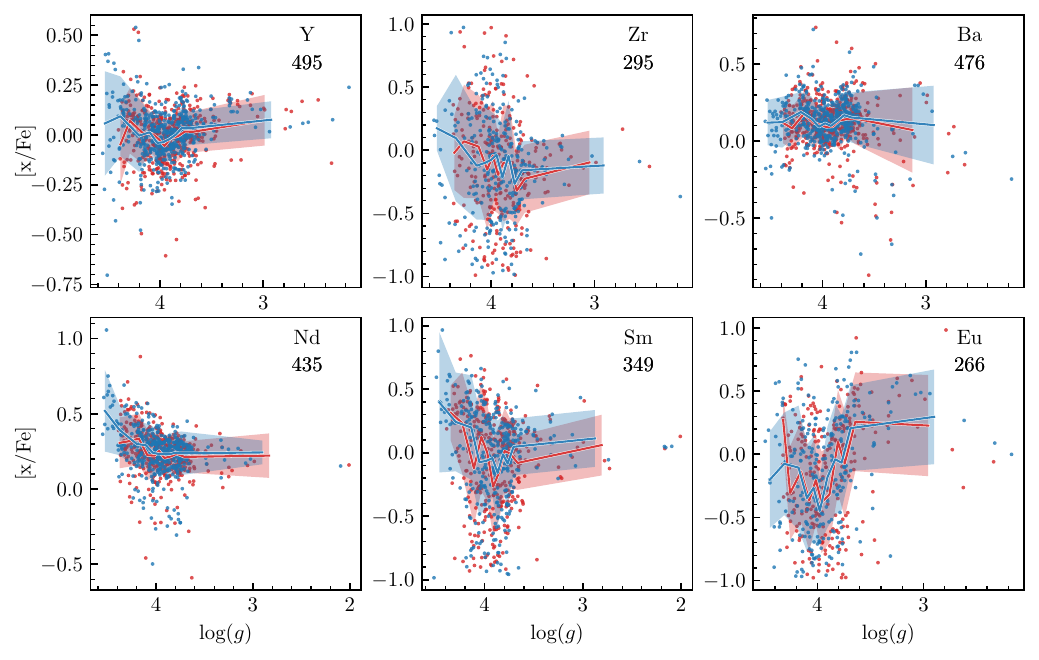}
    \caption{Continuation of Fig. \ref{fig:M67 Abundance spread}}
    \label{fig:M67 Abundance spread page 3}
\end{figure*}
\begin{figure*}
\centering
\includegraphics[height=0.90\textheight,width=\textwidth]{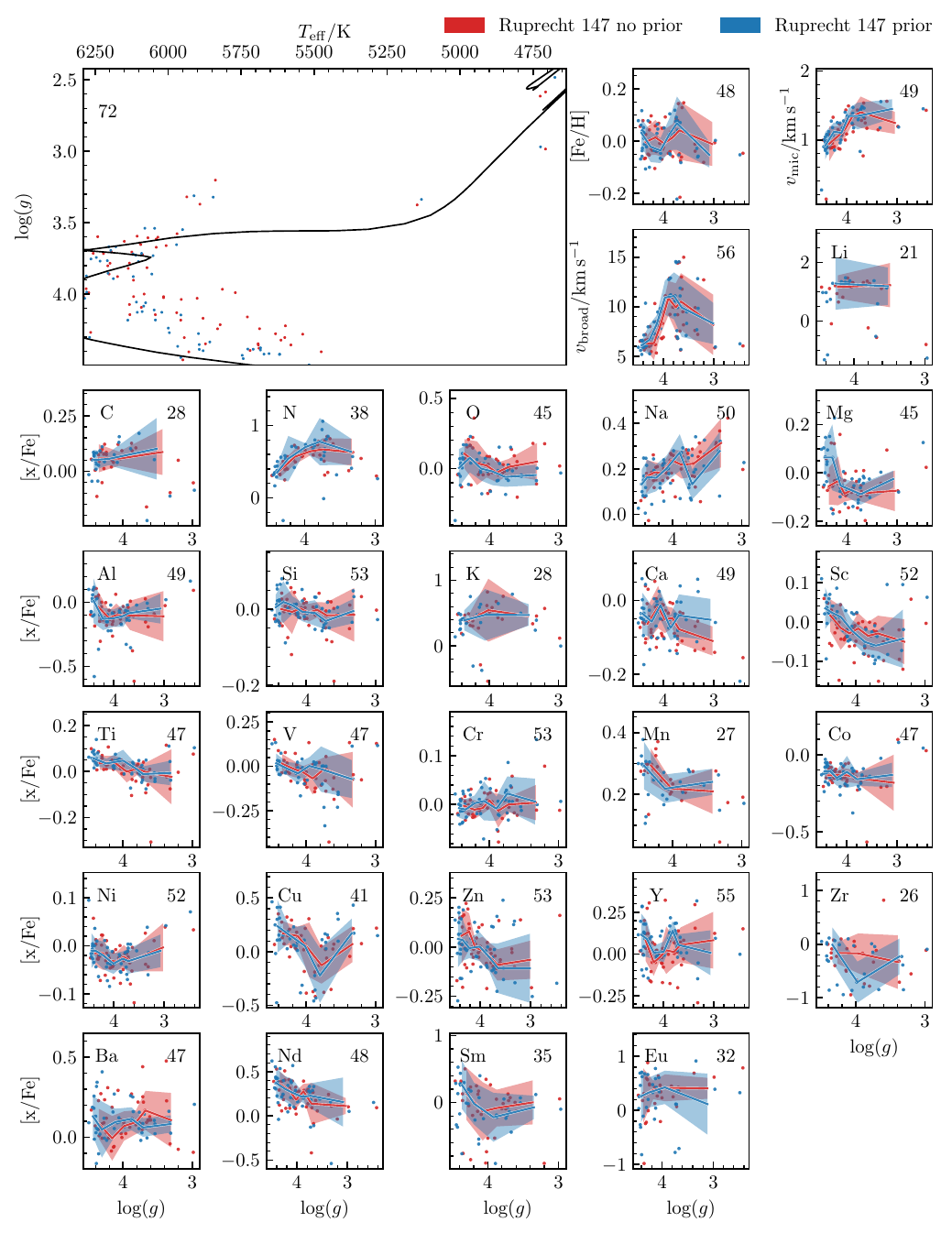}
    \caption{Stellar parameters and abundance spread with respect to $\log(g)$ for Ruprecht 147. Blue dots are the stars that were analysed using the photometric prior and red without the prior. The number on the top right corner of each graph indicates the number of stars present in each graph. The Kiel diagram panel has the best fitting isochrone used in getting the photometric parameters. The solid lines are the median of a bin containing 7 stars, with the coloured area showing $\pm$ 1 $\sigma$.}
    \label{fig:Ruprecht 147 abundance spread}
\end{figure*}

The average spectroscopic values of [Fe/H] did not change significantly when comparing stars fitted with or without the photometric prior. The small change that occurred did not shift the spectroscopic [Fe/H] towards the average metallicity chosen to generate the isochrones.

In clusters with stars with a small range of $\log(g)$, the effect of the photometric prior is large enough that there is no overlap in $\log(g)$ within the population fitted with and without the photometric prior, thus making it hard to compare trends in those clusters. Consequently, we have concentrated our analysis mostly on M67 and Ruprecht 147, which have a wide range of stars in $\log(g)$ and so a large overlap.

Within each open cluster, our abundances fitted with or without the photometric prior do not have a linear trend with respect to $T_{\rm{eff}}$ or $\log(g)$. We attribute this to using non-LTE models to create the training set data and fitting the majority of the wavelength range. We found that we fit most abundances within the ranges set out in the training set of our neural model. [Eu/Fe] and [Zr/Fe] were the elements that were most likely to fit with a value that was outside of our training set. The [Fe/H] of M67 and Ruprecht 147 decreases as $\log(g)$ decreases, which we attribute to systematic errors in our model. The higher-order abundance trends in M67 and Ruprecht 147, we attribute to atomic diffusion. This is because the higher-order trends present themselves as dips around the turn-off point, which is at a $\log(g)$ of $\sim 4$ for M67 and Ruprecht 147 \citep{liu_chemical_2019,gaoGALAHSurveyVerifying2018}. These dips are present in studies that have modelled atomic diffusion in M67 \citep{liu_chemical_2019,gaoGALAHSurveyVerifying2018}.

The higher-order trends within M67 are visible in [Fe/H], [Ni/Fe], [Cu/Fe], [O/Fe], [V/Fe], [Mg/Fe], [Ti/Fe], [Sc/Fe], [Al/Fe], [V/Fe] , [Ba/Fe] and [Mn/Fe]. The higher-order trends are visible in the abundances fitted with and without the photometric prior. For these elements, there is a clear dip at around $\log(g)$=4. The depth varies, with [Ni/Fe] having a magnitude of 0.05 dex, while [Mn/Fe] has a magnitude of 0.1 dex. These depths match the expected dip shown in \citet{liu_chemical_2019}, with shallower dips in [Ni/Fe] with a depth of 0.05. Our [Ca/Fe] abundances stay flat, as predicted by \citet{liu_chemical_2019}.

In M67 there is an increased range in $\log(g)$ in the data fitted using the photometric prior compared to the data fitted without the photometric prior. The increased range in $\log(g)$ enables the observation of how the trends in abundances at a $\log(g)$ larger than 4.2 dex. In most of the abundances, the trends above a $\log(g)$ of 4.2 dex in the fit using the photometric prior are just a continuation of the trends in the fit without using the photometric prior. However, in [Si/Fe] and [Nd/Fe], the trend differs. Furthermore the stars that were analysed without using the photometric prior and were shifted to higher than 4.2 dex in the analysis using the photometric prior, will add scatter to the abundances. As a result, it becomes more challenging to discern higher-order trends in the analysis without the photometric prior.

In Ruprecht 147, the effect of the photometric prior on the spread of $\log(g)$ is also observable. Stars with $4.0<\log(g)$ show that the fit without the photometric prior underestimates $\log(g)$. The ranges of $\log(g)$ of the stars fitted using the prior and without are similar in Ruprecht 147. This stands in contrast to M67 which has a population of stars with a $\log(g)$ of 4.5 in the analysis using the photometric prior which is not present in the analysis without the photometric prior, as illustrated in Fig. \ref{fig:m67 logg}. An atomic diffusion dip at around $\log(g)$ of $4.1$ dex is visible in [Fe/H], [Al/Fe], [Mg/Fe], [Ni/Fe], [Cu/Fe], and [Mn/Fe]. However, in [Fe/H], [Al/Fe], and [Mg/Fe], the dip was only present in the abundances that were fitted using the photometric prior.

The abundances calculated using the photometric prior in the M67 and Ruprecht 147 open clusters were both more accurate than the abundances calculated without using the photometric prior. In M67, the increase in accuracy was caused by the population's $\log(g)$ of $4.2-4.55$ dex, which only occurs when fitting using the photometric prior; however, other than the stars that get shifted to this population, the rest of the population's $\log(g)$ did not change significantly. The new population causes the effects of atomic diffusion to be more visible, as we could see the effect of atomic diffusion on stars with a $\log(g)$$>4.2$. Unlike in M67 in Ruprecht 147, most of the stellar population shifted its $\log(g)$ when fitted using the photometric prior and without. This caused larger differences in the abundance trends. We could see evidence of atomic diffusion more clearly using the photometric prior compared to without. These two clusters show that using photometric priors causes better $\log(g)$ compared to the analysis without using the photometric priors, thus enabling us to see atomic diffusion.

A comparison of our stellar parameters and abundances fitted using the photometric prior (blue) and GALAH DR4 (pink) in M67 and Ruprecht 147 can be seen in Fig. \ref{fig:M67 Abundance spread GALAH } and Fig. \ref{fig:Ruprecht 147 abundance spread GALAHDR4}, and for the other clusters they can be seen in Appendix \ref{appendix: abundances}. Both data set uses the same quality cuts, but with the GALAH DR4 data has an additional quality cut of $\texttt{flag\_sp}==0$. The medians and standard deviations of the stellar parameter and abundances in Fig. \ref{fig:M67 Abundance spread GALAH } and Fig. \ref{fig:Ruprecht 147 abundance spread GALAHDR4} differs to Fig. \ref{fig:M67 Abundance spread} and Fig. \ref{fig:Ruprecht 147 abundance spread} slightly. This is because the Fig. \ref{fig:M67 Abundance spread} and Fig. \ref{fig:Ruprecht 147 abundance spread} the data has to pass the quality cuts for both the data analysed with and without the photometric prior, but in Fig. \ref{fig:M67 Abundance spread GALAH } and Fig. \ref{fig:Ruprecht 147 abundance spread GALAHDR4} it only has to pass the quality cuts of the data analysed with the photometric prior.

Comparing our results to GALAH DR4, we found that GALAH DR4 $T_{\rm{eff}}$ and $\log(g)$ followed the Kiel diagram of the best fitting isochrone used in getting the photometric prior better than our work. This could be caused by how in GALAH DR4, a different neural network is used to calculate stellar parameters and abundances depending on its $T_{\rm{eff}}$, $\log(g)$, and [Fe/H]. i.e., a different neural network is used to analyse Sun-like stars, red clump stars, etc. Unlike in this work, where we used a single neural network per band for all the stars analysed. The difference between our $\log(g)$ and GALAH DR4 $\log(g)$ could be due to GALAH DR4 using astrometry from \textit{Gaia} and photometry from 2MASS to calculate $\log(g)$, which is different from this work. A 0.1 dex decrease in [Fe/H] was found in most clusters. The absolute average values of [Mn/Fe], [K/Fe], and [Ni/Fe] are significantly different. For Ni, Mn, and K, this can be explained by this work not using a non-LTE model for these specific elements when generating spectra for training our neural network. We also found significant differences in the trends of Carbon and Nitrogen; this is explained by the differences in the methods of measuring Carbon. GALAH DR4 uses molecular features such as the $\rm{C_2}$ and CN molecules, and this work uses weak Carbon and Nitrogen lines. Thus leading to the better abundances in GALAH DR4 for those two elements. This work's lithium spread is also much higher than GALAH DR4. When comparing the abundances of stars in M67 with $\log(g)$ above 4.4 dex, we can see that the trends of [Si/Fe], [Ni/Fe], [Cu/Fe], [Sc/Fe], [Ti/Fe] and [Cr/Fe] are significantly different. Atomic diffusion is much more visible in this work in [Mg/Fe], [Si/Fe], [Ni/Fe], [O/Fe], [Sc/Fe], [Ti/Fe], [Cu/Fe], [Sm/Fe], [Ba/Fe], and [Al/Fe] in M67. In Ruprecht 147, the difference is larger, with atomic diffusion only visible in this work in [Cu/Fe], [Co/Fe] and [Ni/Fe]. The spread of Calcium in all open clusters is lower in this work compared to GALAH DR4.

\begin{figure*}
    \centering
    \includegraphics[height=0.90\textheight,width=\textwidth]{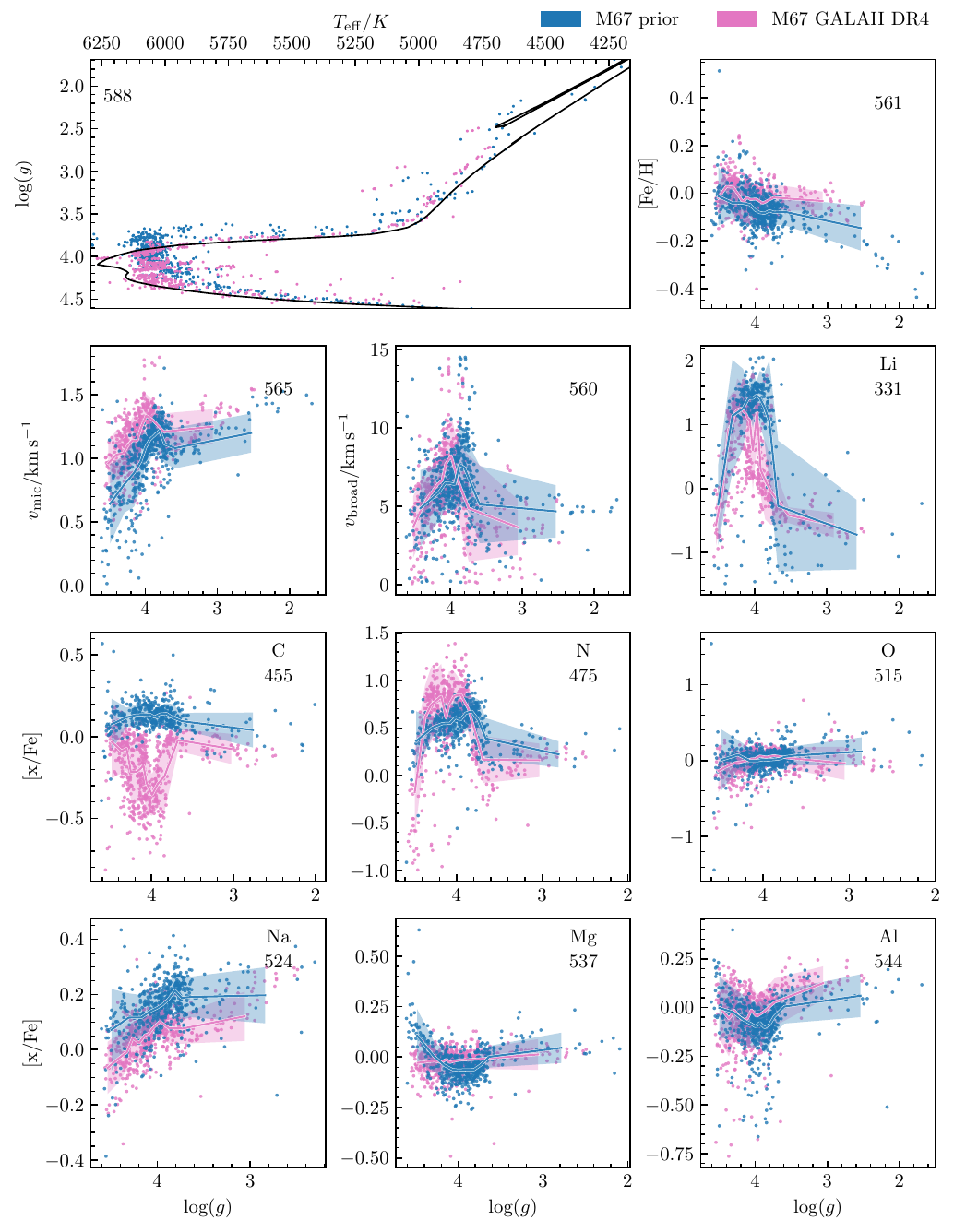}
    \caption{Stellar parameters and abundance spread with respect to $\log(g)$ for M67. Blue dots are the stars that were analysed using the photometric prior and green dots are from GALAH DR4. The number on the top right corner of each graph indicates the number of stars present used in the photometric prior graph. The Kiel diagram panel has the best fitting isochrone used in getting the photometric parameters. The solid lines are the median of a bin containing 10\% of the stars in the cluster, except for the Li panel where it is 5\% of the stars in the cluster. The coloured area shows $\pm$ 1 $\sigma$.}
    \label{fig:M67 Abundance spread GALAH }
\end{figure*}
\begin{figure*}
\addtocounter{figure}{-1}
    \centering
    \includegraphics[height=0.96\textheight,width=\textwidth]{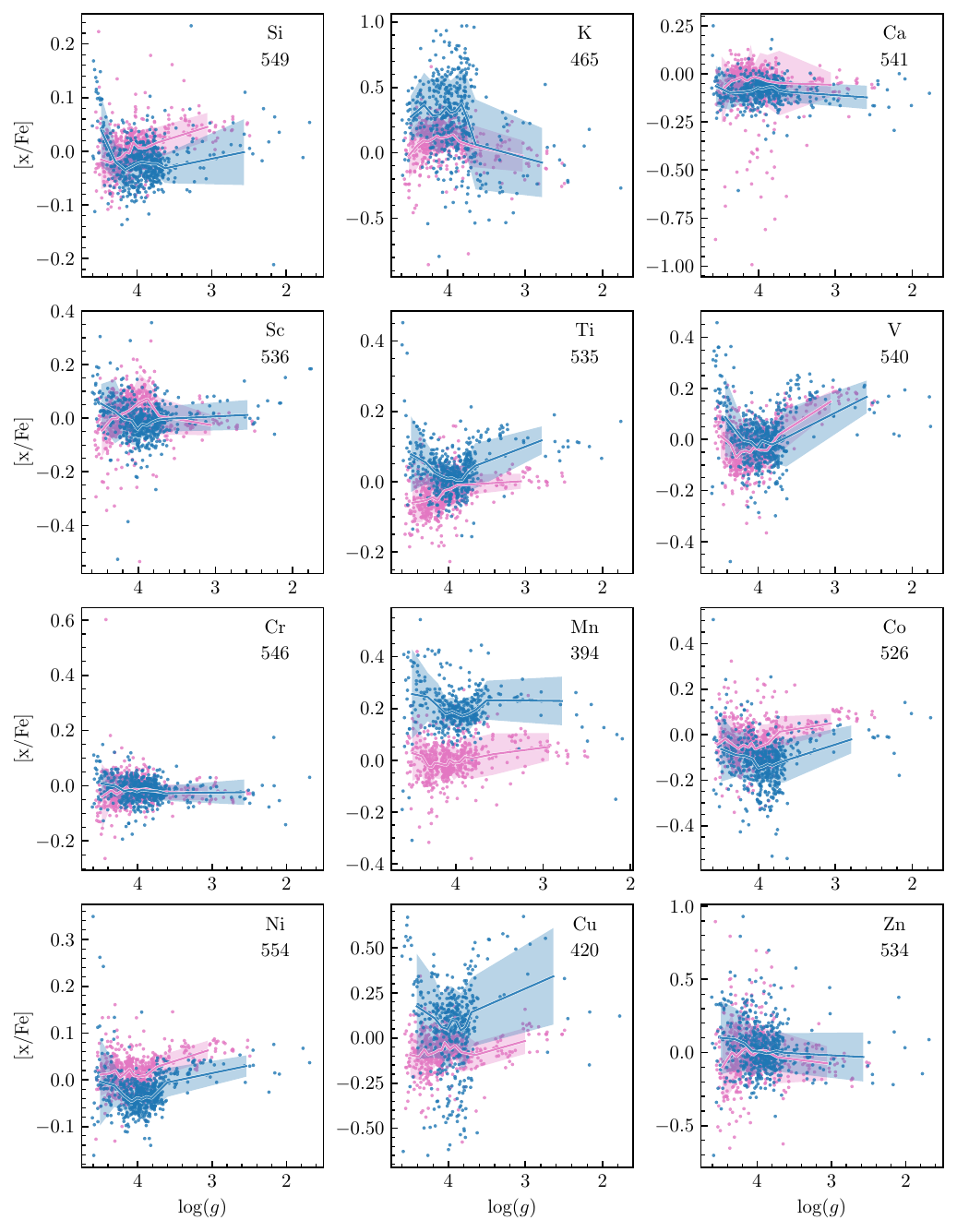}
    \caption{Continuation of Fig. \ref{fig:M67 Abundance spread GALAH }}
    \label{fig:M67 Abundance spread GALAH page2}
\end{figure*}
\begin{figure*}
\addtocounter{figure}{-1}
    \centering
    \includegraphics[width=\textwidth]{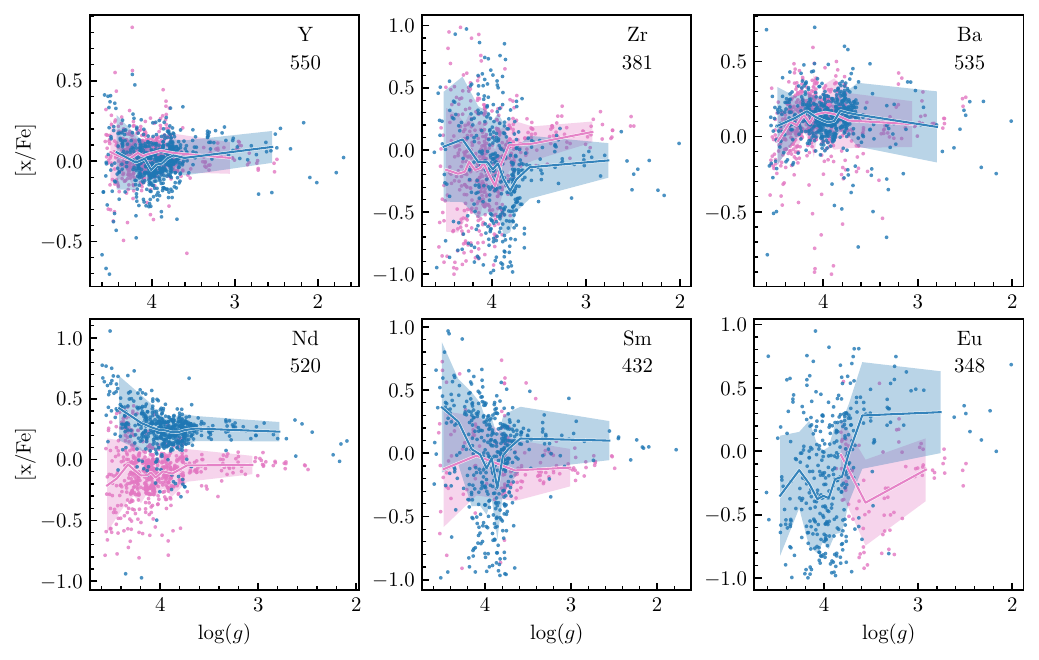}
    \caption{Continuation of Fig. \ref{fig:M67 Abundance spread GALAH }}
    \label{fig:M67 Abundance spread GALAH page3}
\end{figure*}
\begin{figure*}
    \centering
    \includegraphics[height=0.90\textheight,width=\textwidth]{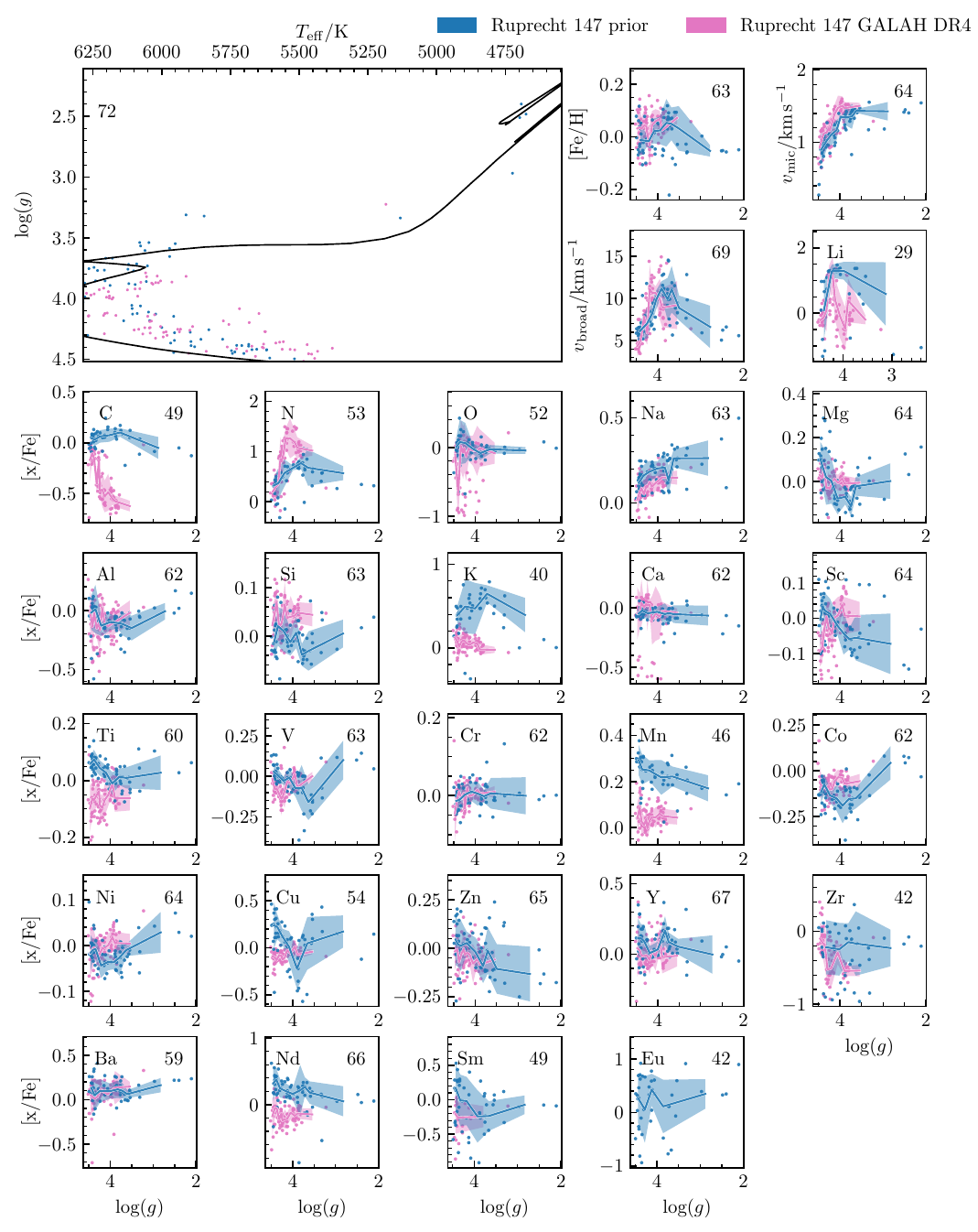}
    \caption{Stellar parameters and abundance spread with respect to $\log(g)$ for Ruprecht 147. Blue dots are the stars that were analysed using the photometric prior and green dots are from GALAH DR4. The number on the top right corner of each graph indicates the number of stars present used in the photometric prior graph. The Kiel diagram panel has the best fitting isochrone used in getting the photometric parameters. The solid lines are the median of a bin containing 7 stars, with the coloured area showing $\pm$ 1 $\sigma$.}
    \label{fig:Ruprecht 147 abundance spread GALAHDR4}
\end{figure*}

\section{Spectroscopic precision}\label{spectroscopic uncertainties}

MCMC provides an opportunity to more meticulously sample the probability space while fitting, as it gives the exact shape of the PDF compared to methods such as using the covariance matrix. Thus, we have explored how well MCMC estimates precision. In this work we only measure the precision, defined as the statistical variability of a measurement. We do not measure accuracy, which is defined as how close a measurement is to the true value.

\begin{figure}
    \centering
    \includegraphics{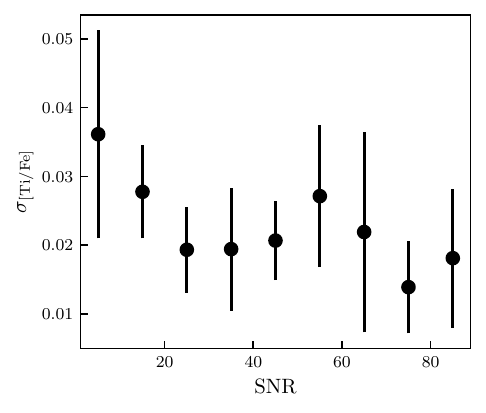}
    \includegraphics{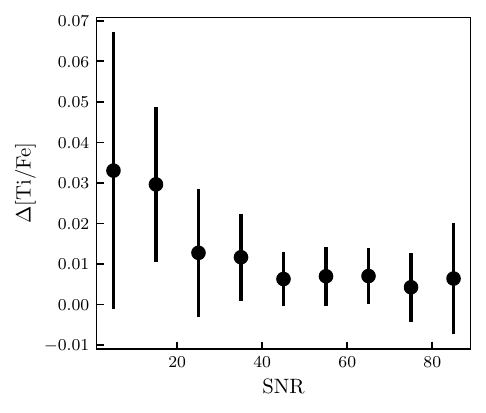}
    \caption{The top panel shows the effect of changing the SNR of a star on the precision of [Ti/Fe]. The bottom graph shows the effect of changing the SNR of a star on the scatter of [Ti/Fe]}
    \label{fig:SNR role}
\end{figure}

We first investigated the role that the SNR of spectra has on the precision and the scatter of the fitted spectroscopic parameter. We investigated this by picking 10 spectra with an SNR of higher than 100 with $6100 \, K>T_{\rm{eff}}>5000 \, K$ and $3.9$ dex $ >\log(g)>2.9$ dex. We added Gaussian noise to the spectra to create spectra with an SNR ranging from 10 to 90 in steps of 10. We fitted them without using the photometric prior. To investigate the effect of SNR on the average precision, we took the average precision of [Ti/Fe] in bins of 10 in SNR, as shown in the top panel in Fig. \ref{fig:SNR role}. In these spectra the deepest blended Titanium lines are $\sim 0.25$ and $\sim 0.15$ in normalised flux for non-blended lines. When the SNR is above $\sim$ 30, there is a minimal decrease in uncertainty with the increase in SNR. To examine the change in scatter, we first calculated the mean [Ti/Fe] for each star. Subsequently, we measure the absolute difference between the mean [Ti/Fe] and the corresponding fitted [Ti/Fe] value at different SNRs. Lastly, we calculate the average change in [Ti/Fe] values within bins of 10 SNR units. The average change is shown in the bottom panel in Fig. \ref{fig:SNR role}. At all SNRs, the average change is smaller than the average uncertainty. Under an SNR of 30, there is a significant increase in uncertainty compared to the higher SNR spectra. The effect of SNR on the other fitted parameters is similar to [Ti/Fe] except for the abundances of elements with the weakest lines.

Table \ref{tab:spreads and uncertainties} shows the effect of the prior on the precision of individual abundances. These precisions are the average precision of the spectra that fitted with more than 400 individual samples. We do not see worse precision of the abundances fitted using the photometric prior compared to abundances fitted without using the photometric prior. This is because even though the photometric prior's spread is much larger than the value of the precision in the spectroscopic fit without using the photometric prior, their most probable values differ from each other. Therefore, the overlap of the probability space is smaller. This is shown in the precision of $\log(g)$. The difference between the photometric $\log(g)$ and the $\log(g)$ obtained using spectroscopy without using the photometric prior is large. Hence, we see a decrease in the uncertainty in $\log(g)$ when fitting using spectroscopy without the photometric prior. As the precision in $T_{\rm{eff}}$ and $\log(g)$ do not change significantly, we do not observe a significant change in the precision of the abundances when fitting with the photometric prior compared to without using the photometric prior for the rest of the parameters.

To validate our spectroscopic precision calculated using MCMC, we have estimated precision in two additional ways. The first is to use the scatter of abundances as a measure of precision. To take into account the trends caused by effects such as atomic diffusion, we have selected stars within a narrow range of $\log(g)$ where their abundances should not change significantly. We selected the range between $4.0<\log(g)<4.1$, as the abundances should theoretically change by less than 0.005 dex \citep{liu_chemical_2019}. Thus, by calculating the scatter of abundances in that section, we can compare the precision from MCMC and the scatter. We decided to only use M67 as the abundance trends calculated in \citet{liu_chemical_2019} are specifically made for the ages and metallicity of M67, and as we are only using a small range of $\log(g)$, all the other open clusters do not have enough fitted members in this range. To reduce the scatter introduced by the range of SNR, we have only used stars with an SNR of $35-45$, decreasing the role the SNR will have on the scatter of abundances. Using those cuts we have 88 stars to calculate the spread. The spread of abundances is shown in Table \ref{tab:spreads and uncertainties}. We found that the median abundance precision estimated by MCMC is underestimated by 30\% compared to precision measured from the scatter in M67. Thus MCMC is more optimistic in precision compared to the precision calculated using M67 scatter.

The second method to evaluate our abundance precision uses the scatter in abundances derived from repeat observations of stars. We selected repeat observations where their SNR only changes by 10\% between the observations. We found that the median precision estimated by MCMC is underestimated by 9\% compared to the precision estimated by the repeated measurements. However, if we only compare abundances, the precision is only underestimated by 1\%. To compare precisions we are using the median value rather than the mean as there is an outlier with the estimate of the precision of [Li/Fe]. We found that these two results show that the abundance precision calculated using MCMC is accurate, as their value is comparable to the repeat observation. The precision calculated by MCMC for $T_{\rm{eff}}$, [Fe/H] and $v_{\rm{mic}}$ is underestimated. Thus MCMC is more optimistic in the precision of these parameters compared to the precision calculated in the repeat observations.

In GALAH DR4, we calculated the total uncertainty of all fitted parameters by adding in quadrature, its precision, and its accuracy. Note that accuracy is only available for $T_{\rm{eff}}$, $\log(g)$, $v_{\rm{mic}}$, [Fe/H], and $v_{\rm{broad}}$. In this work, we only measure precision, so we will not compare accuracy. GALAH measures the precision of the fitted parameters in two main ways. The first is by using repeat observations and seeing how the fitted values differ in this observation. The second method is by using the precision derived from the covariance matrix calculated by \texttt{SME} while fitting. Whichever of the two methods produces the biggest value becomes precision. The average precision, accuracy, and total uncertainty for the stars used in this work and GALAH DR4 are shown in table \ref{tab:spreads and uncertainties}. We find that GALAH DR4 precision is 25\% worse compared to the precision calculated by MCMC, 20\% better compared to our precision calculated using repeat observations, and 17\% better compared to the precision calculated by scatter in M67.

MCMC provides a more meticulous way to measure precision as it gives the exact shape of the probability distribution compared to using a covariance matrix like in GALAH. We find that GALAH estimates its precision well in abundances and other stellar parameters by combining both the precision acquired using repeat observations and its fitting procedure. GALAH has a good estimate of both abundance and the stellar parameters' precision.

\section{Discussion}\label{discussion}

Calculating accurate and precise photometric $T_{\rm{eff}}$ and $\log(g)$ was one of the key reasons we were able to measure accurate abundances for open cluster stars. One of the ways we increased the accuracy and precision of the photometric $T_{\rm{eff}}$ and $\log(g)$ measurements was to use a cluster distance prior. This method will only be valid when determining distances for stars that are inside the open cluster core. However, with the increased accuracy and precision of astrometry provided by \textit{Gaia} DR3, \citet{hunt_improving_2023,meingastExtendedStellarSystems2021,tarricqStructuralParameters3892022} have shown the possibility of determining members that are in the tidal tail of the open cluster. With these members, our method of determining distance might not be applicable as the members can be many pc away from the cluster centre.

Our method of using the cluster distance prior relies on having an accurate membership of each open cluster. Upper Scorpius and Orion (which contains the open cluster ASCC 16) contain multiple groups of open clusters with a very complex spatial structure. In Upper Scorpius, we did not pick a specific group inside; thus, the cluster distance prior did not succeed in decreasing the uncertainty of the stars in that cluster. In these complex regions, we note the importance of having radial velocities to be able to calculate accurate memberships to the open clusters \citep{kosDiscovery21Myr2019,miret-roigStarFormationHistory2022}.

The complex relationship between $T_{\rm{eff}}$ and $\log(g)$ in the photometric prior shown in this work is the result of how varying ages, metallicities, and extinction affect the shape of the isochrone. \citet{fuNewParsecData2018} has shown that the $\alpha-$element abundance can have a significant effect on the shape of the isochrone. Further studies should also be done into how changing the $\alpha$-element abundances in stellar models will affect the shape of the isochrone, thus affecting the shape of the joint $T_{\rm{eff}}$ and $\log(g)$ probability distribution.

We have chosen our stars to be far from the galactic plane reducing the effect of the reddening. We have also assumed that the extinction is the same for all stars in the open cluster. However, if we would like to more accurately fit isochrone to the HR diagram of these stars, especially those that are closer to the galactic plane. In that case, we should take into account the effects of differential extinction.

The accuracy of generating synthetic spectra is greatly affected by the line list from which the synthetic spectra are generated from. As discussed in \citet{jofre_accuracy_2019}, many transition data ($\log(gf)$ values) are not known accurately, and the line list available is incomplete with transitions missing. It is crucial to have an accurate and complete line list since we are fitting the majority of the wavelength range, rather than just the areas with the strongest non-blended lines. We also do not quantify the uncertainty that is associated with inaccurate atomic characteristics, such as the wavelength, transition probabilities, and properties of the atomic states responsible for these atomic transitions.

To increase the accuracy of generating synthetic spectra, we would also need to increase the accuracy of stellar models. Effects such as magnetic fields and stellar spots are not taken into account in this work. Depending on the spectral type, magnetic fields, and stellar spots can affect the equivalent width and depth of the absorption lines. The strength of magnetic fields and stellar spots varies during the activity cycle. Thus, measurements of $T_{\rm{eff}}$, [Fe/H], $v_{\rm{mic}}$, and chemical abundances will vary through the activity cycle \citep{spinaHowMagneticActivity2020}. When not taken into account, the magnetic fields and stellar spots can increase the scatter of the measured abundances.

Another effect that we did not take into account is the effect of diffuse interstellar bands (DIBs). These bands can affect the observed spectra at visible and near-infrared wavelengths. These effects although small ($\sim 0.5$ per cent of normalised flux \citep{vogrincicGALAHSurveyNew2023}) are still measurable in high precision spectroscopy. Further studies should use catalogues such as \citet{vogrincicGALAHSurveyNew2023} to correctly model DIBs when generating synthetic spectra.

These inaccuracies when generating synthetic spectra could be one of the reasons why we overestimated the precision of $T_{\rm{eff}}$, $\log(g)$, $v_{\rm{broad}} $, and $v_{\rm{mic}}$ using MCMC compared to the repeat precision measurements. With further improvements to the accuracy of generating synthetic spectra, we can reduce the need to mask certain wavelengths of the spectra, allowing us to use more of the wavelength range.

With the continued use of Bayesian statistics in fitting spectra, as demonstrated in \citet{buder_galah_2021,gent_sapp_2022}, works like this show that MCMC provides an accurate way to measure the probability space of the fitted spectra. However, MCMC is very computationally heavy, making it impractical to use in large surveys compared to our work. In GALAH DR4, an average spectrum takes 2 minutes to fit, compared to our 4 hours. In our work, the two functions that consume the most time are \texttt{scipy's} fast Fourier transform, which is used when equalising the resolution of the synthetic spectra with the observed spectra. The second function is \texttt{Numpy's} \texttt{einsum}, which is used to multiply vectors and matrices to generate the synthetic spectra. These two functions are well optimised, so the way to increase the sampling speed must come from somewhere else. Hamiltonian Monte Carlo (HMC) is a more efficient sampling method compared to MCMC. This is because HMC uses the probability gradient at each point to calculate its next step. This causes the auto-correlation time to be much shorter, so fewer samples are needed to acquire one independent sample. This has not been used in fitting spectra, as the method to generate spectra has traditionally not been able to be analytically differentiated. However, with the use of neural networks to generate spectra, we can analytically differentiate the spectra with respect to its training variables, thus getting the gradient of the probability space while fitting analytically. In further studies, this fact should be used to increase the efficiency of the fitting procedure.

\citet{liu_chemical_2019,dotterMESAIsochronesStellar2016a} and our work have further shown that the idea of chemically tagging stars with a wide range of $\log(g)$ and $T_{\rm{eff}}$ will only be possible with a greater understanding of how abundances in stars change as they age. This is because effects such as atomic diffusion are large enough that without taking them into account, it will be impossible to accurately group stars that come from the same star forming region. The effects of atomic diffusion on abundances depend on many variables, such as the age of the star, which element is being studied, the metallicity of the star, and the current evolutionary stage of the star. This complex relationship means that modelling the effect of atomic diffusion is a non-trivial thing to do for a large range of stars. Our work shows the necessity of having spectroscopic observations of stars in different evolutionary stages in each open cluster to be able to see the effect.

The effects of atomic diffusion are clear when looking at the higher-order trends of abundances with respect to $\log(g)$. This further highlights the importance of getting an accurate $\log(g)$; without the photometric prior, our $\log(g)$ would not have been accurate enough for us to see the effects of atomic diffusion. A better understanding of atomic diffusion will also allow us to measure the precision of abundances in open clusters, as we will be able to measure the scatter of abundances compared to the theoretical abundance trend lines.

\section{Conclusions}\label{congclusion}
In this paper, we examined the effects of using a joint $T_{\rm{eff}}$ and $\log(g)$ photometric prior on spectroscopic $T_{\rm{eff}}$, $\log(g)$, $v_{\rm{mic}}$, $ v_{\rm{broad}}$, and 26 elemental abundances of stars in open clusters. We used photometry and astrometry from the \textit{Gaia} DR3 catalogue. We selected members in the core of the open cluster, which were selected by simple cuts in position and proper motion space. To generate more accurate and precise photometric priors, we first reduced the uncertainty of the absolute magnitude of the stars by reducing the uncertainty in their distance measurement. We achieved this by assuming that the stars are normally distributed around the cluster centre. We computed the joint probability distribution of the location of the centre of each cluster and the size of each cluster. Using this prior, we recalculated the distances of the stars, achieving a reduction in uncertainties by a factor of 31\% compared to geometric distances in \textit{Gaia} DR3. The reduction of the uncertainties is important as it allows us to fit isochrones more precisely to the HR diagrams of the open clusters and consequently it allows us to calculate more precise $T_{\rm{eff}}$ and $\log(g)$ from each isochrone.

To create a joint $T_{\rm{eff}}$ and $\log(g)$ photometric prior, we first fitted isochrones to HR diagrams of the open clusters to see which ranges of ages, metallicities, and extinctions would encompass the majority of stars in the regions sensitive to the change in the isochrone parameters in the HR diagram. By using the absolute magnitude of individual stars, we calculated $T_{\rm{eff}}$ and $\log(g)$ and their precision for that star from an isochrone. By calculating the ranges of ages, metallicities, and extinctions for each open cluster, we were able to sample the joint $T_{\rm{eff}}$ and $\log(g)$ probability spaces. There was a complex relationship between $T_{\rm{eff}}$ and $\log(g)$ that differed depending on where the star was located on the HR diagram. We concluded that the shape of the $T_{\rm{eff}}$ and $\log(g)$ joint probability distribution was primarily due to the size of the ranges of ages, metallicities, and extinctions rather than the contribution of the precision of $T_{\rm{eff}}$ and $\log(g)$ from individual isochrones (i.e., equation \ref{eqn:errors isochrone individual teff}).

We used a single neural network generated using \textit{Payne} for each spectral band and trained the neural network on spectra that are representative of the population of stars in the open clusters. We fitted GALAH spectra of 1979 stars across nine open clusters with and without using the joint photometric prior. We found that using photometric prior led to a more accurate and precise $\log(g)$. This allowed us to see higher-order trends like the signature of atomic diffusion in the abundances analysed using the photometric prior. The atomic diffusion was visible in [Fe/H], [Ni/Fe], [Cu/Fe], [O/Fe], [V/Fe], [Mg/Fe], [Ti/Fe], [Sc/Fe], [Al/Fe], [V/Fe], [Ba/Fe] and [Mn/Fe] in M67 and [Fe/H], [Al/Fe], [Mg/Fe], [Ni/Fe], [Cu/Fe], and [Mn/Fe] in Ruprecht 147. However, in [Fe/H], [Al/Fe], and [Mg/Fe], the dip was only present in the abundances that were fitted using the photometric prior.

We validated the uncertainty generated by MCMC using a variety of methods. We compared the precision to the spread of abundances in M67 in a small range of $\log(g)$. We also compared our precision to the precision calculated using repeated measurements. We found that the estimate of the precision of abundances is well estimated by MCMC, but tends to inflate the precision values of $T_{\rm{eff}}$, $\log(g)$, $v_{\rm{mic}}$, and $v_{\rm{broad}} $. We investigated the precision of GALAH values calculated for DR4 by comparing it to our measures of precision. GALAH's method of picking the largest precision between the precision calculated using repeat observations and the precision calculated from its fitting procedure provides a reliable value for precision.

\section*{Acknowledgements}

This work made use of the Fourth Data Release of the GALAH survey (Buder et al. 2023, in preparation). The GALAH survey is based on data acquired through the Australian Astronomical Observatory, under programs: A/2013B/13 (GALAH pilot survey); A/2014A/25, A/2015A/19, A2017A/18 (GALAH survey, Phase 1); A2018A/18 (Open clusters with HERMES); A2019A/1 (Hierarchical star formation in Ori OB1); A2019A/15 (GALAH survey, Phase 2); A/2015B/19, A/2016A/22, A/2016B/10, A/2017B/16, A/2018B/15 (HERMESTESS programme); and A/2015A/3, A/2015B/1, A/2015B/19, A/2016A/22, A/2016B/12, A/2017A/14 (HERMES K2-follow-up programme). We acknowledge the traditional owners of the land on which the Anglo-Australian Telescope stands, the Gamilaraay people, and pay our respects to elders past and present. This paper includes data that have been provided by AAO Data Central (\href{datacentral.org.au}{datacentral.org.au}). This work has made use of the VALD database, operated at Uppsala University, the Institute of Astronomy RAS in Moscow, and the University of Vienna. KLB, JK, and GT are grateful for the financial support of the Slovenian Research Agency (research core funding No. P10188). SLM acknowledges funding support from the UNSW Scientia program and from the Australian Research Council through Discovery Project grant 220102254. This work has made use of data from the European Space Agency (ESA) mission Gaia (\href{https://www.cosmos.esa.int/gaia}{https://www.cosmos.esa.int/gaia}), processed by the Gaia Data Processing and Analysis Consortium (DPAC; \href{https://www.cosmos.esa.int/web/gaia/dpac/consortium}{https://www.cosmos.esa.int/web/gaia/dpac/consortium}). Funding for the DPAC has been provided by national institutions, in particular the institutions participating in the Gaia Multilateral Agreement.
\section*{Data Availability}

Data from this paper can be accessed upon reasonable request through private communication. Please contact Kevin Beeson at klbeeson@gmail.com for data inquiries.



\bibliographystyle{mnras}
\bibliography{Elemental_abundances_in_open_clusters_using_photometric_priors} 




\appendix
\onecolumn
\section{\textit{Gaia} DR3 ADQL quality cuts}\label{section:ADQL code}
\texttt{WHERE CONTAINS(POINT('ICRS',gaiaedr3.gaia\_source.ra,gaiaedr3.gaia\_source.dec),\\ CIRCLE('ICRS',Centre\_of\_Cluster,Radious\_around\_cluster))=1\\
          AND  (gaiaedr3.gaia\_source.visibility\_periods\_used>=7)\\
          AND pmra IS NOT NULL\\
          AND pmdec IS NOT NULL\\
          AND bp\_rp IS NOT NULL\\
          AND phot\_g\_mean\_flux\_over\_error>25\\
          AND phot\_rp\_mean\_flux\_over\_error>10\\
          AND phot\_bp\_mean\_flux\_over\_error>10\\
          AND phot\_bp\_rp\_excess\_factor < 1.3+0.06*power(phot\_bp\_mean\_mag-phot\_rp\_mean\_mag,2)\\
          AND phot\_bp\_rp\_excess\_factor > 1.0+0.015*power(phot\_bp\_mean\_mag-phot\_rp\_mean\_mag,2)\\
          AND  astrometric\_chi2\_al/(astrometric\_n\_good\_obs\_al-5)<1.44*greatest(1,exp(-0.4*(phot\_g\_mean\_mag-19.5)))\\
          AND parallax> Min\_parallax AND parallax < max\_parallax}
\section{Line list refrences}\label{appendix: line list}
The line list refrences are: \citet{DHWL,ZLLZ,LWHS,DLSC,LWCS,K09,BBSB,K07,K14,BWL,BPM,BA-J,K08,JLIY,K10,AJG,AMS,APH,AS,CB,CBcor,PQWB,WLHK,MFW,ATJL,NLLN,GNEL,B-WPNP,SLWLL,NHEL,NI,LNAJ,SPN,ILW,OL,LNWLX,IAN,IWD,NZL,NIJL,NWL,SLS,DLSSC,FW,FL,K09,LD,ABH,APR,ASa,BKM,BKM,BKP,BKV,BKPb,AJGa,AZ,BGHL,WSL,JJL,K12,XSCL,K13,RU,K11,BBPL,WLa,K99,GC,PRT,GCBHA,CFNS,KMath,WGTG,VGH,DIKH,BDMQ,FMW,MFW,NIST10,CDROM18, AZS, FFa, FFb, FTb, GUES, K, KP, LV, LZ, SCH, SLa, WSM, Wb, BWLa,K06,BLQS,JBL, MSL, BMWL, KK, Wcor,PGHcor, B, BSScor, BGHR, HLGN,RJ, BJP, T83av,KG, KSG, K75, KP, LCK, ROIG, SLa, WSM,	LN, GUES, KP, WSM,QPB,ISAN, RRKB, WTCR,RWJ,K03,K12,XSQG,GhShB, YVCHY,BMPcor, BHN, RHL,QPBM,HLGBW, PN, BBEHL,INTP, KP, MSb, WSM,K04,BLb, PGBH, SLd,ASb, AYM, BBB, BBCB, BBD, BDF, BIEMb,BM, BPS, BRO, BSMB, BW, CHY, CJ, CSD, DBG,DC, GUES, DM, DPC, DRA, EDHE, FAW, FIX,FOS, FTa, GAN, GL, GLBD, GO, HIBa, HIBb,IL, KBC, KP, LAU, LAW, LCV, LG, LGa,LI, LKIP, LN, LWa, La, Lb, M, MBO,MCa, MCb, MRB, MWRB, NORM, NSW, NSa, NSb,NSc, NUSa, NUSb, PDKI, PILK, PST, RSS, SEN,SMW, TA, TZ, Tid, WAlt, WEI, WSG, WSM, Wa, Wd,RPU,SR, S, SN, ABH, Sh, SG}

\section{Technical fitting details}
\subsection{Masking}\label{appendix: masks}

The mask is composed of three different components. The first one masks out the areas of the spectra that we would never like to fit; this includes, for example, areas around the Hydrogen Balmer lines H$\alpha$ and H$\beta$ lines. We mask these lines because our stellar atmosphere models do not accurately model them. This mask is a modified version of the same mask used in GALAH DR4. As we are using different neural networks compared to GALAH DR4 to synthesise spectra, We added extra wavelengths to mask by comparing our synthetic spectra to some high signal to noise ratio (SNR) GALAH spectra and concluded that areas with large differences in normalised flux will be areas that we cannot synthesise accurately. This can be caused by missing or bad line data (e.g., bad atomic transition values) or our atmosphere model failing to calculate accurately how the spectra will behave. The second mask bounds the area that we will always want to include in the fit, i.e., areas around important lines. These will be areas where there are strong absorption lines that are not blended, or they will be the only area to detect certain absorption lines of a specific element. The last mask is created after the first round of Bayesian fitting (the rounds of fitting are explained in the next paragraph). Using the best-fitting spectra from the first round, we take the difference between them and the observed spectra. Wherever the difference in normalised flux is above 0.3 and 5 times the difference in flux over uncertainty, we do not think that that point in the spectrum will ever converge, so we mask it out. The cut off points are set high to allow areas with strong lines to converge. If the normalised flux is above 1.05, we also mask it out, regardless of the previous criterion. This can occur, for example, if there are problems in the reduction of spectra where strong telluric emission lines or cosmic rays are not filtered out in their entirety.
\subsection{Auto-correlation}\label{appendix:auto-correlation}

Emcee uses the method outlined in \citet{goodmanEnsembleSamplersAffine2010} to calculate auto-correlation time. However when auto-correlation is calculated using short chains (chains below 50 times the auto-correlation length), the method used in emcee will underestimate the auto-correlation time. As we are using 60 walkers, the 400 independent samples are obtained long before an individual chain reaches more than 50 auto-correlation time (if it is above 50 by definition, we will have at least 3000 samples, which is far too much for our needs). By fitting an auto-regression model to our chain, we can get a more accurate auto-correlation time with much shorter chains. We can see that the auto-correlation time calculated using emcee converges when the iterations are around 50 times the auto-correlation time. The auto-regression model gives a far better estimate of the auto-correlation time using a shorter chain. Therefore, we chose to use the auto-regression model to estimate our auto-correlation time. The difference between the converged auto-correlation time estimated by emcee and the auto-regression was typically small enough for our purposes. 

\section{HR diagrams}\label{appendix: all HR diagrams}
\begin{figure*}
\includegraphics[scale=1]{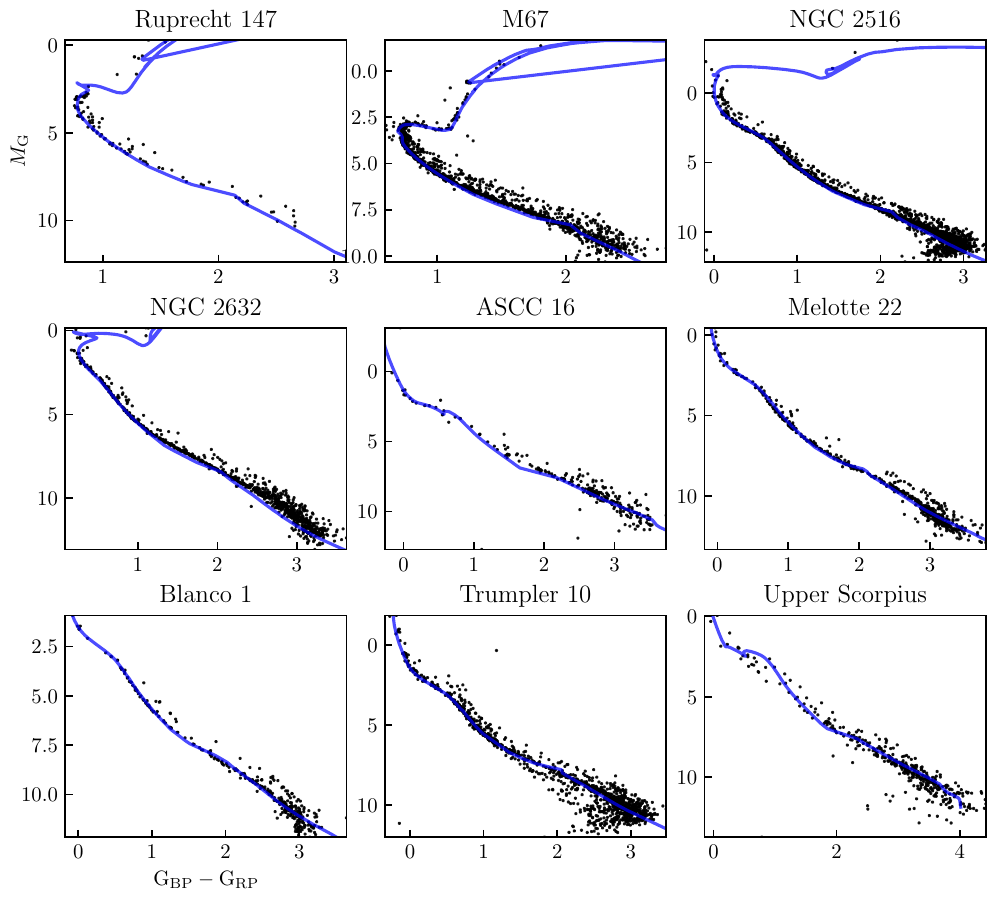}
\caption{HR diagram of all the open clusters studied in this work. The best fitting padova isochrone is shown in by the blue line.}
\end{figure*}
\section{Checks of the validity of the ranges of the chosen ages, metallicity and extinctions}\label{appendix: validity of ranges}

To check the validity of the ranges of ages, metallicities, and extinctions, three main approaches were used. The first one is to fit the Padova isochrones with the metallicity, ages, and extinctions we selected from fitting the isochrones to the HR diagrams from stars in the \textit{Gaia} catalogue to the stars in the Panstarss \citep{chambers_pan-starrs1_2019} and Allwise \citep{cutri_vizier_2021} catalogues. The approach is done to see if there is much bias when deriving metallicity, ages, and extinction from the \textit{Gaia} catalogue. Using the member stars identified using the values provided by the \textit{Gaia} catalogue and the cross-match catalogue provided by \citet{marreseGaiaDataRelease2019}; we found the corresponding stars in the ALLWISE and Pan-STARRS catalogue. The total amount of stars found in each catalogue is shown in table \ref{table:allwise and panstarrs}. The ALLWISE catalogue had 83 \% of the stars found in using the \textit{Gaia} catalogue. The stars that are missing in the ALLWISE catalogue did not correspond to a specific magnitude cut-off. The Pan-STARRS catalogue did not include NGC 2516 and Trumpler 10, excluding these clusters Pan-STARRS had 68 \% of the stars in the \textit{Gaia} catalogue. The missing stars in the Pan-STARRS catalogue are mainly stars with a magnitude higher than $\sim 12$ mag in the G band. We found that there were no multiple matches in these catalogues.

\begin{figure}
    \centering
    \includegraphics{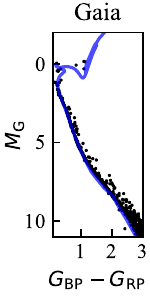}
    \includegraphics{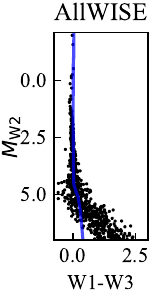}
    \includegraphics{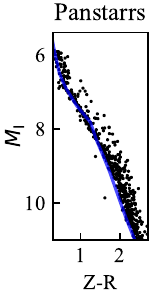}
    \caption{Isochrone fits to NGC 2632 using different photometric surveys. }
    \label{fig: comparison different catalogs}
\end{figure}

These two catalogues give us enough stars in the main sequence to see if the fit is consistent. A comparison of the fit on NGC 2632 is shown in Fig. \ref{fig: comparison different catalogs}. In the figure, we can see that the isochrone fits the HR diagram generated from the stars in the Panstarrs catalogue as well as the HR diagram generated from the stars in the \textit{Gaia} catalogue, but the HR diagram from the Allwise catalogue only contains stars in the lower part of the main sequence; these stars generally fit much worse than the rest of the main sequence. This is caused by limitations in the Padova isochrones. Overall, we found that the isochrones fitted well in the Panstarss and Allwise catalogues.

\begin{figure}\label{fig:NGC 2516 isochrone comparison}
    \centering
	\includegraphics{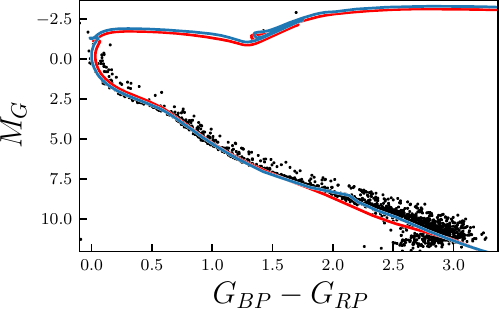}
	\caption{HR diagram of NGC 2516 showing the difference of using Padova isochrone (blue) and MIST isochrone (red). The parameters used to synthesize the isochrones are: Age$= 10^{8.30}$ $\rm{years}$, metallicity $=0.20$ dex, extinction $0.30$ mag.}
    \label{fig:Isochrone comparison}
\end{figure}

The second approach is to compare the Padova isochrones to MESA isochrones \& Stellar Tracks (MIST) \citep{choiMesaIsochronesStellar2016a,dotterMESAIsochronesStellar2016a}. When generating an isochrone of the same age, metallicity, and extinction, the MIST isochrone and Padova will differ in shape. The difference is shown in Fig. \ref{fig:Isochrone comparison}. We found that the best-fitting MIST isochrones fit worse than the Padova isochrones. Thus, we chose to use the Padova isochrones.

The third approach is to compare the fitted ages, metallicities, and extinctions to the BO19 catalogue. We have six overlapping open clusters in this catalogue (Blanco 1, Melotte 22, NGC 2516, NGC 2632, M67, and ASCC 16). All of our selected age and extinction ranges for our open clusters overlapped with BO19. The NGC 2516, NGC 2632, and M67 metallicities ranges in BO19 do not overlap with our ranges, with the largest metallicity difference being NGC 2632, which is 0.08 dex higher than BO19. Because of the complete overlap in our age and extinction ranges with BO19 and the small differences in metallicities, we are confident in using our ranges.

\begin{table}\caption{The number of members calculated using the \textit{Gaia}, ALLWISE, and Pan-STARRS catalogues for each open cluster.}
\begin{tabular}{llll}
\hline
Cluster        & $N_{\textit{Gaia}}$ & $N_{\rm{ALLWISE}}$ & $N_{\rm{Pan-STARRS}}$ \\ \hline\hline
Marianne Williamson
ASCC 16        & 247     & 232     & 225                            \\
Blanco 1       & 328     & 313     & 288                           \\
Melotte 22     & 552     & 532     & 445                           \\
NGC 2516       & 1686    & 1330    & 0                               \\
NGC 2632       & 913     & 864     & 740                           \\
M67            & 1389    & 1244    & 1342                          \\
Ruprecht 147   & 118     & 93      & 72                             \\
Trumpler 10    & 1190    & 680     & 0                               \\
Upper Scorpius & 401     & 371     & 357                          \\\hline
\end{tabular}\label{table:allwise and panstarrs}
\end{table}

\section{Photometric diagnostic plot}\label{appendix: photometric diagnostic plots}
\begin{figure}
\includegraphics[scale=1]{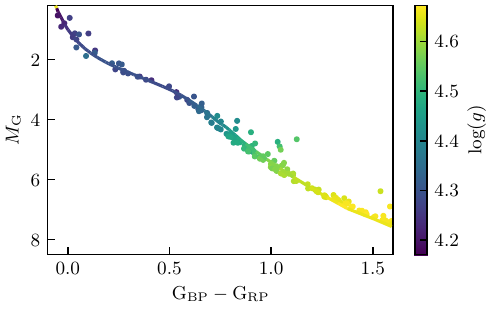}
\caption{The line is the best fitting isochrone for Melotte 22. Each point represents a star with its colour representing its calculated $\log(g)$ value. }
\label{fig:Melotte 22 logg}
\end{figure}

Fig. \ref{fig:Melotte 22 logg} is one of the diagnostic plots to show that we can get the accurate photometric $T_{\rm{eff}}$ and $\log(g)$ from isochrones. We can see that the stars $\log(g)$ follow closely the $\log(g)$ of the isochrone.

\section{Abudances comparison}\label{appendix: abundances}

Given the limited number of observed member stars in these clusters, we plotted all points in both the data set reduced using photometric priors and without that passed their own individual quality cuts.

\begin{figure*}
\includegraphics[height=0.90\textheight,width=\textwidth]{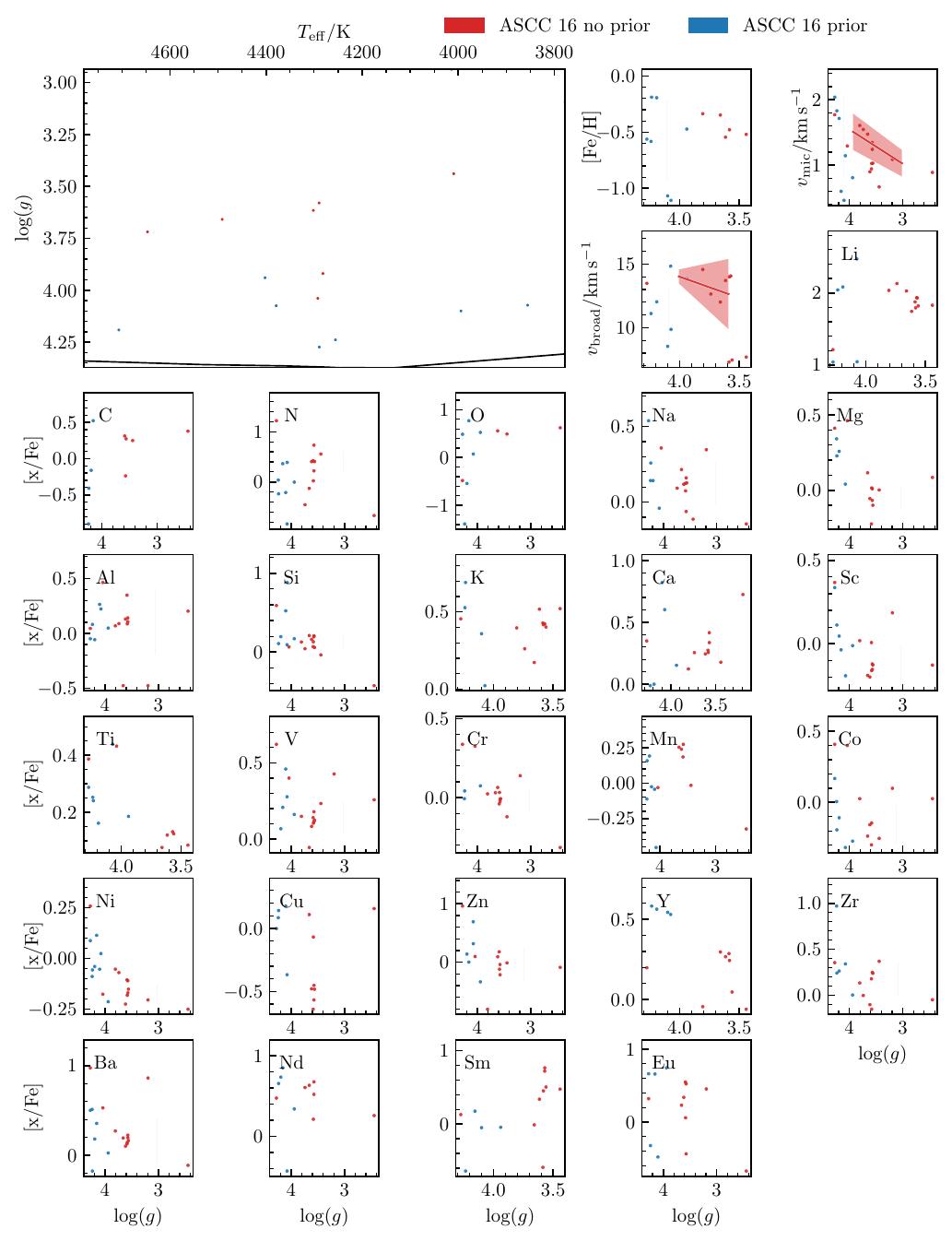}
\caption{Stellar parameters and abundance spread with respect to $\log(g)$ for ASCC 16. Blue dots are the stars that were analysed using the photometric prior and red without the prior. The Kiel diagram panel has the best fitting isochrone used in getting the photometric parameters. The solid lines are the median of a bin containing 7 stars, with the coloured area showing $\pm$ 1 $\sigma$.}
\end{figure*}
\begin{figure*}
\includegraphics[height=0.90\textheight,width=\textwidth]{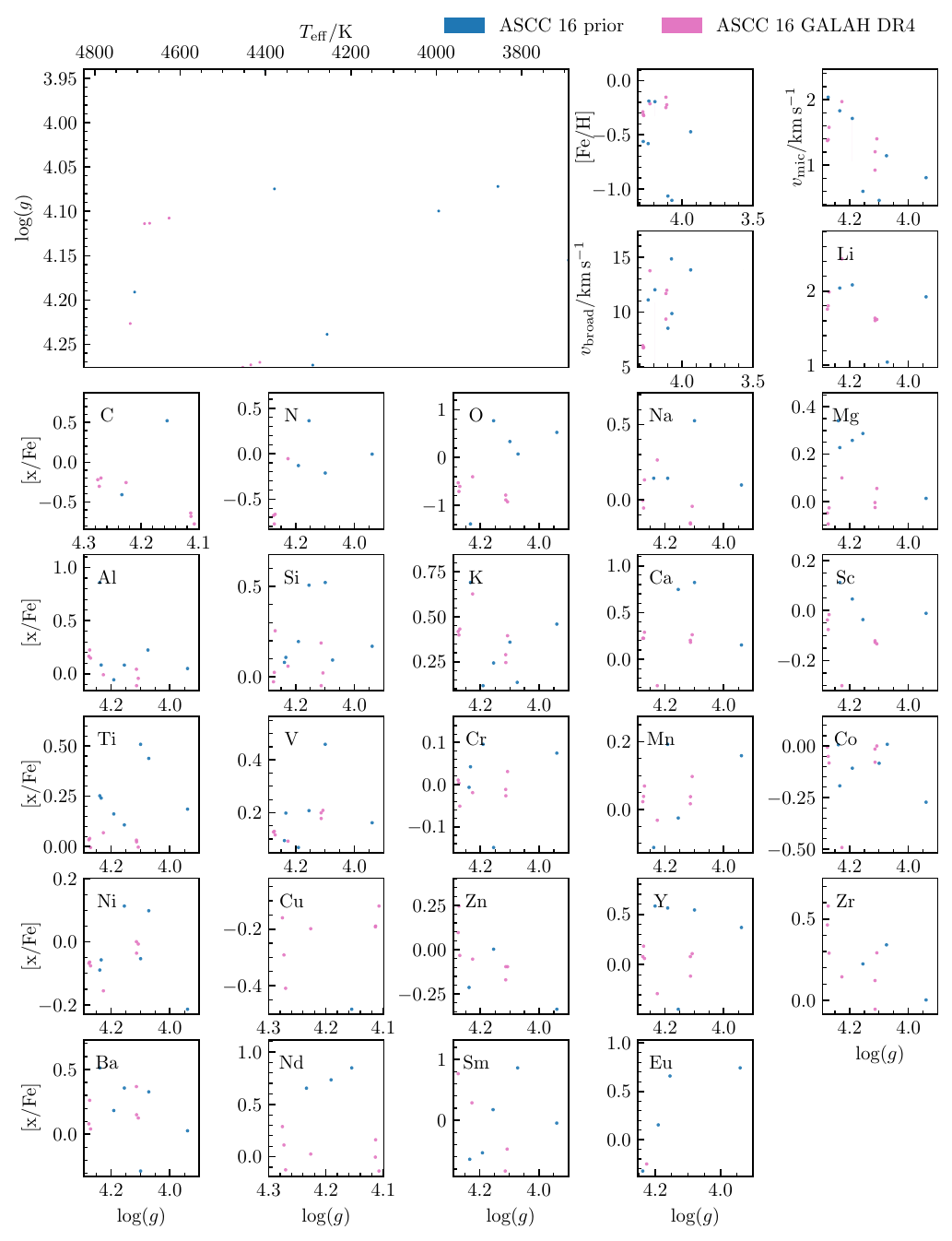}   
\caption{Stellar parameters and abundance spread with respect to $\log(g)$ for ASCC 16. Blue dots are the stars that were analysed using the photometric prior and green dots are from GALAH DR4. The Kiel diagram panel has the best fitting isochrone used in getting the photometric parameters. The solid lines are the median of a bin containing 7 stars, with the coloured area showing $\pm$ 1 $\sigma$.}
\end{figure*}
\begin{figure*}
\includegraphics[height=0.90\textheight,width=\textwidth]{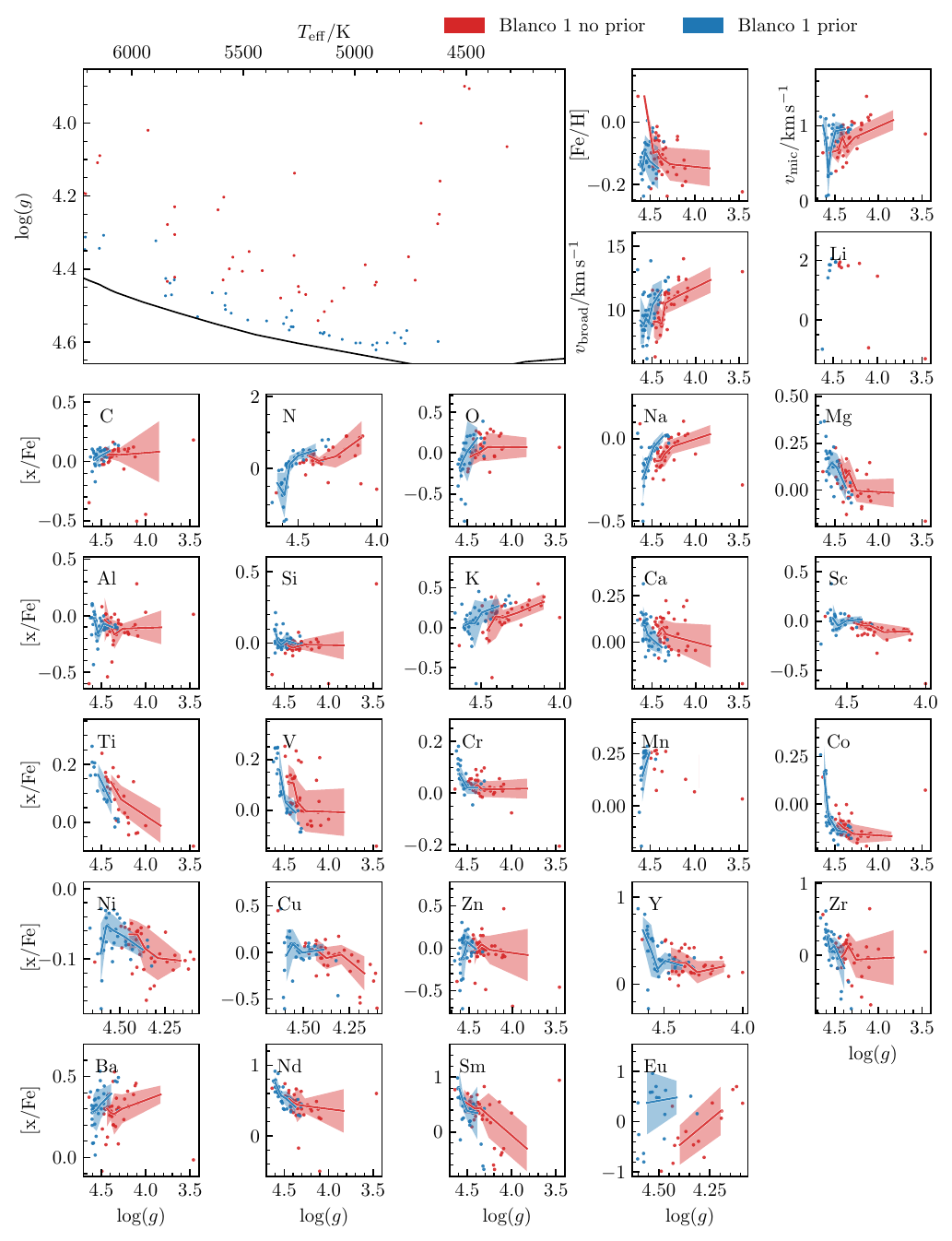}
\caption{Stellar parameters and abundance spread with respect to $\log(g)$ for Blanco 1. Blue dots are the stars that were analysed using the photometric prior and red without the prior. The Kiel diagram panel has the best fitting isochrone used in getting the photometric parameters. The solid lines are the median of a bin containing 7 stars, with the coloured area showing $\pm$ 1 $\sigma$.}
\end{figure*}
\begin{figure*}
\includegraphics[height=0.90\textheight,width=\textwidth]{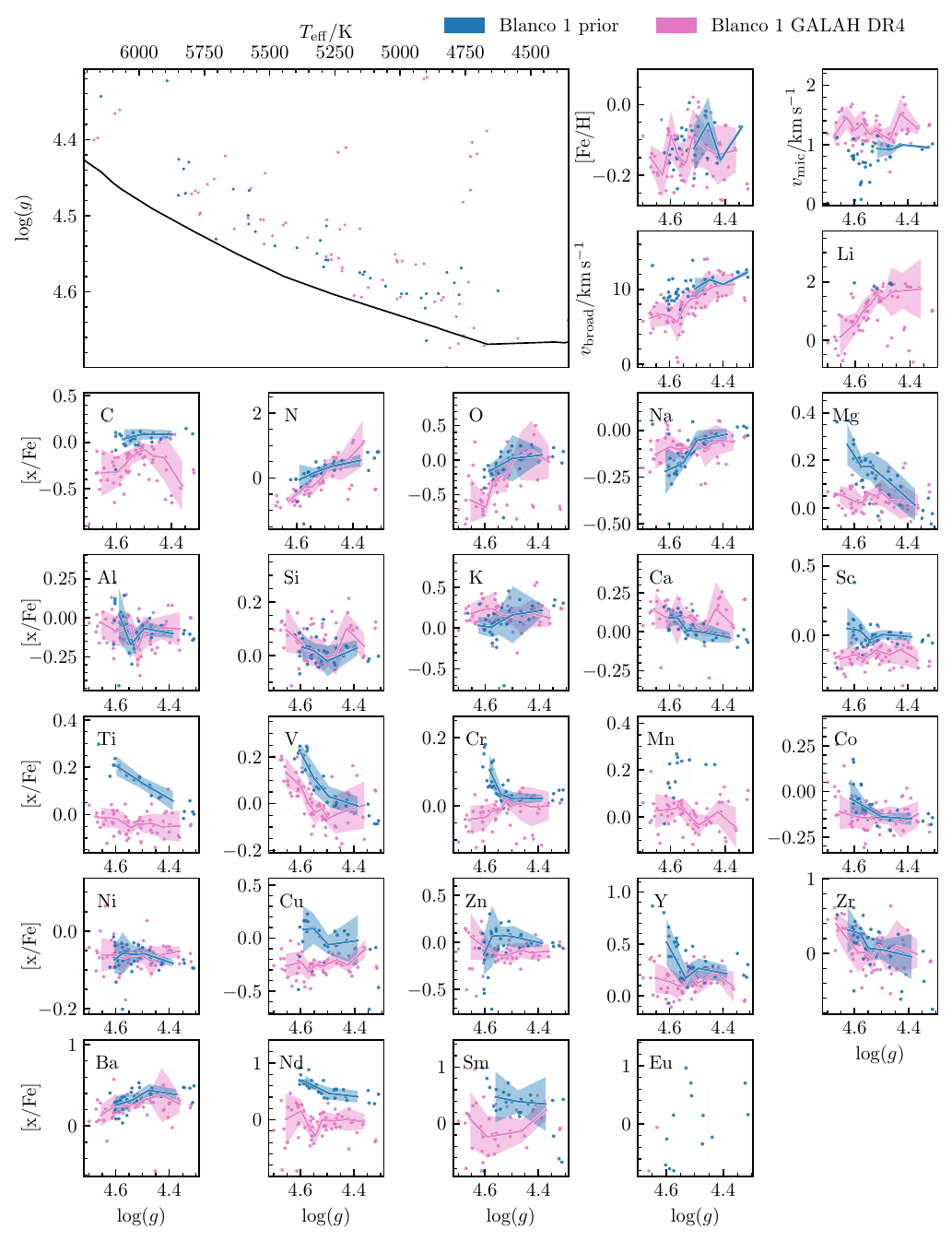}
\caption{Stellar parameters and abundance spread with respect to $\log(g)$ for Blanco 1. Blue dots are the stars that were analysed using the photometric prior and green dots are from GALAH DR4. The Kiel diagram panel has the best fitting isochrone used in getting the photometric parameters. The solid lines are the median of a bin containing 7 stars, with the coloured area showing $\pm$ 1 $\sigma$.}
\end{figure*}
\begin{figure*}
\includegraphics[height=0.90\textheight,width=\textwidth]{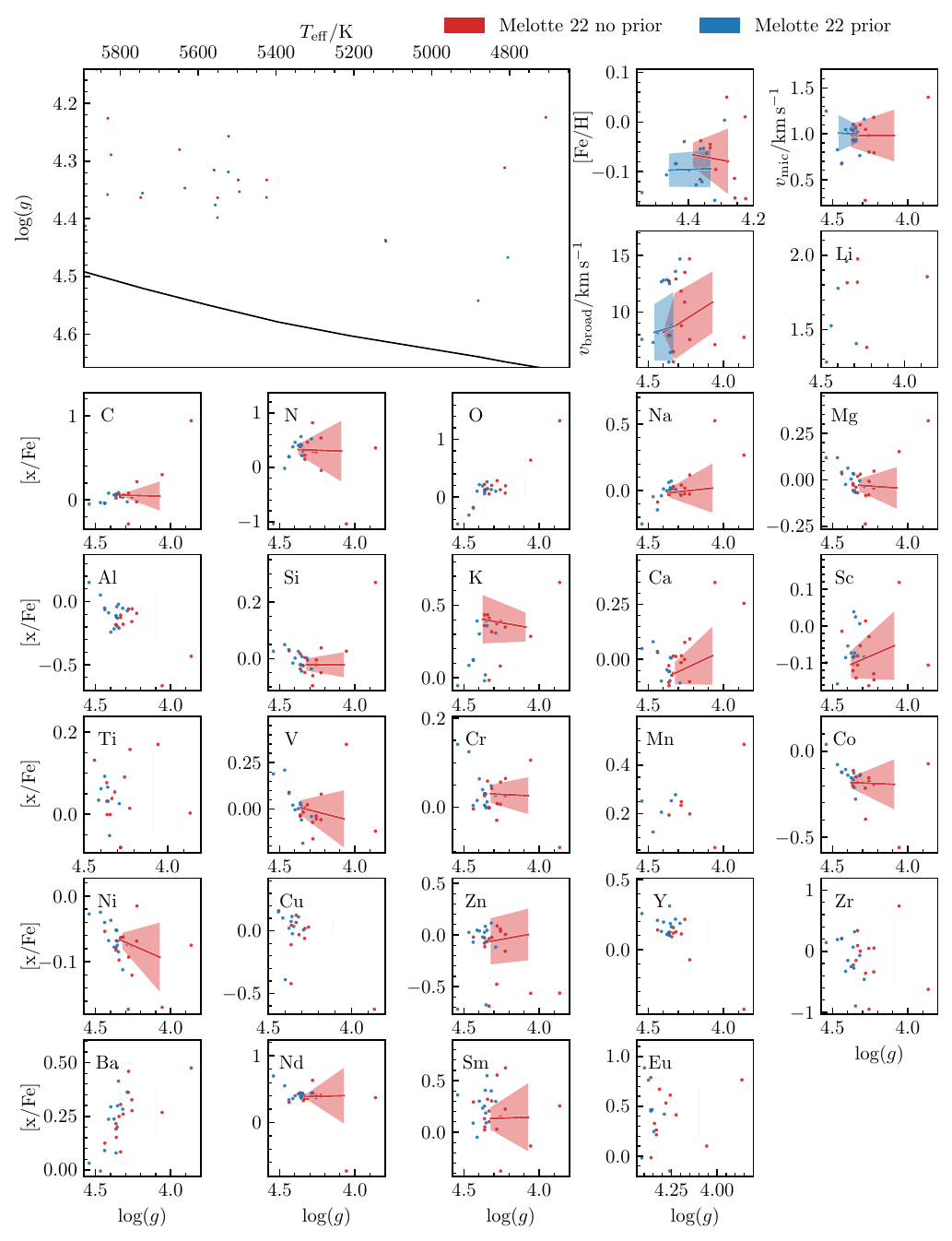}
\caption{Stellar parameters and abundance spread with respect to $\log(g)$ for Melotte 22. Blue dots are the stars that were analysed using the photometric prior and red without the prior. The Kiel diagram panel has the best fitting isochrone used in getting the photometric parameters. The solid lines are the median of a bin containing 7 stars, with the coloured area showing $\pm$ 1 $\sigma$.}
\end{figure*}
\begin{figure*}
\includegraphics[height=0.90\textheight,width=\textwidth]{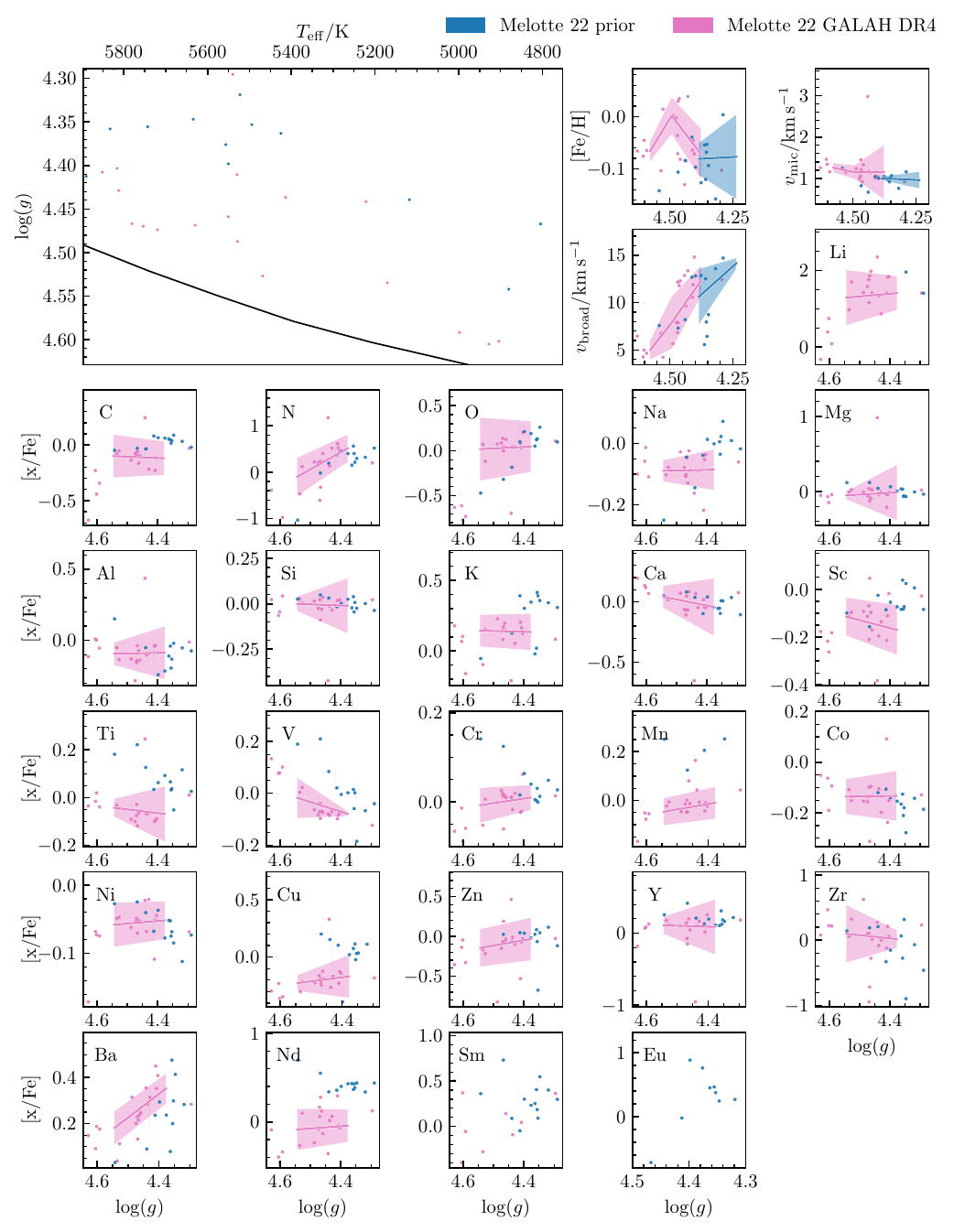}
\caption{Stellar parameters and abundance spread with respect to $\log(g)$ for Melotte 22. Blue dots are the stars that were analysed using the photometric prior and green dots are from GALAH DR4. The Kiel diagram panel has the best fitting isochrone used in getting the photometric parameters. The solid lines are the median of a bin containing 7 stars, with the coloured area showing $\pm$ 1 $\sigma$.}
\end{figure*}
\begin{figure*}
\includegraphics[height=0.90\textheight,width=\textwidth]{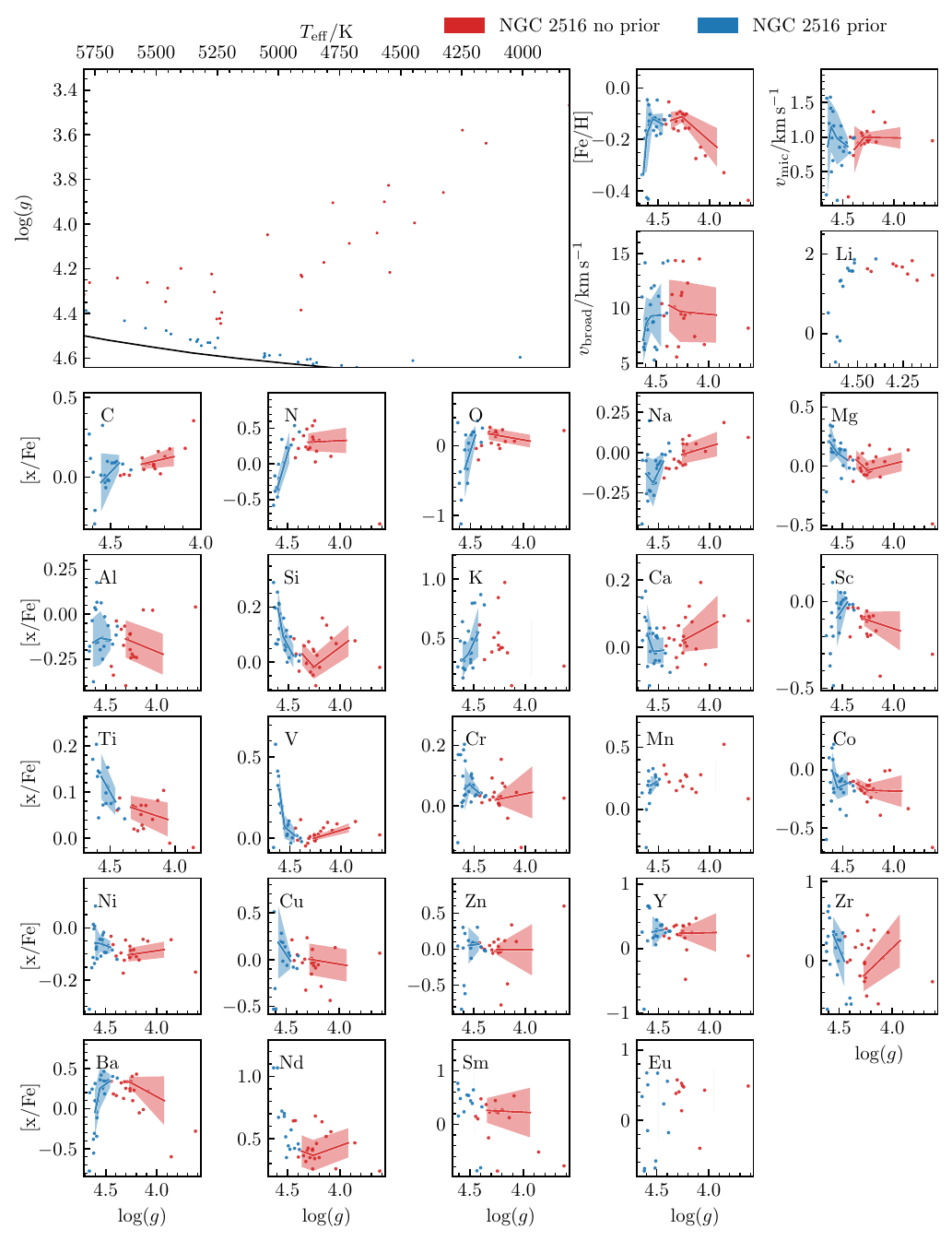}
\caption{Stellar parameters and abundance spread with respect to $\log(g)$ for NGC 2516. Blue dots are the stars that were analysed using the photometric prior and red without the prior. The Kiel diagram panel has the best fitting isochrone used in getting the photometric parameters. The solid lines are the median of a bin containing 7 stars, with the coloured area showing $\pm$ 1 $\sigma$.}
\end{figure*}
\begin{figure*}
\includegraphics[height=0.90\textheight,width=\textwidth]{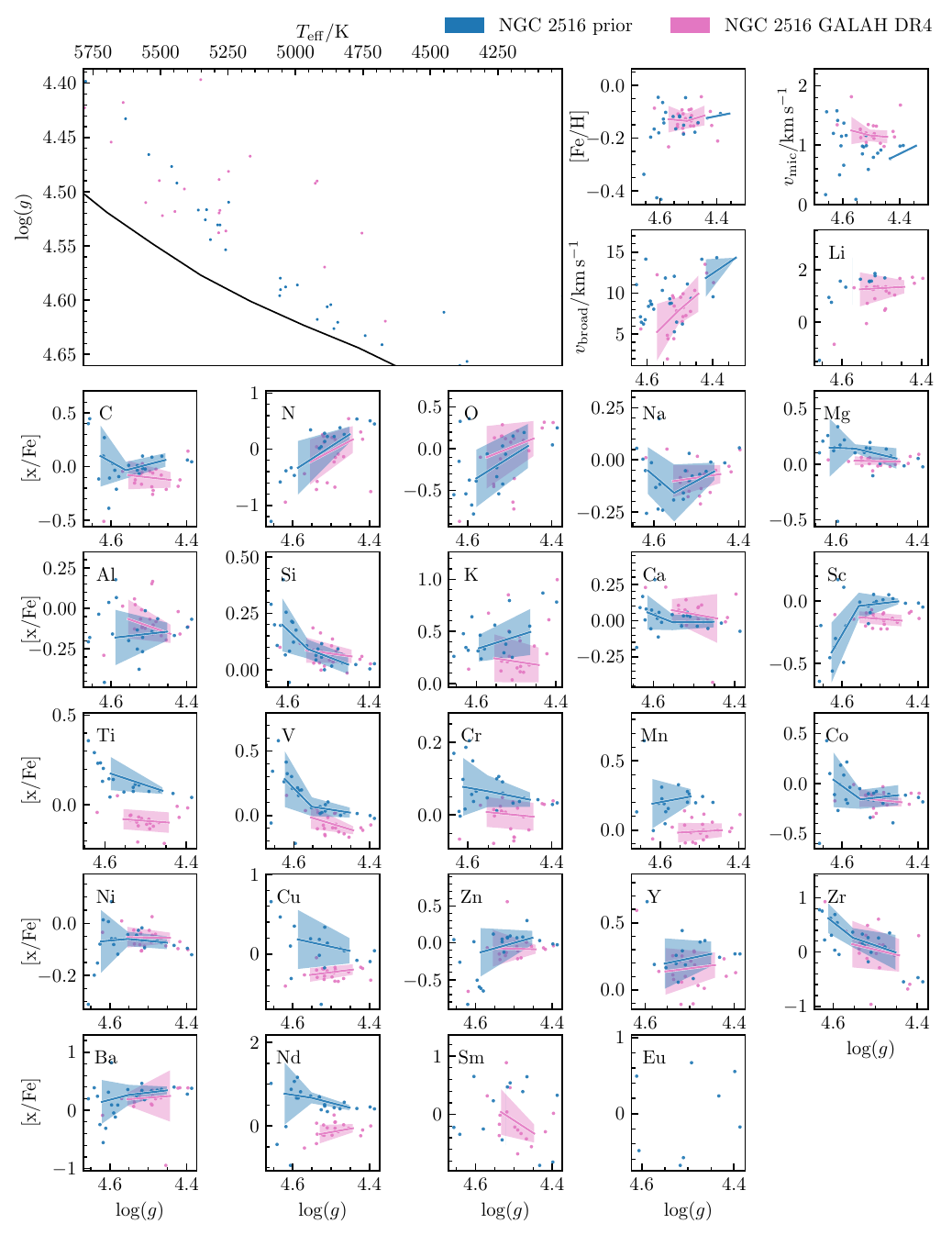}
\caption{Stellar parameters and abundance spread with respect to $\log(g)$ for NGC 2516. Blue dots are the stars that were analysed using the photometric prior and green dots are from GALAH DR4. The Kiel diagram panel has the best fitting isochrone used in getting the photometric parameters. The solid lines are the median of a bin containing 7 stars, with the coloured area showing $\pm$ 1 $\sigma$.}
\end{figure*}
\begin{figure*}
\includegraphics[height=0.90\textheight,width=\textwidth]{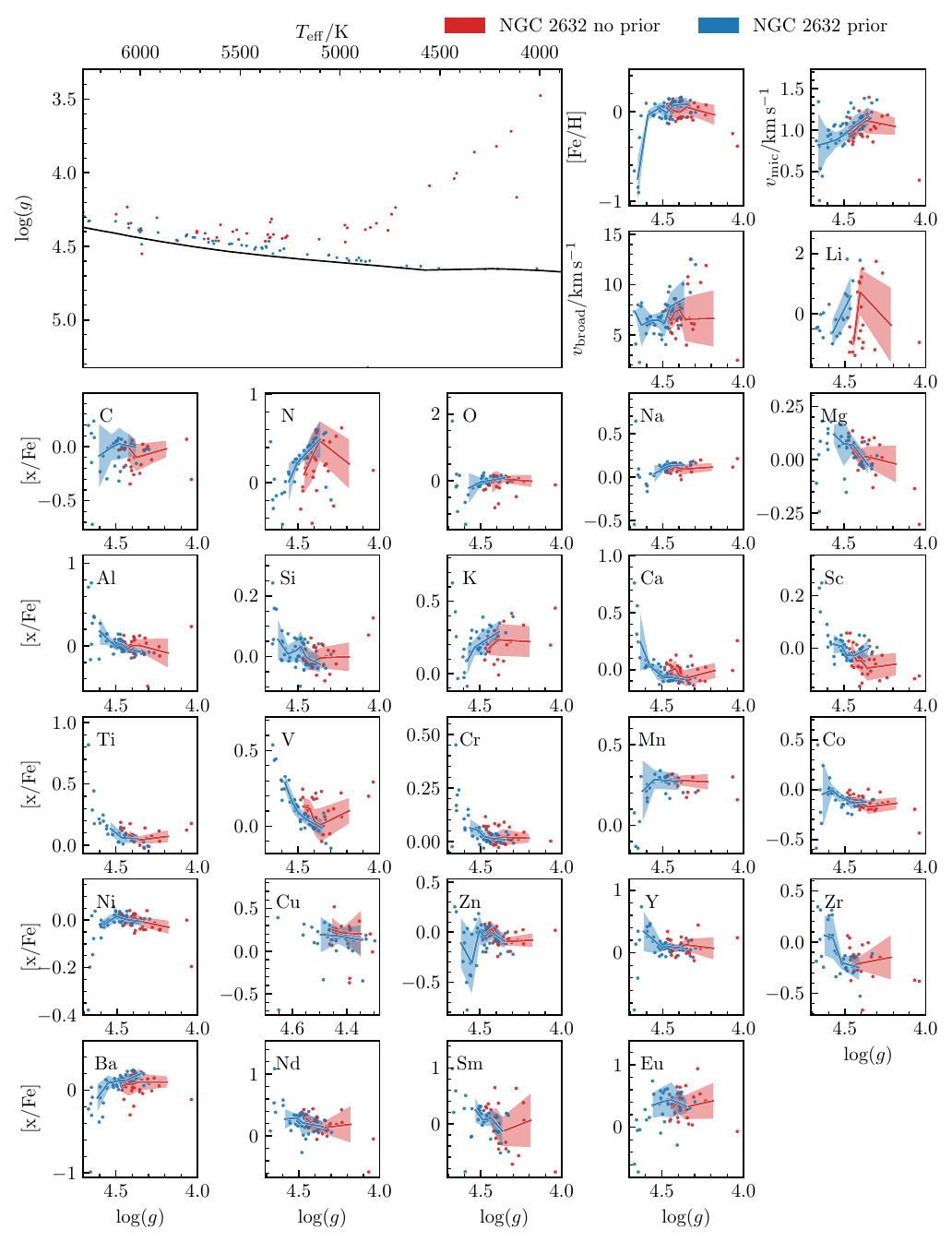}
\caption{Stellar parameters and abundance spread with respect to $\log(g)$ for NGC 2632. Blue dots are the stars that were analysed using the photometric prior and red without the prior. The Kiel diagram panel has the best fitting isochrone used in getting the photometric parameters. The solid lines are the median of a bin containing 7 stars, with the coloured area showing $\pm$ 1 $\sigma$.}
\end{figure*}
\begin{figure*}
\includegraphics[height=0.90\textheight,width=\textwidth]{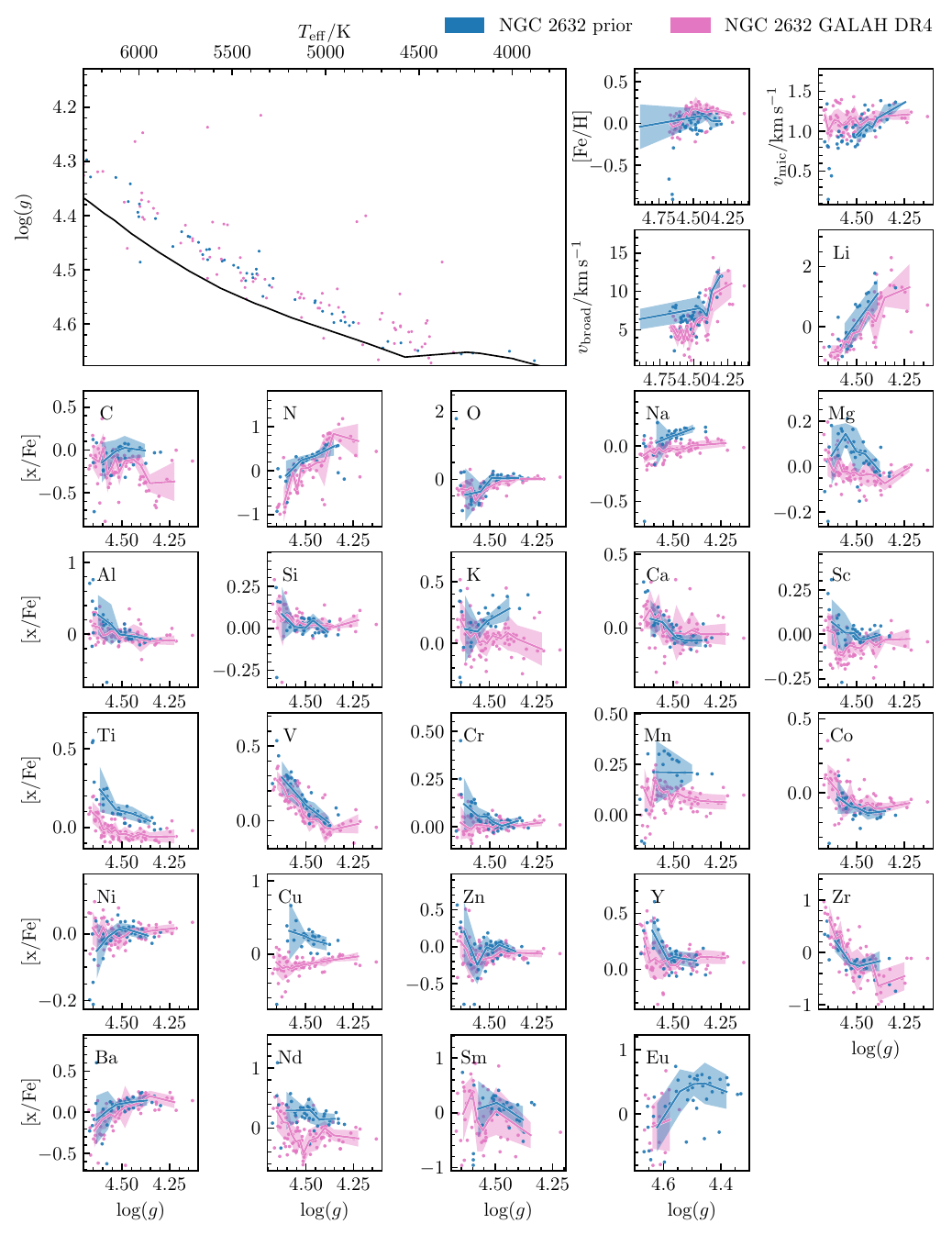}
\caption{Stellar parameters and abundance spread with respect to $\log(g)$ for NGC 2632. Blue dots are the stars that were analysed using the photometric prior and green dots are from GALAH DR4. The Kiel diagram panel has the best fitting isochrone used in getting the photometric parameters. The solid lines are the median of a bin containing 7 stars, with the coloured area showing $\pm$ 1 $\sigma$.}
\end{figure*}
\begin{figure*}
\includegraphics[height=0.90\textheight,width=\textwidth]{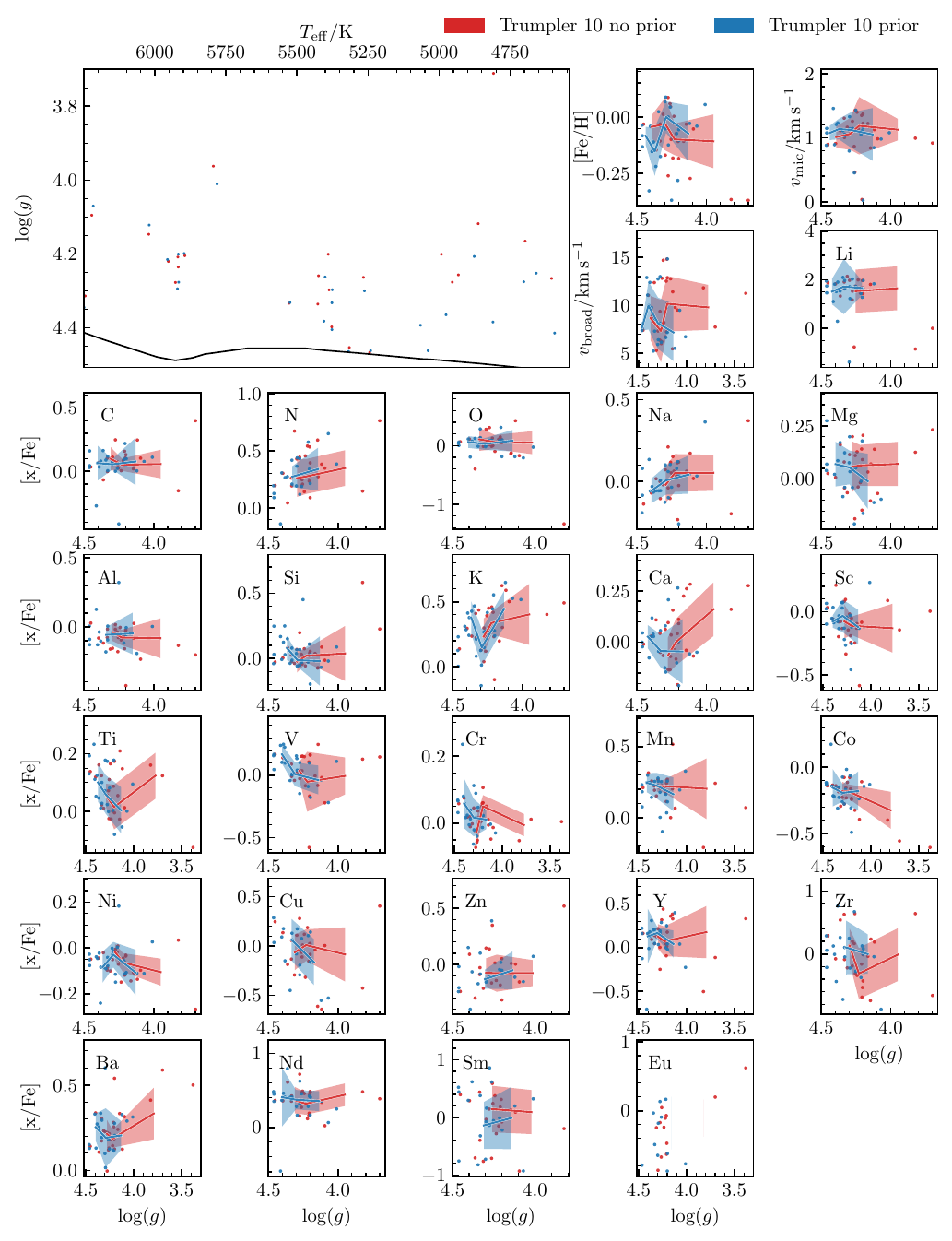}
\caption{Stellar parameters and abundance spread with respect to $\log(g)$ for Trumpler 10. Blue dots are the stars that were analysed using the photometric prior and red without the prior. The Kiel diagram panel has the best fitting isochrone used in getting the photometric parameters. The solid lines are the median of a bin containing 7 stars, with the coloured area showing $\pm$ 1 $\sigma$.}
\end{figure*}
\begin{figure*}
\includegraphics[height=0.90\textheight,width=\textwidth]{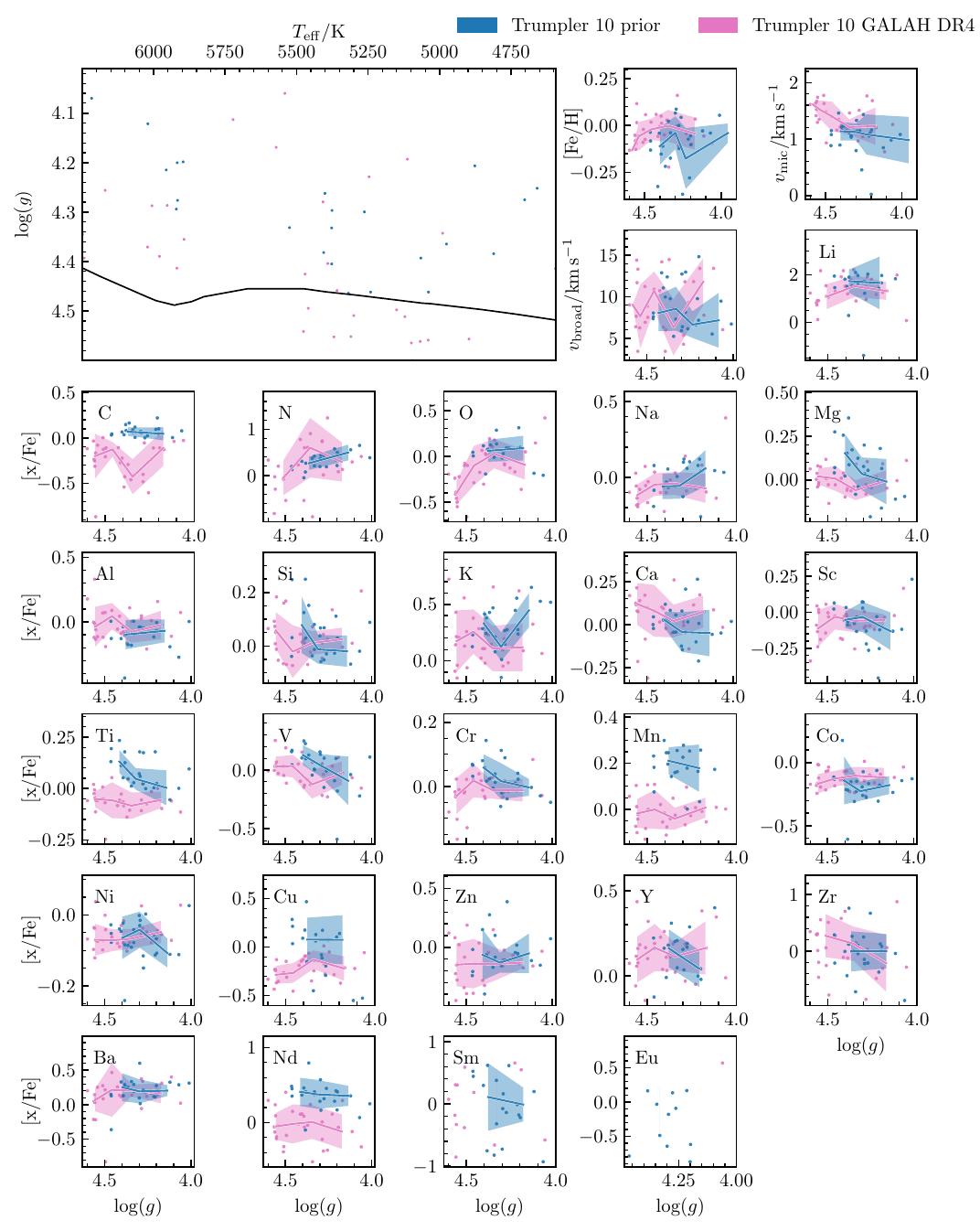}
\caption{Stellar parameters and abundance spread with respect to $\log(g)$ for Trumpler 10. Blue dots are the stars that were analysed using the photometric prior and green dots are from GALAH DR4. The Kiel diagram panel has the best fitting isochrone used in getting the photometric parameters. The solid lines are the median of a bin containing 7 stars, with the coloured area showing $\pm$ 1 $\sigma$.}
\end{figure*}


\bsp	
\label{lastpage}
\end{document}